\documentclass[pdftex,twocolumn,epjc3_preprint,runningheads]{svjour3}

\usepackage[colorlinks,citecolor=blue,urlcolor=blue,linkcolor=blue,breaklinks=true]{hyperref}
\usepackage[dvipsnames]{xcolor}
              
\pdfoutput=1

\usepackage[T1]{fontenc}
\usepackage{lmodern}
\usepackage{calc}
\usepackage{graphicx}
\usepackage{booktabs}
\usepackage{textcomp}
\usepackage{xspace}
\usepackage{relsize}
\usepackage{amssymb}
\usepackage{amsmath}
\usepackage{listings}
\usepackage{microtype}
\usepackage{multirow}
\usepackage{tabularx}
\usepackage{array}
\usepackage{placeins}
\usepackage{cuted}
\usepackage{soul} 
\usepackage{fixltx2e}
\usepackage{slashed}
\usepackage{bm}
\usepackage[numbers,sort&compress]{natbib}
\usepackage[labelfont=bf,font=small]{caption}
\usepackage[skip=-2pt]{subcaption}
\usepackage[clockwise,figuresright]{rotating}
\usepackage{tikz}
\usepackage[normalem]{ulem}
\usepackage[utf8]{inputenc}

\usepackage{etoolbox}
\AfterEndEnvironment{strip}{\leavevmode}

\allowdisplaybreaks

\newcolumntype{L}{>{\raggedright\let\newline\\\arraybackslash\hspace{0pt}}X}
\newcolumntype{R}{>{\raggedleft\let\newline\\\arraybackslash\hspace{0pt}}X}
\newcolumntype{C}{>{\centering\let\newline\\\arraybackslash\hspace{0pt}}X}

\newcommand{\gambitinstitute}[1]{\expandafter\csname #1\endcsname\label{#1}}
\newcommand{\gi}[1]{\gambitinstitute{#1}\and}
\newcommand{\last}[1]{\gambitinstitute{#1}}

\makeatletter

\newcommand{\preprintnumber}[1]{\gdef\@preprintnumber{\begin{flushright}{#1}\end{flushright}}}

\g@addto@macro\bfseries{\boldmath}
\makeatother

\let\underscore\_
\renewcommand{\_}{\discretionary{\underscore}{}{\underscore}}

\makeatletter
\let\orgdescriptionlabel\descriptionlabel
\renewcommand*{\descriptionlabel}[1]{%
  \let\orglabel\label
  \let\label\@gobble
  \phantomsection
  \protected@edef\@currentlabel{#1}%
  \let\label\orglabel
  \orgdescriptionlabel{#1}%
}
\makeatother

\newcommand\postnewlinemarker{\hbox{\ensuremath{\hookrightarrow}}}
\newcommand\cpp[1]{{\lstinline!#1!}}  
\newcommand\cpppragma[1]{{\CPPcommentstyle#1}}
\newcommand\yaml[1]{{\lstset{style=yaml}\lstinline!#1!\lstset{style=cpp}}}

\newcommand\term[1]{{\lstset{style=terminal}\lstinline!#1!\lstset{style=cpp}}}
\newcommand\termalt[1]{{\lstset{style=terminalalt}\lstinline!#1!\lstset{style=cpp}}}
\newcommand\fortran[1]{{\lstset{style=fortran}\lstinline!#1!\lstset{style=cpp}}}
\newcommand\py[1]{{\lstset{style=python}\lstinline!#1!\lstset{style=cpp}}}
\newcommand\customtilde{{\raisebox{0.2ex}{\scalebox{0.6}{\boldmath$\sim$}}}}
\newcommand\mathematica[1]{{\lstset{style=Mathematica}\lstinline!#1!\lstset{style=cpp}}}
\newcommand\guminline[1]{{{\lstset{style=gum}\lstinline!#1!}}}
\newcommand\textinline[1]{{{\lstset{style=text}\lstinline!#1!}}}

\def\be{\begin{equation}}
\def\ee{\end{equation}}
\def\ba{\begin{eqnarray}}
\def\ea{\end{eqnarray}}
\newcommand{\bea}{\begin{eqnarray}}
\newcommand{\eea}{\end{eqnarray}}

\lstnewenvironment{lstlistingyaml}{\lstset{style=yaml}}{\lstset{style=cpp}}
\lstnewenvironment{lstlistingterm}{\lstset{style=terminal}}{\lstset{style=cpp}}
\lstnewenvironment{lstlistingfortran}{\lstset{style=fortran}}{\lstset{style=cpp}}
\lstnewenvironment{lstcpp}{\lstset{style=cpp}}{\lstset{style=cpp}}
\lstnewenvironment{lstcppalt}{\lstset{style=cppalt}}{\lstset{style=cpp}}
\lstnewenvironment{lstcppnum}{\lstset{style=cppnum}}{\lstset{style=cpp}}
\lstnewenvironment{lstyaml}{\lstset{style=yaml}}{\lstset{style=cpp}}
\lstnewenvironment{lstgum}{\lstset{style=gum}}{\lstset{style=cpp}}
\lstnewenvironment{lstterm}{\lstset{style=terminal}}{\lstset{style=cpp}}
\lstnewenvironment{lsttermalt}{\lstset{style=terminalalt}}{\lstset{style=cpp}}
\lstnewenvironment{lsttext}{\lstset{style=text}}{\lstset{style=cpp}}
\lstnewenvironment{lstfortran}{\lstset{style=fortran}}{\lstset{style=cpp}}
\lstnewenvironment{lstpy}{\lstset{style=python}}{\lstset{style=cpp}}
\lstnewenvironment{lstmathematica}{\lstset{style=mathematica}}{\lstset{style=cpp}}

\newcommand{\tmpname}{}
\newcommand{\tmplistingname}{}
\makeatletter
\newif\ifATOlabelname
\lst@Key{labelname}{Listing}{\def\ATOlabelname{#1}\global\ATOlabelnametrue}
\makeatother
\lstnewenvironment{lstcpplabel}[1][]{
  \lstset{style=cpp,#1} 
  \ifATOlabelname
    \renewcommand{\tmpname}{\lstlistingname}
    \renewcommand{\tmplistingname}{\lstlistlistingname}
    \renewcommand{\lstlistingname}{\ATOlabelname}
    \renewcommand{\lstlistlistingname}{List of \lstlistingname s}
  \fi
}{
  \renewcommand{\lstlistingname}{\tmpname}
  \renewcommand{\lstlistlistingname}{\tmplistingname}
  \lstset{style=cpp}
}
\definecolor{solarized@base03}{HTML}{002B36}
\definecolor{solarized@base02}{HTML}{073642}
\definecolor{solarized@base01}{HTML}{586e75}
\definecolor{solarized@base00}{HTML}{657b83}
\definecolor{solarized@base0}{HTML}{839496}
\definecolor{solarized@base1}{HTML}{93a1a1}
\definecolor{solarized@base2}{HTML}{EEE8D5}
\definecolor{solarized@base3}{HTML}{FDF6E3}
\definecolor{solarized@yellow}{HTML}{B58900}
\definecolor{solarized@orange}{HTML}{CB4B16}
\definecolor{solarized@red}{HTML}{DC322F}
\definecolor{solarized@magenta}{HTML}{D33682}
\definecolor{solarized@violet}{HTML}{6C71C4}
\definecolor{solarized@blue}{HTML}{268BD2}
\definecolor{solarized@cyan}{HTML}{2AA198}
\definecolor{solarized@green}{HTML}{859900}
\definecolor{darkred}{HTML}{550003}
\definecolor{darkgreen}{HTML}{00AA00}
\definecolor{orchid}{HTML}{AF06F5}

\newcommand\YAMLstringstyle{\footnotesize\color{solarized@green}\mdseries}
\newcommand\YAMLkeystyle{\footnotesize\color{solarized@blue}\ttfamily}
\newcommand\YAMLvaluestyle{\footnotesize\color{blue}\mdseries}
\newcommand\ProcessThreeDashes{\llap{\color{cyan}\mdseries-{-}-}}

\newcommand\CPPcommentstyle{\color{solarized@violet}\footnotesize\ttfamily}
\newcommand\CPPdirectivestyle{\color{solarized@magenta}\footnotesize\ttfamily}
\newcommand\termplainstyle{\footnotesize\ttfamily}

\newcommand\YAMLcommentstyle{\color{solarized@orange}\ttfamily}

\newcommand\processLongMacroDelimiter
{%
\CPPdirectivestyle%
\#define%
}

\lstdefinestyle{cpp}
{
  language=C++,
  basicstyle=\footnotesize\ttfamily,
  basewidth={0.53em,0.44em}, 
  numbers=none,
  tabsize=2,
  breaklines=true,
  escapeinside={@}{@},
  showstringspaces=false,
  numberstyle=\tiny\color{solarized@base01},
  keywordstyle=\color{solarized@orange},
  stringstyle=\color{solarized@red}\ttfamily,
  identifierstyle=\color{solarized@blue},
  commentstyle=\CPPcommentstyle,
  directivestyle=\CPPdirectivestyle,
  emphstyle=\color{solarized@green},
  frame=single,
  rulecolor=\color{solarized@base2},
  rulesepcolor=\color{solarized@base2},
  literate={~} {\customtilde}1,
  moredelim=*[directive]\ \ \#,
  moredelim=*[directive]\ \ \ \ \#
}

\lstdefinestyle{cppalt}
{
  language=C++,
  basicstyle=\footnotesize\ttfamily,
  basewidth={0.53em,0.44em}, 
  numbers=none,
  tabsize=2,
  breaklines=true,
  escapeinside={*@}{@*},
  showstringspaces=false,
  numberstyle=\tiny\color{solarized@base01},
  keywordstyle=\color{solarized@orange},
  stringstyle=\color{solarized@red}\ttfamily,
  identifierstyle=\color{solarized@blue},
  commentstyle=\CPPcommentstyle,
  directivestyle=\CPPdirectivestyle,
  emphstyle=\color{solarized@green},
  frame=single,
  rulecolor=\color{solarized@base2},
  rulesepcolor=\color{solarized@base2},
  literate={~}{\customtilde}1,
  moredelim=**[is][\processLongMacroDelimiter]{BeginLongMacro}{EndLongMacro} 
}

\lstdefinestyle{cppnum}
{
  language=C++,
  basicstyle=\footnotesize\ttfamily,
  basewidth={0.53em,0.44em}, 
  numbers=none,
  tabsize=2,
  breaklines=true,
  escapeinside={@}{@},
  numberstyle=\tiny\color{solarized@base01},
  showstringspaces=false,
  keywordstyle=\color{solarized@orange},
  stringstyle=\color{solarized@red}\ttfamily,
  identifierstyle=\color{solarized@blue},
  commentstyle=\CPPcommentstyle,
  directivestyle=\CPPdirectivestyle,
  emphstyle=\color{solarized@green},
  frame=single,
  rulecolor=\color{solarized@base2},
  rulesepcolor=\color{solarized@base2},
  literate={~} {\customtilde}1,
  moredelim=*[directive]\ \ \#,
  moredelim=*[directive]\ \ \ \ \#
}

\lstdefinestyle{python}
{
  language=Python,
  basicstyle=\footnotesize\ttfamily,
  basewidth={0.53em,0.44em},
  numbers=none,
  tabsize=2,
  breaklines=true,
  escapeinside={@}{@},
  showstringspaces=false,
  numberstyle=\tiny\color{solarized@base01},
  keywordstyle=\color{blue},
  stringstyle=\color{orange}\ttfamily,
  identifierstyle=\color{darkred},
  commentstyle=\color{purple},
  emphstyle=\color{green},
  frame=single,
  rulecolor=\color{solarized@base2},
  rulesepcolor=\color{solarized@base2},
  literate = {~}{\customtilde}1
             {\ as\ }{{\color{blue}\ as\ \color{black}}}3
             {.set}{{\color{black}.}{\color{darkred}set}}4
}

\lstdefinestyle{fortran}
{
  language=Fortran,
  basicstyle=\footnotesize\ttfamily,
  basewidth={0.53em,0.44em},
  numbers=none,
  tabsize=2,
  breaklines=true,
  escapeinside={@}{@},
  showstringspaces=false,
  numberstyle=\tiny\color{solarized@base01},
  keywordstyle=\color{blue},
  stringstyle=\color{orange}\ttfamily,
  identifierstyle=\color{Periwinkle},
  commentstyle=\color{purple},
  emphstyle=\color{green},
  morekeywords={and, or, true, false},
  frame=single,
  rulecolor=\color{solarized@base2},
  rulesepcolor=\color{solarized@base2},
  literate={~}{\customtilde}1
}

\lstdefinestyle{terminal}
{
  language=bash,
  basicstyle=\termplainstyle,
  numbers=none,
  tabsize=2,
  breaklines=true,
  escapeinside={@}{@},
  frame=single,
  showstringspaces=false,
  numberstyle=\tiny\color{solarized@base01},
  keywordstyle=\color{solarized@orange},
  stringstyle=\color{solarized@red}\ttfamily,
  identifierstyle=\color{black},
  commentstyle=\color{solarized@violet},
  emphstyle=\color{solarized@green},
  frame=single,
  rulecolor=\color{solarized@base2},
  rulesepcolor=\color{solarized@base2},
  morekeywords={gambit, cmake, make, mkdir, gum, python, wget, tar, cp, pippi, mpirun},
  deletekeywords={test},
  literate = {/gambit}{{/}{\color{black}}gambit}6
             {gambit/}{{\color{black}}gambit{/}}6
             {gum/}{{\color{black}}gum{/}}4
             {/include}{{/}{\color{black}}include}8
             {cmake/}{{\color{black}}cmake/}6
             {.cmake}{{.}{\color{black}}cmake}6
             {.gum}{{.}{\color{black}}gum}6
             {.tar}{{.}{\color{black}}tar}4
             {source/}{{\color{black}}source{/}}7
             { type}{{\color{black}}{}type}5
             {~}{\customtilde}1
             {math}{{\color{solarized@orange}}math}4
}

\lstdefinestyle{terminalalt}
{
  language=bash,
  basicstyle=\footnotesize\ttfamily,
  numbers=none,
  tabsize=2,
  breaklines=true,
  escapeinside={*@}{@*},
  frame=single,
  showstringspaces=false,
  numberstyle=\tiny\color{solarized@base01},
  keywordstyle=\color{solarized@orange},
  stringstyle=\color{solarized@red}\ttfamily,
  identifierstyle=\color{black},
  commentstyle=\color{solarized@violet},
  emphstyle=\color{solarized@green},
  frame=single,
  rulecolor=\color{solarized@base2},
  rulesepcolor=\color{solarized@base2},
  morekeywords={gambit, cmake, make, mkdir},
  deletekeywords={test},
  literate = {\ gambit}{{\ }{\color{black}}gambit}7
             {/gambit}{{/}{\color{black}}gambit}6
             {gambit/}{{\color{black}}gambit{/}}6
             {/include}{{/}{\color{black}}include}8
             {cmake/}{{\color{black}}cmake/}6
             {.cmake}{{.}{\color{black}}cmake}6
             {~}{\customtilde}1
}

\lstdefinestyle{text}
{
  language={},
  basicstyle=\footnotesize\ttfamily,
  identifierstyle=\color{black},
  numbers=none,
  tabsize=2,
  breaklines=true,
  escapeinside={*@}{@*},
  showstringspaces=false,
  frame=single,
  rulecolor=\color{solarized@base2},
  rulesepcolor=\color{solarized@base2},
  literate={~}{\customtilde}1
}

\lstdefinestyle{yaml}
{
  language=bash,
  escapeinside={@}{@},
  keywords={true,false,null},
  otherkeywords={},
  keywordstyle=\color{solarized@base0}\bfseries,
  basicstyle=\footnotesize\color{black}\ttfamily,
  identifierstyle=\YAMLkeystyle,
  sensitive=false,
  commentstyle=\YAMLcommentstyle,
  morecomment=[l]{\#},
  morecomment=[s]{/*}{*/},
  stringstyle=\YAMLstringstyle\ttfamily,
  moredelim=**[s][\YAMLkeystyle]{,}{:},   
  moredelim=**[l][\YAMLvaluestyle]{:},    
  morestring=[b]',
  morestring=[b]",
  literate =    {---}{{\ProcessThreeDashes}}3
                {>}{{\textcolor{solarized@red}\textgreater}}1
                {gtr}{\textgreater}1
                {grt}{\textgreater}1
                {|}{{\textcolor{solarized@red}\textbar}}1
                {\ -\ }{{\mdseries\color{black}\ -\ \negmedspace}}3
                {\}}{{{\color{black} \}}}}1
                {\{}{{{\color{black} \{}}}1
                {[}{{{\color{black} [}}}1
                {]}{{{\color{black} ]}}}1
                {~}{\customtilde}1,
  breakindent=0pt,
  breakatwhitespace,
  columns=fullflexible
}

\lstdefinestyle{gum}
{
  language=bash,
  escapeinside={@}{@},
  keywords={true,false,null,all},
  otherkeywords={},
  keywordstyle=\color{solarized@base02}\bfseries,
  basicstyle=\footnotesize\color{black}\ttfamily,
  identifierstyle=\color{solarized@magenta},
  sensitive=false,
  commentstyle=\color{solarized@cyan}\ttfamily,
  morecomment=[l]{\#},
  morecomment=[s]{/*}{*/},
  stringstyle=\footnotesize\color{solarized@base01}\mdseries\ttfamily,
  moredelim=**[l][\footnotesize\color{solarized@base02}\mdseries]{:},    
  morestring=[b]',
  morestring=[b]",
  literate =    {---}{{\ProcessThreeDashes}}3
                {grt}{{\textcolor{solarized@magenta}\textgreater}}1
                {gtr}{{\textcolor{solarized@base02}\textgreater}}1
                {/>}{{\textcolor{solarized@magenta}\textgreater}}1
                {/<}{{\textcolor{solarized@magenta}\textless}}1
                {lss}{{\textcolor{solarized@base02}\textless}}1
                {pls}{{\textcolor{solarized@magenta}+}}1
                {mns}{{\textcolor{solarized@magenta}-}}1
                {|}{{\textcolor{solarized@base02}\textbar}}1
                {\ -\ }{{\mdseries\color{solarized@base02}\ -\ \negmedspace}}3
                {\}}{{{\color{solarized@base02} \}}}}1
                {\{}{{{\color{solarized@base02} \{}}}1
                {[}{{{\color{solarized@base02} [}}}1
                {]}{{{\color{solarized@base02} ]}}}1
                {~}{\customtilde}1,
  breakindent=0pt,
  breakatwhitespace,
  columns=fullflexible
}

\lstdefinestyle{mathematica}
{
  language={Mathematica},
  basicstyle=\footnotesize\ttfamily,
  basewidth={0.53em,0.44em},
  numbers=none,
  tabsize=2,
  breaklines=true,
  postbreak=,
  escapeinside={@}{@},
  numberstyle=\tiny\color{black},
  showstringspaces=false,
  numberstyle=\tiny\color{solarized@base01},
  keywordstyle=\color{solarized@orange},
  stringstyle=\color{solarized@red}\ttfamily,
  identifierstyle=\color{solarized@orange}\ttfamily,
  commentstyle=\color{solarized@gray}\ttfamily,
  directivestyle=\color{solarized@orange}\ttfamily,
  emphstyle=\color{solarized@green},
  frame=single,
  rulecolor=\color{solarized@base2},
  rulesepcolor=\color{solarized@base2},
  literate={~} {\customtilde}1,
  moredelim=*[directive]\ \ \#,
  moredelim=*[directive]\ \ \ \ \#,
  mathescape=false
}

\newcommand{\doublecross}[2]{\hyperref[#2]{\textbf{#1}}}
\newcommand{\doublecrosssf}[2]{\hyperref[#2]{\textbf{\textsf{#1}}}}

\newcommand{\startglossary}{\section{Glossary}\label{glossary}Here we explain some terms that have specific technical definitions in \GB.\begin{description}}
\newcommand{\finishglossary}{\end{description}}

\newcommand{\bcode}{\begin{lstlisting}}
\newcommand{\ecode}{\end{lstlisting}}
\newcommand{\metavarf}[1]{\textit{\color{darkgreen}\footnotesize\textrm{#1}}}
\newcommand{\metavars}[1]{\textit{\color{darkgreen}\scriptsize\textrm{#1}}}
\newcommand{\metavar}{\metavarf}

\newcommand{\eV}{\ensuremath{\text{e}\mspace{-0.8mu}\text{V}}\xspace}
\newcommand{\keV}{\text{k\eV}\xspace}

\newcommand{\GeV}{\text{G\eV}\xspace}
\newcommand{\TeV}{\text{T\eV}\xspace}

\newcommand{\fb}{\text{fb}\xspace}

\newcommand{\invfb}{\ensuremath{\fb^{-1}}\xspace}

\newcommand{\MSBar}{\overline{MS}}

\newcommand{\BR}{\ensuremath{\mathrm{BR}}\xspace}

\newcommand{\gambit}{\textsf{GAMBIT}\xspace}

\newcommand{\colliderbit}{\textsf{ColliderBit}\xspace}
\newcommand{\flavbit}{\textsf{FlavBit}\xspace}
\newcommand{\specbit}{\textsf{SpecBit}\xspace}
\newcommand{\decaybit}{\textsf{DecayBit}\xspace}

\newcommand{\scannerbit}{\textsf{ScannerBit}\xspace}

\newcommand{\BOSS}{\textsf{BOSS}\xspace}
\newcommand{\GB}{\gambit}

\newcommand{\buckfast}{\textsf{BuckFast}\xspace}

\newcommand{\pythiaeight}{\textsf{Pythia\,8}\xspace}

\newcommand{\rivet}{\textsf{Rivet}\xspace}
\newcommand{\yoda}{\textsf{YODA}\xspace}
\newcommand{\contur}{\textsf{Contur}\xspace}

\newcommand\FlexibleSUSY{\textsf{FlexibleSUSY}\xspace}

\newcommand\SOFTSUSY{\textsf{SOFTSUSY}\xspace}

\newcommand\HDECAY{\textsf{HDECAY}\xspace}

\newcommand\SDECAY{\textsf{SDECAY}\xspace}
\newcommand\SUSYHIT{\textsf{SUSY-HIT}\xspace}

\newcommand\SARAH{\textsf{SARAH}\xspace}

\newcommand\diver{\textsf{Diver}\xspace}

\newcommand\xx{\raisebox{0.2ex}{\smaller ++}\xspace}
\newcommand\Cpp{\textsf{C\xx}\xspace}

\newcommand\Zenodo{\textsf{Zenodo}\xspace}

\newcommand\beq{\begin{equation}}
\newcommand\eeq{\end{equation}}

\newcommand{\subparagraph}{} 
\journalname{Eur.\ Phys.\ J.\ C}
\smartqed

\makeatletter
\patchcmd{\ttlh@hang}{\parindent\z@}{\parindent\z@\leavevmode}{}{}
\patchcmd{\ttlh@hang}{\noindent}{}{}{}
\makeatother

\usepackage{siunitx}

\usepackage[title]{appendix}

\usepackage{amsmath} 
\usepackage[displaymath, mathlines,switch]{lineno} 

\newcommand{\GEWMSSM}{$\tilde{G}$-EWMSSM\xspace}

\renewcommand{\MSBar}{\ensuremath{\overline{\text{MS}}}}

\newcommand{\gravitino}{\ensuremath{\tilde G}}
\newcommand{\mg}{\ensuremath{m_{3/2}}}

\newcommand{\bfsf}[1]{\textbf{\textsf{#1}}}

\begin{document}

\preprintnumber{TTP23-009, KCL-PH-TH/2023-21, gambit-physics-23, MCnet-23-05, ADP-23-08/T1217, CERN-TH-2023-043} 

\title{Collider constraints on electroweakinos in the presence of a light gravitino}

\author{The GAMBIT Collaboration: 
Viktor Ananyev\thanksref{oslo} \and 
Csaba Bal{\'a}zs\thanksref{monash} \and \\
Ankit Beniwal\thanksref{kings} \and 
Lasse Lorentz Braseth\thanksref{oslo} \and 
Andy Buckley\thanksref{glasgow} \and \\
Jonathan Butterworth\thanksref{ucl} \and 
Christopher Chang\thanksref{uq} \and 
Matthias Danninger\thanksref{sfu} \and \\
Andrew Fowlie\thanksref{xjtlu} \and 
Tom\'{a}s E.\ Gonzalo\thanksref{kitTTP,a} \and 
Anders Kvellestad\thanksref{oslo,b} \and \\
Farvah Mahmoudi\thanksref{lyon,cern} \and 
Gregory D.\ Martinez\thanksref{ucla} \and
Markus T.\ Prim\thanksref{bonn} \and \\
Tomasz Procter\thanksref{glasgow} \and 
Are Raklev\thanksref{oslo} \and
Pat Scott\thanksref{qb} \and
Patrick St\"{o}cker\thanksref{aachen} \and \\
Jeriek Van den Abeele\thanksref{oslo,telenor} \and 
Martin White\thanksref{adelaide} \and
Yang Zhang\thanksref{zhengzhou,casbejing}
}

\institute{
    \gi{oslo}
    \gi{monash}
    \gi{kings}
    \gi{glasgow}
    \gi{ucl}
    \gi{uq}
    \gi{sfu}
    \gi{xjtlu}
    \gi{kitTTP}
    \gi{lyon}
    \gi{cern}
    \gi{ucla}
    \gi{bonn}
    \gi{qb}
    \gi{aachen}
    \gi{adelaide}
    \gi{zhengzhou}
    \gi{telenor}
    \last{casbejing}
}

\thankstext{a}{tomas.gonzalo@kit.edu}
\thankstext{b}{anders.kvellestad@fys.uio.no}

\titlerunning{Electroweakino scenarios with a light gravitino}
\authorrunning{The GAMBIT Collaboration}

\date{Received: date / Accepted: date}

\maketitle

\begin{abstract}

Using the \textsf{GAMBIT} global fitting framework, we constrain the MSSM with an eV-scale gravitino as the lightest supersymmetric particle, and the six electroweakinos (neutralinos and charginos) as the only other light new states.  We combine 15 ATLAS and 12 CMS searches at 13\,TeV, along with a large collection of ATLAS and CMS measurements of Standard Model signatures. This model, which we refer to as the $\tilde G$-EWMSSM, exhibits quite varied collider phenomenology due to its many permitted electroweakino production processes and decay modes. Characteristic $\tilde G$-EWMSSM signal events have two or more Standard Model bosons and missing energy due to the escaping gravitinos. While much of the $\tilde G$-EWMSSM parameter space is excluded, we find several viable parameter regions that predict phenomenologically rich scenarios with multiple neutralinos and charginos within the kinematic reach of the LHC during Run 3, or the High Luminosity LHC. In particular, we identify scenarios with Higgsino-dominated electroweakinos as light as 140\,GeV that are consistent with our combined set of collider searches and measurements.
The full set of $\tilde G$-EWMSSM parameter samples and \textsf{GAMBIT} input files generated for this work is available via \textsf{Zenodo}.

\end{abstract}

\tableofcontents


\section{Introduction}\label{sec:introduction}

Although supersymmetry (SUSY) was not invented to address shortcomings of the Standard Model (SM) of particle physics or cosmology, it addresses them in various aspects.  Inflation, dark matter, the cosmic matter-antimatter asymmetry, neutrino masses, patterns of fermion families, gauge and Yukawa couplings, naturalness, and more, can all be accommodated if supersymmetry is a symmetry of nature that is broken near the TeV scale; see for example Refs.~\cite{Nilles:1983ge,Haber:1984rc,Martin:1997ns,Chung:2003fi,Feng:2013pwa} for reviews.  Consequently, a major goal of the Large Hadron Collider (LHC) is to search for superpartners.  So far, the LHC experiments have found no concrete evidence for SUSY and the impact of the null results in simple SUSY scenarios has been well explored~(see e.g.\ the global fits in Refs.~\cite{Ruiz06,arXiv:1111.6098,Fowlie:2012im,Fittino12,Fittinocoverage,MastercodeCMSSM, MasterCodemSUGRA,Bagnaschi:2017tru,Costa:2017gup,CMSSM,MSSM}). For example, in our previous work~\cite{EWMSSM}, we investigated the collider constraints on the electroweakino sector of the Minimal Supersymmetric Standard Model (MSSM). Gravitinos, however, are an interesting and often ignored possibility in SUSY collider phenomenology.

The gravitino is the spin-3/2 superpartner of the spin-2 graviton. Its existence is a necessary consequence of supergravity~\cite{
nath1975generalized,deser1976consistent,freedman1976properties,arnowitt1975superfield}, a local supersymmetry that implies gravity~\cite{Akulov:1975ax,Wess:1977fn,Zumino:1979et,Stelle:1978ye,Stelle:1978yr,Ferrara:1976ni}.
The gravitino acquires mass through the super-Higgs mechanism and the mass is set solely by the scale of supersymmetry breaking; $\mg \sim \langle F \rangle / M_P$ for $F$-term supersymmetry breaking~\cite{Deser:1977uq,Cremmer:1978iv,Cremmer:1978hn} where $M_P$ is the Planck mass. In gravity-mediated supersymmetry breaking~\cite{Chamseddine:1982jx,Barbieri:1982eh,Ibanez:1982ee,Hall:1983iz,Ellis:1982wr,AlvarezGaume:1983gj,Brignole:1997wnc}, the soft-breaking masses are of order $m_\text{soft} \sim \langle F \rangle / M_P \sim \mg$, so that the gravitino can lie anywhere in the supersymmetric mass spectrum. In gauge-mediated supersymmetry breaking (GMSB)~\cite{Dine:1981za,Dimopoulos:1981au,Alvarez-Gaume:1981abe,Nappi:1982hm,Dine:1993yw,Dine:1994vc,Dine:1995ag,Kolda:1997wt}, on the other hand, the soft-breaking masses are of order $m_\text{soft} \sim \langle F \rangle / M_\text{mess}$, where $M_\text{mess}$ is the scale of the messengers mediating SUSY breaking.  Consequently, the gravitino mass is Planck-scale suppressed by $M_\text{mess} / M_P$ relative to the masses of the other superpartners. Thus, in GMSB the gravitino is expected to be the lightest supersymmetric particle (LSP). 

Motivated by GMSB, in this work we consider the electroweakino sector and an approximately massless gravitino LSP, with the other superpartners decoupled. The next-to-lightest supersymmetric particle (NLSP) must then be a neutralino or a chargino, though the latter is unusual in the MSSM parameter space~\cite{Kribs:2008hq,Bomark:2013nya}. The electroweakinos, $\tilde{\chi}^0_{1,2,3,4}$ and $\tilde{\chi}^\pm_{1,2}$, may decay to a gravitino and an SM particle. Naively, one might expect this to proceed slowly through gravitational interactions. However, as the gravitino acquires goldstino interactions through the super-Higgs mechanism~\cite{fayet1977mixing,Fayet:1979qi}, the decay may be prompt when $\mg \lesssim 1\,\keV$~\cite{Ambrosanio:1996jn}. The neutralino decays $\tilde{\chi}^0 \to \{h, H, A, Z\} \, \gravitino$ and the chargino decays $\tilde{\chi}^{\pm} \to \{H^{\pm}, W^{\pm}\} \, \gravitino$ could be kinematically allowed depending on the mass spectrum, whereas the neutralino decays $\tilde{\chi}^0 \to \gamma \, \gravitino$ are guaranteed to be allowed and dominate for the lightest neutralino, $\tilde{\chi}^0_1$, across much of parameter space~\cite{Dimopoulos:1996vz}. We thus assume that the electroweakinos may decay promptly through any kinematically open channel to an SM particle and a gravitino.  

Direct LHC production of gravitino pairs, or associated production of a gravitino and another superpartner, can only reach detectable rates if $\mg \lll 1\,\eV$~\cite{Maltoni:2015twa,Brignole:1998me}. For scenarios with electroweakinos within LHC reach and an eV scale gravitino, which is the focus of our study, the dominant gravitino production mode is through the prompt decay of the NLSP. This gives rise to distinctive collider signatures, such as two gravitinos that carry away missing energy and two energetic photons. Whilst the NLSP always decays promptly to a gravitino, an eV scale gravitino implies that the heavier electroweakinos decay predominantly to lighter ones~\cite{Ambrosanio:1996jn}, unless the mass degeneracy between the electroweakinos is severe (see below). Production of heavier electroweakinos will therefore typically result in multi-step decay chains that terminate with the decay of the NLSP to the gravitino.

The phenomenological impacts of electron-positron collider~\cite{Fayet:1986zc,Dicus:1990dy,Stump:1996wd,Dimopoulos:1996vz,Brignole:1997sk,Lopez:1996ey,Ellis:1996aa,Ambrosanio:1997rv}, Tevatron~\cite{Dicus:1989gg,Dimopoulos:1996va,Ambrosanio:1996jn,Brignole:1998me,Matchev:1999ft,Baer:1999tx,Dimopoulos:1996vz,Dimopoulos:1996fj,SUSYWorkingGroup:2000ooo,Meade:2009qv} and LHC~\cite{Kim:2017pvm,Kim:2019vcp,Dutta:2017jpe,Lu:2017oee,Gu:2020ozv,Arbey:2015vlo,Maltoni:2015twa,Asano:2011ri,Roszkowski:2004jd,Cahill-Rowley:2012ydr} searches on these scenarios have been previously studied. Reference~\cite{Kim:2017pvm}, for example, establishes limits on the electroweakino sector using light gravitino pair-production via electroweakino decay in the context of GMSB in the MSSM. This study shows that while LHC searches specifically designed for such scenarios are important, other LHC searches and measurements provide useful complementary constraints. 
Using the \gambit software~\cite{gambit,gambit_addendum}, we here go beyond previous works by performing the first global fit of electroweakinos in the presence of a light gravitino. We include up-to-date results from LHC Run 2, described in Sec.~\ref{sec:simulated_LHC_searches}, and for the first time in a global fit we check that our models are allowed by a suite of measurements of SM-like final states using \contur~\cite{Butterworth:2016sqg,Buckley:2021neu}; see Sec.~\ref{sec:contur} for further details. Lastly, we include constraints from the Large Electron-Positron collider (LEP); see Sec.~\ref{sec:lep}. We do not include Tevatron searches as these constraints are in general superseded by LHC results, and performing event simulations for Tevatron searches in addition to LHC searches would greatly increase the computational expense of our study.

Whilst a gravitino LSP could play the role of dark matter~(DM), and there are strong constraints that we do not consider~\cite{Asaka:2000zh,Feng:2004mt,Ellis:2003dn,Steffen:2006hw}, each of these requires some additional assumptions. It was originally thought that to avoid over-closing the Universe it must be that $\mg \lesssim 1\,\keV$~\cite{Pagels:1981ke}. Although this constraint is weakened when one considers inflation~\cite{Khlopov:1984pf,Ellis:1984eq}, non-thermal production of gravitinos and the NLSP decays to gravitinos are both constrained by the measured abundance of DM. There are, furthermore, constraints from cosmic structure~\cite{Viel:2005qj} and big-bang nucleosynthesis~\cite{Moroi:1993mb,Jedamzik:2009uy}, however, the latter does not apply to our scenario where the NLSP decays promptly.
We choose not to include constraints from the dark matter properties of the gravitino in this work, in order to explore electroweakinos more generally without making any limiting assumptions about cosmology.

A recent motivation for studying the possibility of light electroweakinos in this scenario is the surprising result from the CDF measurement of the $W$ boson mass~\cite{CDF:2022hxs}, which gives a value considerably above both the SM prediction and above existing experimental results. See Ref.~\cite{Workman:2022ynf} for a review of the SM value and a summary of the experimental status.  
Light electroweakinos, in particular light winos and Higgsinos, are known to result in significant positive corrections to the $W$ mass~\cite{Heinemeyer:2006px,Bagnaschi:2022qhb,Yang:2022gvz}. 
However, given the current uncertainty about the interpretation of the new result and its compatibility with other recent measurements, e.g.\ Ref.~\cite{LHCb:2021bjt}, we will not use this as a constraint on our model.

\section{Model}
\label{sec:model}

The model under consideration in this study is a variant of the MSSM where all supersymmetric states except the electroweakinos and a quasi-massless gravitino are decoupled. This model, henceforth \GEWMSSM, differs from the model in our previous study~\cite{EWMSSM} by the addition of the light gravitino. As discussed in the introduction, a very light gravitino can be motivated in certain supersymmetry breaking scenarios, e.g.\ gauge mediation. 

The general neutralino can be any linear combination of the neutral gauginos ($\tilde{B}$, $\tilde{W}^0$), and the neutral Higgsinos ($\tilde{H}^0_u$, $\tilde{H}^0_d$),
\begin{linenomath*}
\begin{align}\label{eq:neutralino}
    \tilde{\chi}_{i}^{0}=N_{i1}\tilde{B}+N_{i2}\tilde{W}^{0}+N_{i3}\tilde{H}_{d}^{0}+N_{i4}\tilde{H}_{u}^{0},
\end{align}
\end{linenomath*}
where $N_{ij}$ are the mass eigenvectors indicating the weight of each field component in the gauge basis, $(\psi^0)^{T} = (\tilde{B}, \tilde{W}^0, \tilde{H}^0_d, \tilde{H}^0_u)$. The corresponding bilinear terms in the Lagrangian density are
\begin{linenomath*}
\begin{equation}
 \mathcal{L}_{\text{$\tilde{\chi}^{0}$-mass}} = -\frac{1}{2}(\psi^0)^T M_N \psi^0 + \text{c.c.}
\end{equation}
\end{linenomath*}
where the neutralino mass matrix, $M_N$, is given by
\begin{linenomath*}
\begin{align}
 M_N &= \left(
 \begin{matrix}
  M_1 & 0 & -\tfrac{1}{2}g^\prime vc_\beta & \tfrac{1}{2}g^\prime vs_\beta \\
  0 & M_2 & \tfrac{1}{2}gvc_\beta & -\tfrac{1}{2}gvs_\beta\\
  -\tfrac{1}{2}g^\prime vc_\beta & \tfrac{1}{2}gvc_\beta  & 0 & -\mu \\
  \tfrac{1}{2}g^\prime vs_\beta & -\tfrac{1}{2}gvs_\beta & -\mu & 0
 \end{matrix} \right),
\end{align}
\end{linenomath*}
and $M_1$, $M_2$ and $\mu$ are the gaugino and Higgsino soft-breaking bilinear couplings, respectively, which are free parameters in our model. Further, we have $s_\beta = \sin\beta$ and $c_\beta = \cos\beta$, $g$ and $g^\prime$ are the $SU(2)_L$ and $U(1)_Y$ gauge couplings, and $v$ is the electroweak VEV. Amongst these, only the ratio $\tan\beta = v_u/v_d$ is not fixed by data and remains an additional free parameter in our model.

The general chargino eigenstates correspond to the charged Higgsinos ($\tilde{H}^+_u$, $\tilde{H}^-_d$), and gauginos ($\tilde{W}^+$, $\tilde{W}^-$). The corresponding bilinear terms in the Lagrangian density are
\begin{linenomath*}
\begin{equation}
 \mathcal{L}_{\text{$\tilde{\chi}^{\pm}$-mass}} = -\frac{1}{2}(\psi^\pm)^T M_C \psi^\pm + \text{c.c.}
\end{equation}
\end{linenomath*}
where the chargino mass matrix, $M_C$, is given by
\begin{linenomath*}
\begin{align}
 M_C &= \left(
 \begin{matrix}
  0 & X^T \\
  X & 0
 \end{matrix} \right), \quad \text{with}\quad  X = \left(
 \begin{matrix}
  M_2 & \tfrac{1}{\sqrt{2}} g v s_\beta \\
  \tfrac{1}{\sqrt{2}} gv c_\beta & \mu 
 \end{matrix} \right). \notag
\end{align}
\end{linenomath*} 

\begin{figure*} 
  \centering
  \includegraphics[height=0.85\columnwidth]{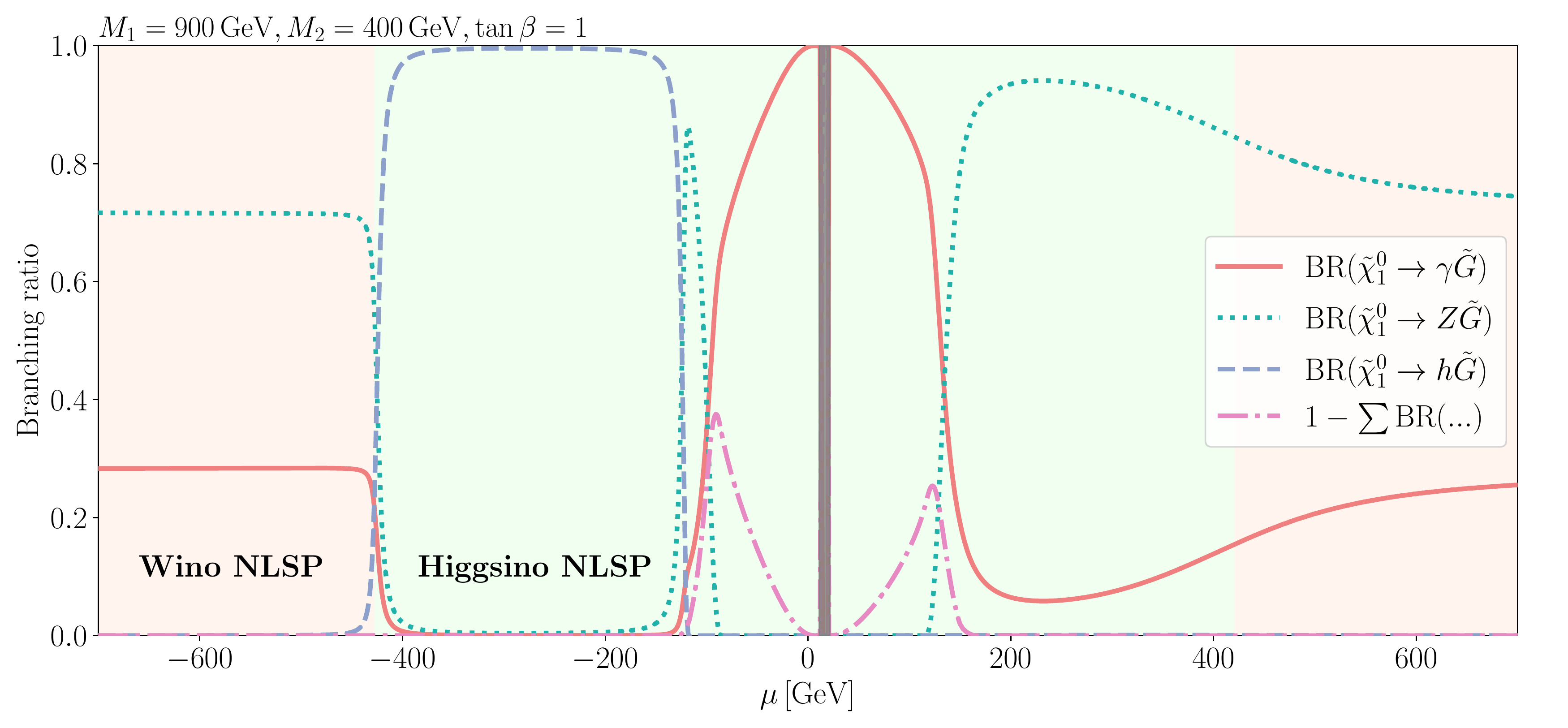}\\
  \includegraphics[height=0.85\columnwidth]{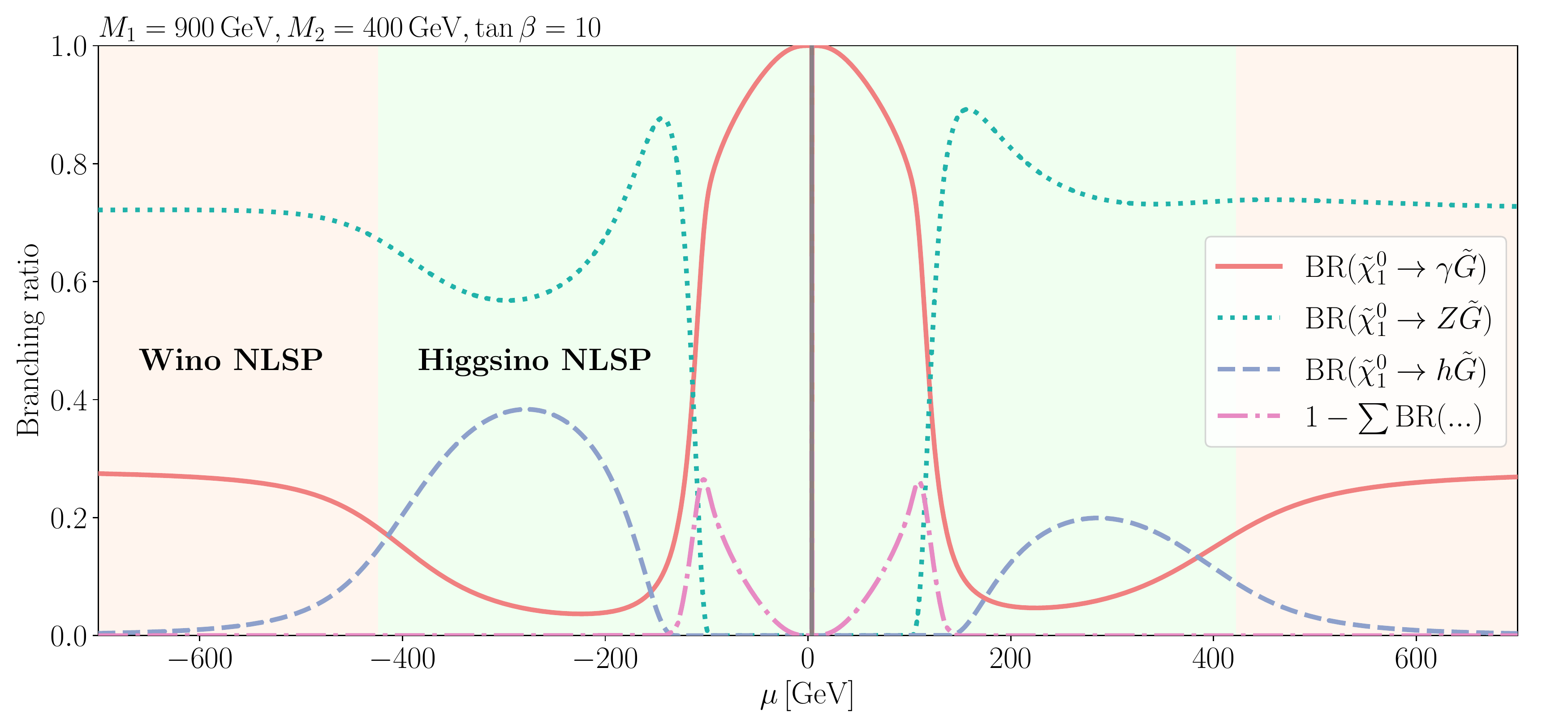}
  \caption{Branching ratios for the lightest neutralino as a function of $\mu$, with $M_1=900$\,GeV, $M_2=400$\,GeV, and $\tan\beta=1$ (top) or $\tan\beta=10$ (bottom). The wino and Higgsino NLSP regions are shown in red and green, respectively. The pink line (dash dot) shows the combined branching ratio for decays to all states other than on-shell $(Z,h,\gamma) + \tilde{G}$. The thin, grey bar marks a parameter region where $m_{\tilde\chi_1^\pm} < m_{\tilde\chi_1^0}$.} 
  \label{fig:neutralinoBR}
\end{figure*}

The gravitino mass $m_{3/2}$ depends on the dynamics of the supersymmetry  breaking, but for the purpose of our study we fix it to $m_{3/2} = 1$\,\eV, similar to what is commonly assumed in ATLAS and CMS searches, see for example Ref.~\cite{CMS:2019oou}. In terms of the collider phenomenology, this makes the gravitino effectively massless and ensures prompt decays of the NLSP. We do not set the mass to exactly zero since the limit $m_{3/2} \to 0$ corresponds to no supersymmetry breaking. The exact choice for the small gravitino mass has very little impact on the results as long as $m_{3/2} \neq 0$. The one small exception is for a wino-like chargino around the $W$ mass or lower, where the gravitino mass may dictate whether the chargino decays directly to the gravitino or via the neutralino NLSP. However, scenarios with such light charginos are in any case heavily constrained, independent of this decay.

Since we do not consider direct production of gravitinos, where the cross section would be low and the signature difficult to disentangle from backgrounds, the LHC phenomenology of this model is dominated by the production and decay of the light electroweakinos. The hierarchy of $M_1$, $M_2$ and $\mu$, and to some extent the value of $\tan\beta$, determines their gaugino and Higgsino components, production cross sections and branching ratios.

A chargino NLSP will decay promptly to the gravitino and a (possibly off-shell) $W$ boson. However, having a chargino NLSP is only possible in narrow regions of parameter space; see Fig.~\ref{fig:neutralinoBR} for an example. Throughout most of parameter space the lightest neutralino is the NLSP. In general, a neutralino NLSP has three possible decay modes: $\tilde{\chi}_{1}^{0}\rightarrow\{\gamma,Z,h\}\,\tilde{G}$. In the limit, $m_{3/2}\ll m_{\{\tilde{\chi},Z,h\}}$, the decay widths take the form~\cite{Feng:2004mt,Covi:2009bk}:
\begin{linenomath*}
\begin{align}
    \Gamma(\tilde{\chi}_{1}^{0}\rightarrow \gamma\tilde{G})&=|N_{11}c_W+N_{12}s_W|^{2}\,\mathcal{R}\,,
    \label{chitogammaG}
    \\
    \begin{split}
    \Gamma(\tilde{\chi}_{1}^{0}\rightarrow Z\tilde{G})&=\left(|-N_{11}s_W+N_{12}c_W|^{2} \right.\\
    &\hspace{0.25cm}\left.+|-N_{13}c_\beta+N_{14}s_\beta|^{2}/2\right)\\
    &\hspace{1cm}\times
    \mathcal{C}(m_{Z},m_{\tilde{\chi}_{1}^{0}})\,\mathcal{R},
    \end{split}
    \label{chitoZG}
    \\
    \begin{split}
    \Gamma(\tilde{\chi}_{1}^{0}\rightarrow h\tilde{G})&=\frac{1}{2}|-N_{13}s_\alpha+N_{14}c_\alpha|^{2}\\
    &\hspace{1cm}\times\mathcal{C}(m_{h},m_{\tilde{\chi}_{1}^{0}})\,\mathcal{R}\,.
    \end{split}
    \label{chitohG}
\end{align}
\end{linenomath*}
Here $s_W$, $c_W$, $s_\alpha$ and $c_\alpha$ are the sines and cosines of the Weinberg angle $\theta_W$ and the mixing angle $\alpha$ between the $CP$-even neutral Higgs states, and
\begin{linenomath*}
\begin{align}
    \mathcal{R}&=\frac{1}{48\pi M_{P}^{2}}\frac{m_{\tilde{\chi}_{1}^{0}}^{5}}{m_{3/2}^{2}}\,,
    \hspace{0.5cm}
    \mathcal{C}(m_{i},m_{\tilde{\chi}_{1}^{0}})&=\Bigg(1-\frac{m_{i}^{2}}{m_{\tilde{\chi}_{1}^{0}}^{2}}\Bigg)^{4}\,.\nonumber
\end{align}
\end{linenomath*}

In Fig.~\ref{fig:neutralinoBR} we show representative branching ratios for the lightest neutralino, using the full expression for the widths from Refs.~\cite{Feng:2004mt,Covi:2009bk,Hasenkamp:2009zz}, including also decay modes through off-shell bosons in the total width.
The plots use values of $\mu$ picked to illustrate the generic behaviour in the different wino NLSP (red) and Higgsino NLSP (green) regions (see below for further discussion), and two different values of $\tan\beta$, which cover the impact of $\tan\beta$ on decays to $Z$ and $h$.
The bino NLSP region (low $M_1$ values) is much simpler and not illustrated since here dominantly $\tilde\chi_1^0\to \gamma\tilde G$, with some small branching ratio to $Z\tilde G$.
We see that the dominant decay mode of the lightest neutralino depends strongly on the relative ordering of the masses $M_1$, $M_2$, and $\mu$, and the size of $\tan\beta$. 

To make our presentation more systematic, we now discuss  the properties of these three major phenomenological regions in terms of the ordering of the gaugino, $M_1$ and $M_2$, and Higgsino, $\mu$, masses.

\textit{Wino NLSP:} With $|M_2| < |M_1|,|\mu|$, the two lightest electroweakinos, $\tilde\chi_1^\pm$ and $\tilde\chi_1^0$, are a charged and neutral wino with relatively large LHC production cross sections. The lightest neutralino decays as $\tilde\chi_1^0\to \{Z,\gamma\}\,\tilde G$, see for example the wino NLSP region (red) of Fig.~\ref{fig:neutralinoBR} with $\mu>M_2$. For the lightest chargino, when $m_{\tilde\chi_1^\pm} \gg m_W$ the small mass difference between the wino-like chargino and neutralino leads to decays directly to the gravitino and an on-shell $W$, $\tilde\chi_1^\pm\to W^\pm\tilde G$. For smaller chargino masses we have instead decays to two fermions (via an off-shell $W$), together with the gravitino or lightest neutralino $\tilde\chi_1^0$.

\textit{Higgsino NLSP:} If instead $|\mu| < |M_1|, |M_2|$, the three lightest electroweakinos, $\tilde\chi_1^0$, $\tilde\chi_2^0$ and $\tilde\chi_1^\pm$, are dominantly Higgsino and have somewhat smaller production cross sections compared to the wino scenario. Pure Higgsinos do not decay to photons at tree level, so in this case the decays $\tilde\chi_1^0\to \{Z,h\}\,\tilde G$ are typically dominant, unless the NLSP mass is so small that the available phase space becomes limiting, or even that these decays go off-shell. In this case decays to photons become important again, especially at low masses, along with three-body final states with two opposite-sign SM fermions at intermediate masses. The relationship between the branching ratios to Higgs and $Z$ final states is determined by the sign of $\mu$ and the value of $\tan\beta$. In particular we note that taking $\mu < 0$ and $\tan\beta \rightarrow 1$ suppresses the $Z \gravitino$ channel, due to cancellation between the $N_{13}$ and $N_{14}$ terms in Eq.~\eqref{chitoZG}. This interplay of decays is again illustrated in Fig.~\ref{fig:neutralinoBR} in the Higgsino NLSP region (green) with $|\mu| < M_2$. The heavier neutralino and the chargino typically decay to the lightest neutralino and SM fermions in three-body decays, instead of the gravitino, due to the generically larger mass differences between the lightest electroweakinos in the Higgsino scenario~\cite{Bomark:2013nya}.

\textit{Bino NLSP:} For $|M_1| < |M_2|, |\mu|$, the NLSP is a mostly bino $\tilde\chi_1^0$ and the direct pair production cross section at the LHC is small. Most of the production is then likely to be from decays of the heavier, wino- or Higgsino-dominated electroweakinos, depending on the hierarchy of $M_2$ and $\mu$. The bino NLSP decays dominantly as $\tilde\chi_1^0\to \gamma\tilde G$.

The overall pattern that can be deduced from the above discussion is that the model predicts events with a pair of bosons picked from $\{h,Z,W,\gamma\}$, along with missing energy from the escaping gravitinos, possibly with one or both bosons being off-shell if the mass of the NLSP is below 125\,GeV. Additional bosons may also be produced from the decays of heavier electroweakinos into the NLSP. In addition to the classic signature of di-photons plus missing energy, we see that this model features events with final state SM fermions from the decays of the massive bosons, meaning that many LHC searches are relevant for the model.

Apart from the addition of the light gravitino LSP, our implementation of the \GEWMSSM model in \gambit is identical to our implementation of the EWMSSM model described in detail in Ref.~\cite{EWMSSM}. In particular, the Higgs mass, which in this study only matters for event kinematics, is set by hand to $125.09\,\GeV$. 

\section{Collider likelihoods}
\label{sec:collider likelihoods}

The total likelihood function explored in our global fit consists of likelihoods for LHC searches for new particles, LHC measurements of SM signatures, and LEP cross-section limits for electroweakino production. We describe each of these likelihoods below.
 
\subsection{LHC searches}
\label{sec:simulated_LHC_searches}

The likelihood contribution from LHC searches is based on passing simulated signal events through our emulations of the 13~\TeV ATLAS and CMS searches in Refs.~\cite{ATLAS:2021yqv,ATLAS:2020syg,ATLAS:2017drc,ATLAS:2017eoo,ATLAS:2021hza,ATLAS:2020aci,ATLAS:2019lff,ATLAS:2017avc,ATLAS:2018tti,ATLAS:2021moa,ATLAS:2021yyr,ATLAS:2019fag,ATLAS:2021ijy,ATLAS:2018nud,ATLAS:2018vzq,CMS:2019zmd,CMS:2017kyj,CMS:2017gbz,CMS:2017jrd,CMS:2018kag,CMS:2020bfa,CMS:2018xqw,CMS:2020cpy,CMS:2021cox-fix,CMS:2017brl,CMS:2019vzo,CMS:2018fon}. Reproducing a collider search to sufficient accuracy can be challenging, e.g.\ due to limited available information about technical details of the analysis, or due to limitations in the tool-chain used for fast event simulation. In some cases we can therefore only incorporate a subset of the signal regions defined by the search. In Appendix~\ref{app:LHC_searches} we provide a short description of each search, and point out which signal regions our signal simulation includes.

For all the included LHC searches we have the background uncertainty for each signal region, but in many cases there is no public information on how these uncertainties are correlated. We then take a conservative approach and, for each search, construct a likelihood function that only uses the signal region $i$ with the best expected sensitivity for the given \GEWMSSM parameter point. Our likelihood function for each of these searches is then constructed from a simple product of a Poisson and a Gaussian factor, 
\begin{linenomath*}
\begin{align}
  \label{eq:search_1SR_like}
  \begin{split}
    \mathcal{L}_{\text{search}}^\text{1SR}(s_i, \gamma_i) 
    =& \left[ \frac{(s_i + b_i + \gamma_i)^{n_i} \, e^{-(s_i + b_i + \gamma_i)}}{n_i!} \right]\\
    & \times \frac{1}{\sqrt{2\pi}\sigma_i} e^{-\frac{\gamma_i^2}{2 \sigma_i^2}} \, , 
  \end{split}
\end{align}
\end{linenomath*}
where $s_i$, $b_i$ and $n_i$ are, respectively, the expected signal yield, expected background yield and observed yield for the given signal region $i$. The Gaussian factor with the nuisance parameter $\gamma_i$ is introduced to account for the uncertainty in the total predicted yield, and we therefore set the width $\sigma_i$ by adding in quadrature the uncertainties of $s_i$ and $b_i$.
For our parameter scans we need a likelihood function that only depends on the predicted signal yield. Thus, for each sampled \GEWMSSM parameter point we profile $\mathcal{L}_{\text{search}}^\text{1SR}(s_i, \gamma_i)$ over the nuisance parameter $\gamma_i$:
\begin{linenomath*}
\begin{align}
  \mathcal{L}_{\text{search}}^\text{1SR}(s_i) \equiv \mathcal{L}_{\text{search}}^\text{1SR}(s_i, \hat{\hat{\gamma_i}}),
  \label{eq:search_1SR_like_profiled}
\end{align}
\end{linenomath*}
where $\hat{\hat{\gamma_i}}$ is the $\gamma_i$ value that maximises $\mathcal{L}_{\text{search}}^\text{1SR}(s_i,\gamma_i)$ for a given $s_i$.

CMS have for a number of their searches published covariance matrices for the background uncertainties, following the \textit{simplified likelihood} approach \cite{Collaboration:2242860,Buckley:2018vdr}. For these searches we can generalise Eq.~\eqref{eq:search_1SR_like} to a likelihood that utilises the information in all signal regions. A search with $n_\text{SR}$ signal regions is then described by the likelihood function
\begin{linenomath*}
\begin{align}
  \label{eq:search_simp_like}
  \begin{split}
    \mathcal{L}_{\text{search}}(\bm{s}, \bm{\gamma})
    =& \prod_{i=1}^{n_\text{SR}} \left[ \frac{(s_i + b_i + \gamma_i)^{n_i} \, e^{-(s_i + b_i + \gamma_i)}}{n_i!} \right]\\
    & \hphantom{\int} \times \frac{1}{\sqrt{\det2\pi\Sigma}} e^{-\frac{1}{2} \bm{\gamma}^T \bm{\Sigma^{-1}} \bm{\gamma}} \,. 
  \end{split}
\end{align}
\end{linenomath*}
Here $\bm{\Sigma}$ is the $n_\text{SR} \times n_\text{SR}$ covariance matrix for the nuisance parameters $\gamma_i$. We construct $\bm{\Sigma}$ by taking the covariance matrix provided by CMS and adding in quadrature our signal yield uncertainties along the diagonal. To obtain a likelihood that only depends on the set of signal yields $\bm{s}$ we, for each \GEWMSSM point, profile $\mathcal{L}_{\text{search}}(\bm{s}, \bm{\gamma})$ over the set of $n_\text{SR}$ nuisance parameters,
\begin{linenomath*}
\begin{align}
  \mathcal{L}_{\text{search}}(\bm{s}) \equiv \mathcal{L}_{\text{search}}(\bm{s}, \hat{\hat{\bm{\gamma}}}).
  \label{eq:search_simp_like_profiled}
\end{align}
\end{linenomath*} We also note that for the searches in Refs.~\cite{ATLAS:2020syg,ATLAS:2019lff,ATLAS:2021moa,ATLAS:2021yyr,ATLAS:2019fag} ATLAS have published the information required to fully utilise all signal regions, through the \textit{full likelihood} framework~\cite{ATLAS:2019oik}. We will make use of these likelihoods in future GAMBIT studies.

The LHC experiments often present results for multiple categories of final states in a single publication, e.g.\ the CMS multilepton search for charginos and neutralinos in Ref.~\cite{CMS:2021cox-fix}, which presents results for searches in 2-lepton, 3-lepton and 4-lepton final states. In these cases we follow the same approach as in~\cite{EWMSSM} and treat the results for the different final states as approximately independent searches, meaning that for each final state category we include a separate likelihood contribution of the form given in Eqs.~\eqref{eq:search_1SR_like_profiled} or~\eqref{eq:search_simp_like_profiled}.\footnote{A new method for identifying non-overlapping combinations of signal regions from large collections of LHC searches was recently presented in Ref.~\cite{Araz:2022vtr}. We plan to implement this method in \gambit and use it in future studies.}

Similar to the approach in Refs.~\cite{EWMSSM,DMEFT,Chang:2022jgo}, we normalise the likelihood function for each LHC search with the corresponding background-only ($\bm{s} = \bm{0}$) likelihood. The log-likelihood contribution from each search therefore takes the form of a log-likelihood difference 
\begin{linenomath*}
\begin{align}
 \label{eq:delta_lnlike}
 \Delta \ln \mathcal{L}_\text{search}(\bm{s}) = \ln \mathcal{L}_\text{search}(\bm{s}) - \ln \mathcal{L}_\text{search}(\bm{s} = \bm{0}).
\end{align}
\end{linenomath*}
Treating the searches as independent, what we consider as the combined log-likelihood from all the LHC searches is 
\begin{linenomath*}
\begin{align}
 \label{eq:delta_lnlike_sum}
 \Delta \ln \mathcal{L}_\text{searches}(\bm{s}) = \sum_{j} \Delta \ln \mathcal{L}_j(\bm{s}),
\end{align}
\end{linenomath*}
where $\Delta \ln \mathcal{L}_j$ is the log-likelihood contribution from search $j$. A positive value for the $\Delta \ln \mathcal{L}_\text{searches}(\bm{s})$ indicates that the combined set of \GEWMSSM signal predictions $\bm{s}(\bm{\theta})$ for parameter point $\bm{\theta}$ gives an overall better agreement with current LHC search results than the background-only assumption does. This happens when the predicted \GEWMSSM signals can help accommodate data excesses in some searches, without conflicting strongly with the results of the other searches.

We will present the result of our global fit as profile likelihood maps in different \GEWMSSM mass planes. For each plane we show the $1\sigma$ (68.3\%) and $2\sigma$ (95.4\%) confidence regions, derived using the likelihood ratio $\mathcal{L}(\bm{\theta}) / \mathcal{L}(\bm{\theta}_\text{best-fit})$,  where $\bm{\theta}_\text{best-fit}$ is the highest-likelihood \GEWMSSM parameter point. Therefore, if the best-fit point can explain some excesses in the search data ($\Delta \ln \mathcal{L}_\text{searches}(\bm{s}) > 0$), the \GEWMSSM parameter regions outside the $2\sigma$ contour should not be considered ``excluded'' in the same sense as for an exclusion limit from an LHC search. Rather, these parameter regions simply provide a significantly worse fit to the combined data compared to that of the best-fit point. It is then interesting to also ask a different question: What \GEWMSSM parameter regions are excluded by the combination of LHC searches, when judged relative to the background-only expectation? A simple way to estimate this is to replace $\Delta \ln \mathcal{L}_\text{searches}(\bm{s})$ in Eq.\ (\ref{eq:delta_lnlike_sum}) with
\begin{linenomath*}
\begin{align}
 \label{eq:capped_lnlike}
 \begin{split}
 \Delta &\ln \mathcal{L}_\text{searches}^{\text{cap}}(\bm{s}) \\
&= \min \left[\Delta \ln \mathcal{L}_\text{searches}(\bm{s}), \Delta \ln \mathcal{L}_\text{searches}(\bm{s} = \bm{0}) \right] \\
 &= \min \left[\Delta \ln \mathcal{L}_\text{searches}(\bm{s}), 0 \right].
 \end{split}
\end{align}
\end{linenomath*}
This log-likelihood penalises \GEWMSSM parameter points that give a joint prediction in worse agreement with data than the background-only prediction, while all other points are assigned the same log-likelihood of 0. We note that the maximum value 
$\Delta \ln \mathcal{L}_\text{searches}^{\text{cap}}(\bm{s}) = 0$ can be obtained in two different ways: The first case is when none of the included searches are sensitive to the given \GEWMSSM parameter point, i.e.\ the limit $\bm{s} \rightarrow \bm{0}$. This is typically what happens for high-mass scenarios, due to small production cross-sections. The second case is when a \GEWMSSM scenario fits the results from some LHC searches sufficiently better than the SM does, enough to offset any likelihood penalty from tensions with other LHC analyses. In Sec.\ \ref{sec:results} we will present results both for the ``full likelihood'' ($\mathcal{L}_\text{searches}$) case and the ``capped likelihood'' ($\mathcal{L}_\text{searches}^{\text{cap}}$) case. This is the same approach as was taken in Refs.~\cite{EWMSSM,DMEFT,Chang:2022jgo}.

\subsection{LHC measurements of SM signatures}\label{sec:contur}

The complexity of the phenomenology of the model means that the possibility that it may produce events which could contribute to
well-measured SM-like final states must also be taken into account. This is the scenario for which
\contur~\cite{Butterworth:2016sqg,Buckley:2021neu} is designed. Via
\rivet~\cite{Bierlich:2019rhm}, \contur has access to an extensive library of measurements from the LHC experiments, mostly corrected for detector effects and thus not requiring explicit detector simulation.
Simulated events are passed through \rivet and projected into the fiducial phase space of the measured cross sections. In the release of \gambit accompanying this paper, we have interfaced \contur and \rivet to the \gambit \colliderbit module.

As binned unfolding of detector effects requires statistically stable bin populations, a $\chi^2$ test has proven indistinguishable from Poisson log-likelihood differences for measurement interpretations. The $\chi^2$ is evaluated and used as the log-likelihood difference between the ``signal-injection'' hypothesis and the SM null hypothesis, in this application assuming the data to be equal to the SM:
\begin{linenomath*}
\begin{align}
 \label{eq:contur_lnlike}
 \begin{split}
 \ln \mathcal{L}_\text{meas}(\bm{s}) 
 &= -\chi^2(\bm{s})/2\\
 &\equiv - \sum_{i \,\in\, \text{active bins}} \left[
 \frac{y_i^\text{s+b}(\bm{s}) - y_i^\text{obs}}{(\Delta y_i)}
 \right]^2 \Big/ 2
 \, ,
 \end{split}
\end{align}
\end{linenomath*}
with the log-likelihood difference then evaluated as 
$\Delta \ln \mathcal{L}_\text{meas}(\bm{s}) = \ln \mathcal{L}_\text{meas}(\bm{s}) - \ln \mathcal{L}_\text{meas}(\bm{s} = 0)$. 
The set of active bins is conservatively selected to avoid acceptance overlaps, as described in Sec.~\ref{sec:software_framework}, and $y_i$ and $\Delta{y_i}$ are the bin values and uncertainties respectively.
The experimental uncertainties are taken into account in the $\chi^2$ construction, but are treated as uncorrelated in the version of \contur (2.3.0) used here.

The set of 13~\TeV analyses used by \contur in this analysis are those described in Refs.~\cite{CMS:2016oae,ATLAS:2017xqp,CMS:2018htd,ATLAS:2016zkp,ATLAS:2017txd,ATLAS:2018sos,ATLAS:2019hau,ATLAS:2021kog,ATLAS:2016zba,CMS:2018dxg,ATLAS:2019rqw,CMS:2021hnp,ATLAS:2017cez,CMS:2018vzn,ATLAS:2018orx,ATLAS:2019ebv,CMS:2018mdf,ATLAS:2019kwg,ATLAS:2017sag,CMS:2019raw,ATLAS:2019zci,ATLAS:2018acq,ATLAS:2020juj,ATLAS:2020nzk,ATLAS:2017zda,ATLAS:2019qet,ATLAS:2019cbr,CMS:2018tdx,ATLAS:2021jgw,CMS:2021lxi,ATLAS:2019hxz,ATLAS:2019rob,ATLAS:2018fwl,CMS:2018mdl,LHCb:2018usb,ATLAS:2020ccu,ATLAS:2021mbt,ATLAS:2020vup,CMS:2016jip,CMS:2019fak,ATLAS:2017bcd,CMS:2020mxy,CMS:2019jjp,CMS:2019eih,ATLAS:2017ble,ATLAS:2020bbn,ATLAS:2019gey,ATLAS:2018nci}. These
cover final states with (multiple) jets, isolated photons and leptons, as well as missing energy. When discussing our results in Sec.\ \ref{sec:results} we will highlight the analyses with the greatest impact.

\subsection{Cross-section limits from LEP searches}\label{sec:lep}

In addition to the above LHC searches and measurements that are implemented at the event level, we include LEP searches and measurements that were published as upper limits on particular electroweakino production cross-sections. See~\cite{ColliderBit,EWMSSM} for general details of our treatment of LEP searches. First, there are searches for electroweakinos that we applied in \cite{EWMSSM} that we re-interpret as searches for gravitinos. Specifically, we consider searches for pair production of charginos that each decay into SM particles and a stable neutralino, $\tilde\chi^\pm \to \text{SM} + \chi$. In our gravitino model, the chargino may decay into SM particles and a gravitino, $\tilde\chi^\pm \to \text{SM} + \gravitino$. This leads to an identical signature as both the gravitino and a stable neutralino only contribute to missing energy.

Second, we include a multi-photon and missing energy search by L3 at $\sqrt{s} = 207\,\GeV$~\cite{L3:2003yon}. In our model, neutralinos can be pair produced at LEP and can each decay to a photon and a gravitino, giving a signature of missing energy and two photons. The number of observed events in the search was less than expected from SM backgrounds, leading to strong constraints on the $e^+ e^- \to \tilde\chi^0_1 \tilde\chi^0_1 \to \gravitino \gravitino \gamma \gamma$ cross section as a function of the gravitino and neutralino masses for masses less than about $\sqrt{s} / 2$. We apply the $95\%$ limits shown in Fig.~6c of \cite{L3:2003yon} following the treatment described in \cite{ColliderBit}. The impact of this constraint on our \GEWMSSM model is limited, however, as our assumption of decoupled selectrons typically leads to a very small $e^+ e^- \to \tilde\chi^0_1 \tilde\chi^0_1$ production cross-section.

\section{Global fit setup}
\label{sec:global_fit_setup}

\subsection{Software framework and event generation}
\label{sec:software_framework}

We perform our study of the \GEWMSSM with the \GB~\textsf{2.4} global fit framework~\cite{gambit,GUM}, utilising the \specbit, \decaybit, \colliderbit and \scannerbit modules~\cite{SDPBit,ColliderBit,ScannerBit}.
To compute the chargino and neutralino mass spectrum at one-loop level, \specbit employs a \FlexibleSUSY \cite{Athron:2014yba,Athron:2017fvs} spectrum generator which uses \SARAH \cite{Staub:2008uz,Staub:2010jh} and routines from \SOFTSUSY~\cite{Allanach:2001kg,Allanach:2013kza}. A more detailed discussion of this spectrum computation is given in~\cite{EWMSSM}.

For this study we have extended \decaybit with the capability to compute decay widths for a neutralino or chargino decaying to final states with a gravitino. The implementation is based on analytical expressions given in Refs.~\cite{Feng:2004mt,Covi:2009bk,Hasenkamp:2009zz}. To compute neutralino and chargino decays into final states with a lighter neutralino or chargino, \decaybit uses \SUSYHIT \textsf{1.5} \cite{Djouadi:2006bz}, which includes the packages \SDECAY \cite{Muhlleitner:2003vg} and \HDECAY \cite{Djouadi:1997yw}.

We simulate LHC events with electroweakino production at $\sqrt{s}=13\,\TeV$ using \colliderbit's parallelised interface to \pythiaeight \cite{Sjostrand:2006za,Sjostrand:2014zea} and native fast detector simulator \buckfast~\cite{ColliderBit}.\footnote{To avoid the additional computational cost of simulating light electroweakino production through decays of SM bosons, we do not consider parameter points with electroweakino masses below $62.5\,\GeV$.} Due to the cost of computing higher-order production cross-sections, we use the cross-sections computed by \pythiaeight at leading-order plus leading-logarithmic (LO+LL) accuracy. As we will see in Sec.~\ref{sec:results}, the lowest-mass scenarios not disfavoured by current results are scenarios where the lightest electroweakinos are Higgsinos with masses around $200\,\GeV$. For such scenarios the production cross-sections at next-to-leading order with next-to-leading-logarithmic corrections (NLO+NLL) can be up to 30\% higher compared the LO+LL cross-sections \cite{Fiaschi:2018hgm}, so this choice is somewhat conservative.

For each parameter point included in our final scan results we generate 16 million LHC events to evaluate the impact of the LHC searches. The main reason that such a high number of events is needed is that for many of the searches we do not have the information needed to allow a proper statistical combination of all the signal regions in the search. As discussed in Sec.~\ref{sec:simulated_LHC_searches}, for these searches the conservative approach is, for each sampled parameter point, to identify the signal region with the best expected sensitivity, and only use this signal region when computing the likelihood contribution from the given search. Many searches will for large parts of the \GEWMSSM parameter space have several signal regions with low and near identical expected sensitivities. Thus, the signal region choice, and through it the likelihood value, becomes highly sensitive to Monte Carlo noise.\footnote{The computational cost of overcoming this problem, also discussed in Refs.~\cite{EWMSSM,Cranmer:2021urp}, is currently a major limiting factor for the proper utilisation of LHC results through full MC simulations in BSM global fits. The severity of the problem is reduced with every new LHC search that is published with enough information to enable a statistical combination of the different signal regions, e.g.\ through the \textit{simplified likelihood}~\cite{Collaboration:2242860,Buckley:2018vdr} or \textit{full likelihood}~\cite{ATLAS:2019oik} approaches.}

As a post-processing step, we generate a further \num{100000} events at each sampled parameter point, which are then passed to first \rivet and then \contur using the new \colliderbit interface. This enables evaluation of whether the parameter point in question would have led to significant but unnoticed collective deviations from the SM expectation in existing measurements. 
Since LHC measurements have much higher acceptances than LHC searches, we here need fewer simulated events to ensure sufficiently small Monte Carlo uncertainties and a stable identification of the most sensitive measurements.
\contur tests the full set of measurements for each parameter point, evaluating the expected likelihood ratio for each  measurement. As is usual with \contur, to account for statistical correlations between measurements and avoid double-counting of BSM effects, these measurements are divided into non-overlapping ``analysis pools'' based upon the run period, experiment and final state. Only the most sensitive measurement from each pool is used, and the set of pool-likelihoods is then combined to provide an overall \contur likelihood, which in \colliderbit is then combined with the likelihoods for the LHC searches and the LEP cross-section limits. The  likelihood provided by \contur in this post-processing step had a significant impact on the final results, which will be discussed in detail in Section \ref{subsec:impact-measurements}.

\subsection{Scanning strategy}
\label{sec:scanning strategy}

With the gravitino mass fixed at $1\,\eV$, the collider phenomenology of our model is determined by the mass parameters $M_1$, $M_2$ and $\mu$, and the dimensionless $\tan\beta$ parameter. We restrict our attention to scenarios where the electroweakino masses are all $\lesssim 1\,\TeV$. This is due to the substantial computational cost of accurately mapping out the profile likelihood function across wide, many-dimensional parameter regions where the likelihood function is mostly flat --- especially when MC event simulation is performed for each scan point. The high detectability of final states with photons and missing energy ensures that current LHC searches can exclude specific scenarios of electroweakino production where the masses of the produced electroweakinos are close to or beyond $1\,\TeV$. These are typically scenarios with production of a dominantly wino chargino-neutralino pair and a large $\BR(\tilde\chi_1^0 \rightarrow \gamma \tilde{G})$~\cite{CMS:2017brl, ATLAS:2018nud}. But as we will see, within the general electroweakino parameter space explored here, there are still large, unconstrained parameter regions with all electroweakinos $\lesssim 1\,\TeV$. 

\begin{table}
\begin{center}
\begin{tabular}{l@{\ }c c}
\hline
Parameter & Range/value & Sampling priors \\
\hline
$M_1(Q)$               & $[-1, 1]\,\TeV$ & hybrid, flat  \\
$M_2(Q)$               & $[0, 1]\,\TeV$  & hybrid, flat  \\
$\mu(Q)$               & $[-1, 1]\,\TeV$ & hybrid, flat  \\
$\tan\beta(m_Z)$       & $[1, 70]$       & log, flat     \\
$\mg$                  & $1\,\eV$        & fixed         \\
\hline
$Q$                    & $3\,\TeV$ & fixed \\
\hline
$\alpha_s^{\MSBar}(m_Z)$ & $0.1181$ & fixed \\
Top quark pole mass      & $171.06\,\GeV$ & fixed \\
Higgs mass               & $125.09\,\GeV$ & fixed \\
\end{tabular}
\caption{\label{tab:parameters} Ranges and scanning priors for the input parameters. The ``hybrid'' prior refers to a prior that is flat on $|x| < 10\,\GeV$, and logarithmic elsewhere.}
\end{center}
\end{table}

In Tab.~\ref{tab:parameters} we summarise our choices for the scan input parameters. The MSSM parametrisation we use is implemented in the \gambit MSSM model hierarchy as \textsf{MSSM11atQ\_mA\_mG} (Appendix~\ref{app:codeextensions}), which has 11 free parameters. For the six parameters not listed in Tab.~\ref{tab:parameters} we use the following fixed values: the trilinear couplings $A_{d_3} = A_{e_3} = A_{u_3} = 0$; the gluino mass parameter $M_3 = 5\,\TeV$; the pseudo-scalar Higgs mass $m_A = 5\,\TeV$; and the squared soft sfermion mass parameters $m_l^2 = m_q^2 = (3\,\TeV)^2$. The parameters are defined at an input scale $Q = 3\,\TeV$. The specific values for these fixed parameters are not important, as they simply ensure that all superpartners except the gravitino and the electroweakinos are decoupled from the collider phenomenology.

In order to obtain accurate profile likelihood maps we must ensure that the parameter space is explored in sufficient detail. We therefore combine the parameter samples from multiple scans using different combinations of the priors (metrics) listed in Tab.~\ref{tab:parameters} to scan the parameters. The ``hybrid'' prior in Tab.~\ref{tab:parameters} combines a logarithmic prior for $|x| > 10\,\GeV$ with a flat prior for $|x| < 10\,\GeV$ ($x = M_1, M_2, \mu$). As the physics is invariant under a global sign change for $M_1$, $M_2$ and $\mu$, we follow the common approach in the literature of restricting $M_2$ to positive values.
All scans are performed with the differential evolution sampler \diver \textsf{1.0.4}~\cite{ScannerBit}, interfaced via \scannerbit. We run \diver in the \textsf{jDE} mode (self-adaptive rand/1/bin evolution), which is based on Ref.~\cite{Brest06}. The final combined data set consists of around $3.1\times10^5$ parameter samples.

\section{Results}
\label{sec:results}

\subsection{Best-fit scenarios}
\label{subsec:best_fit_scenarios}

\begin{figure*} 
  \centering
  \includegraphics[height=0.8\columnwidth]{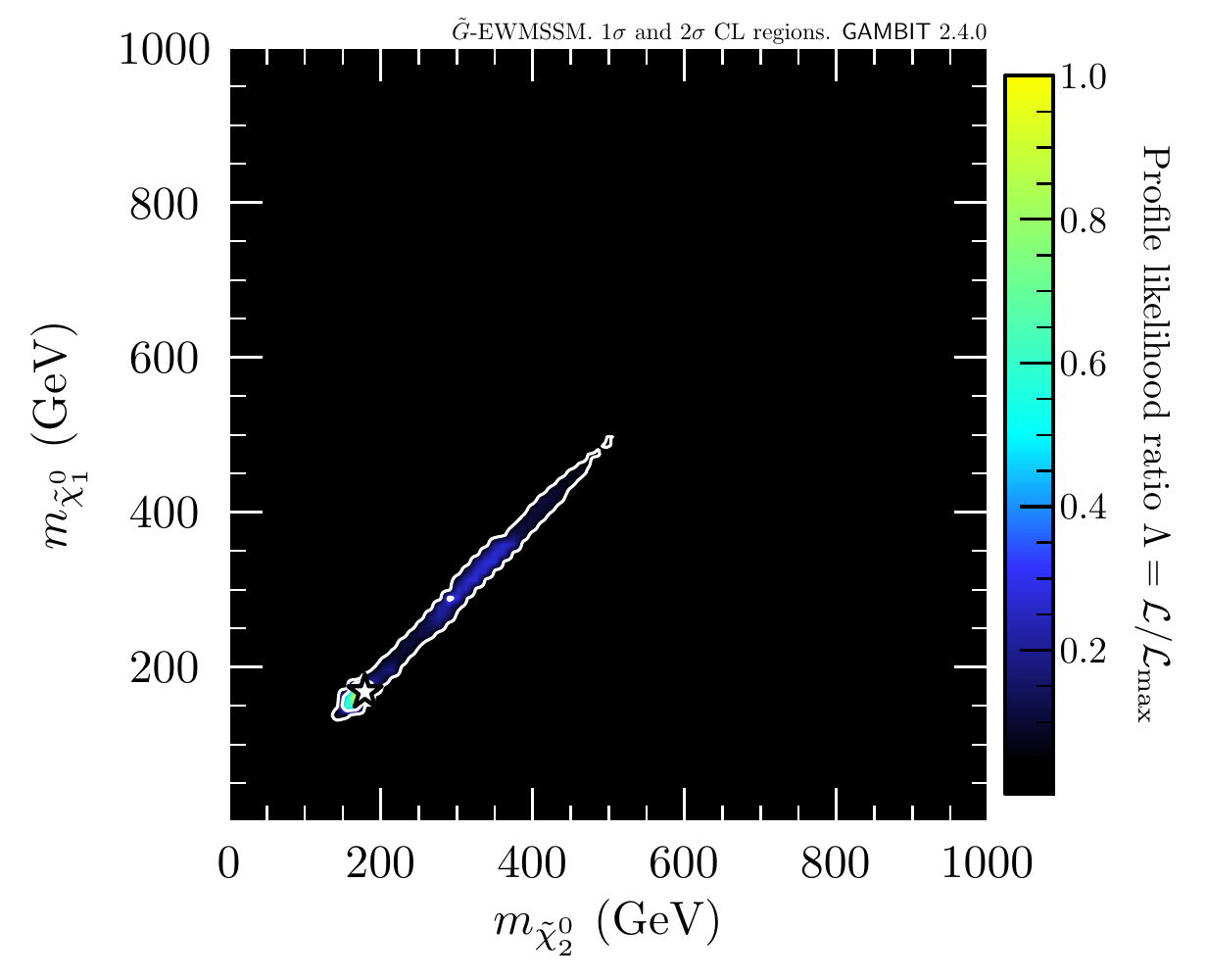}
  \includegraphics[height=0.8\columnwidth]{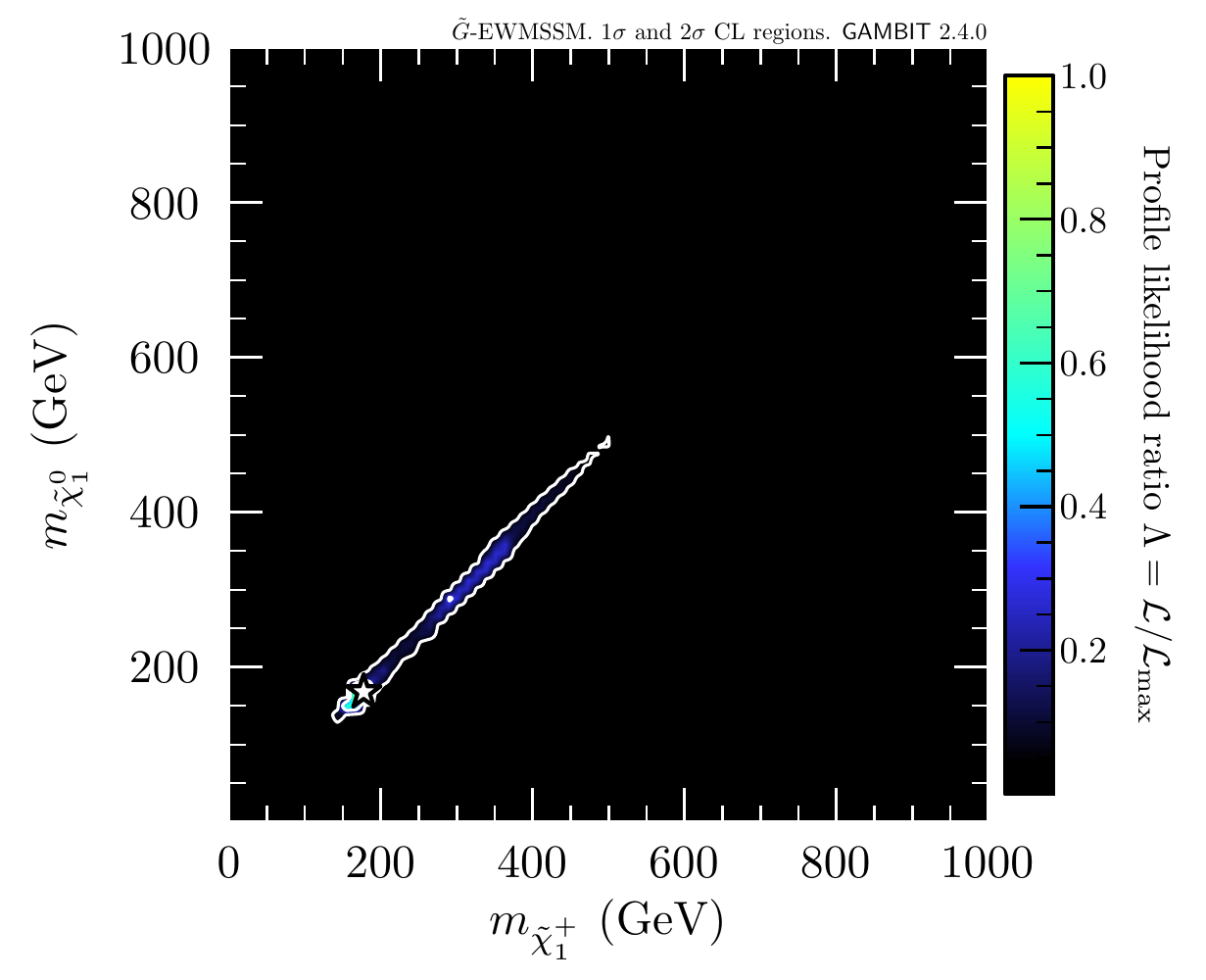}
  \caption{Profile likelihood in the $(m_{\tilde{\chi}_2^0},m_{\tilde{\chi}_1^0})$ plane (left) and in the $(m_{\tilde{\chi}_1^{\pm}},m_{\tilde{\chi}_1^{0}})$ plane (right). The contour lines show the $1\sigma$ and $2\sigma$ confidence regions. The best-fit point is marked by the white star.
  }
  \label{fig:mass_planes_uncapped}
\end{figure*}

In Fig.~\ref{fig:mass_planes_uncapped} we show our fit result in terms of the profile likelihood function across the $(m_{\tilde{\chi}_2^0},m_{\tilde{\chi}_1^0})$ and $(m_{\tilde{\chi}_1^{\pm}},m_{\tilde{\chi}_1^{0}})$ planes. We will present most of our results in one or both of these planes as they are well suited for mapping out the key phenomenological aspects of the high-likelihood scenarios. For reference, in Appendix~\ref{app:input_params} we provide profile likelihood maps in terms of the input parameters.

We find that the \GEWMSSM scenarios in best agreement with current LHC searches and measurements are scenarios where the lightest electroweakinos are dominantly Higgsino, i.e.\ scenarios with $|\mu| < |M_1|, M_2$, corresponding to the Higgsino NLSP region (green) in Fig.\ \ref{fig:neutralinoBR}. As the $\mu$ parameter largely controls the mass of three Higgsino states, these scenarios have near-degenerate masses for $\tilde{\chi}_1^0$, $\tilde{\chi}_2^0$ and $\tilde{\chi}_1^\pm$, explaining why the best-fit region falls along the diagonals of the $(m_{\tilde{\chi}_2^0},m_{\tilde{\chi}_1^0})$ and $(m_{\tilde{\chi}_1^{\pm}},m_{\tilde{\chi}_1^{0}})$ planes. 

For the best-fit point, marked by a white star in Fig.~\ref{fig:mass_planes_uncapped}, the three Higgsinos have masses $m_{\tilde{\chi}_1^0} = 169.9$\,\GeV, $m_{\tilde{\chi}_2^0} = 178.9$\,\GeV and $m_{\tilde{\chi}_1^\pm} = 177.2$\,\GeV. This point further has a pair of wino-dominated $\tilde{\chi}_3^0$ and $\tilde{\chi}_2^\pm$ at $m_{\tilde{\chi}_3^0} = 740.8$\,\GeV and $m_{\tilde{\chi}_2^\pm} = 741.3$\,\GeV, and a dominantly bino $\tilde{\chi}_4^0$ at $m_{\tilde{\chi}_4^0} = 788.1$\,\GeV. The scenarios allowed at $2\sigma$ confidence level (CL) relative to the best-fit point, all predict such a trio of near-degenerate Higgsinos with masses no less than about $140$\,\GeV and no greater than about $500$\,\GeV. 

The scenarios within the $2\sigma$ region in Fig.~\ref{fig:mass_planes_uncapped} are largely scenarios with negative $\mu$ parameter, $|\mu| < M_2, |M_1|$, and $\tan\beta \lesssim 5$, with the highest-likelihood solutions favouring $\tan\beta$ values close to 1.
For such scenarios, the dominant and subdominant decay modes for the lightest neutralino are the $\tilde{\chi}_1^0 \to h \gravitino$ and $\tilde{\chi}_1^0 \to Z \gravitino$ channels, respectively --- see e.g.\ the region around $\mu \sim -300\,\GeV$ in the branching ratio plots in Fig.\ \ref{fig:neutralinoBR}. 
Low branching ratios for decays to $\gamma \gravitino$ final states ensure that the scenarios in the $2\sigma$ region escape the otherwise highly constraining photons + $E_T^\textrm{miss}$ searches. Many of these scenarios also have sizeable branching ratios for $\tilde{\chi}_2^0$ to decay directly to a $\gravitino$ final state, typically through the $\tilde{\chi}_2^0 \to Z \gravitino$ decay mode, rather than decaying exclusively through $\tilde{\chi}_2^0 \to Z^* \tilde{\chi}_1^0$, as often assumed in LHC searches for Higgsino production. Similarly, many scenarios in the higher-mass part of the $2\sigma$ region ($m_{\tilde{\chi}_1^\pm} \gtrsim 300\,\GeV$) have large branching ratios for direct decays of $\tilde{\chi}_1^\pm$ to the gravitino, through $\tilde{\chi}_1^\pm \to W^\pm \gravitino$. 

By tuning the branching ratios $\text{BR}(\tilde{\chi}_{1,2}^0 \to h \gravitino)$ versus $\text{BR}(\tilde{\chi}_{1,2}^0 \to Z \gravitino)$, and $\text{BR}(\tilde{\chi}_1^\pm \to W^\pm \gravitino)$ versus $\text{BR}(\tilde{\chi}_1^\pm \to f f' \tilde{\chi}_1^0)$,\footnote{Here $f$ and $f'$ are SM fermions.} the model can partly fit small excesses in the ATLAS and CMS leptons + $E_T^\textrm{miss}$ searches and the ATLAS $b$-jets + $E_T^\textrm{miss}$ search. (The preference for a small signal contribution in $b$-jet final states in part explains the preference for $\tan\beta \sim 1$, since this increases the branching ratio for $\tilde{\chi}_1^0 \rightarrow h \gravitino$, see Sec.~\ref{sec:model}.) In combination, this produces a weak preference for the lower-mass end of the diagonal in Fig.~\ref{fig:mass_planes_uncapped}, at masses around $170\,\GeV$.\footnote{At the best-fit point, the three dominant contributions to the likelihood come from \textit{i)} a signal region requiring $\geq 3$ $b$-jets, no leptons, $m_{\text{eff}} > 860\,\GeV$ and $E_{T}^{\text{miss}} \in (150,200)\,\GeV$ \cite{ATLAS:2018tti}; \textit{ii)} a signal region requiring 3 leptons, no opposite-sign, same-flavour lepton pairs and $E_{T}^{\text{miss}} > 50\,\GeV$ \cite{ATLAS:2021yyr}; and \textit{iii)} a signal region requiring $\geq 5$ leptons \cite{CMS:2021cox-fix}. Due to the many different final state combinations of leptons and $b$-jets that can arise in the decays of 2--4 on-shell and off-shell $h$, $Z$ and $W^\pm$ bosons, the best-fit parameter point simultaneously predicts small signal contributions in all of these three signal regions.}

We found a preference for low-mass electroweakino scenarios also in our EWMSSM fit in~\cite{EWMSSM}. The EWMSSM parameter regions favoured in that study allow for electroweakino decay chains that produce multiple on-shell $Z$, $h$ and $W$ bosons, and terminate in a bino-dominated $\tilde{\chi}_1^0$ that provides the missing energy signal. The favoured low-mass scenarios in the \GEWMSSM predict a similar collider phenomenology, but now with the gravitino rather than a bino-like neutralino terminating the decay chains. However, in the present study the preference for these low-mass scenarios is weaker, as the previously-observed data excesses are less pronounced in the now larger ATLAS and CMS data sets.

\subsection{Non-excluded scenarios}
\label{subsec:non_exluded_scenarios}

\begin{figure*} 
  \centering
  \includegraphics[height=0.8\columnwidth]{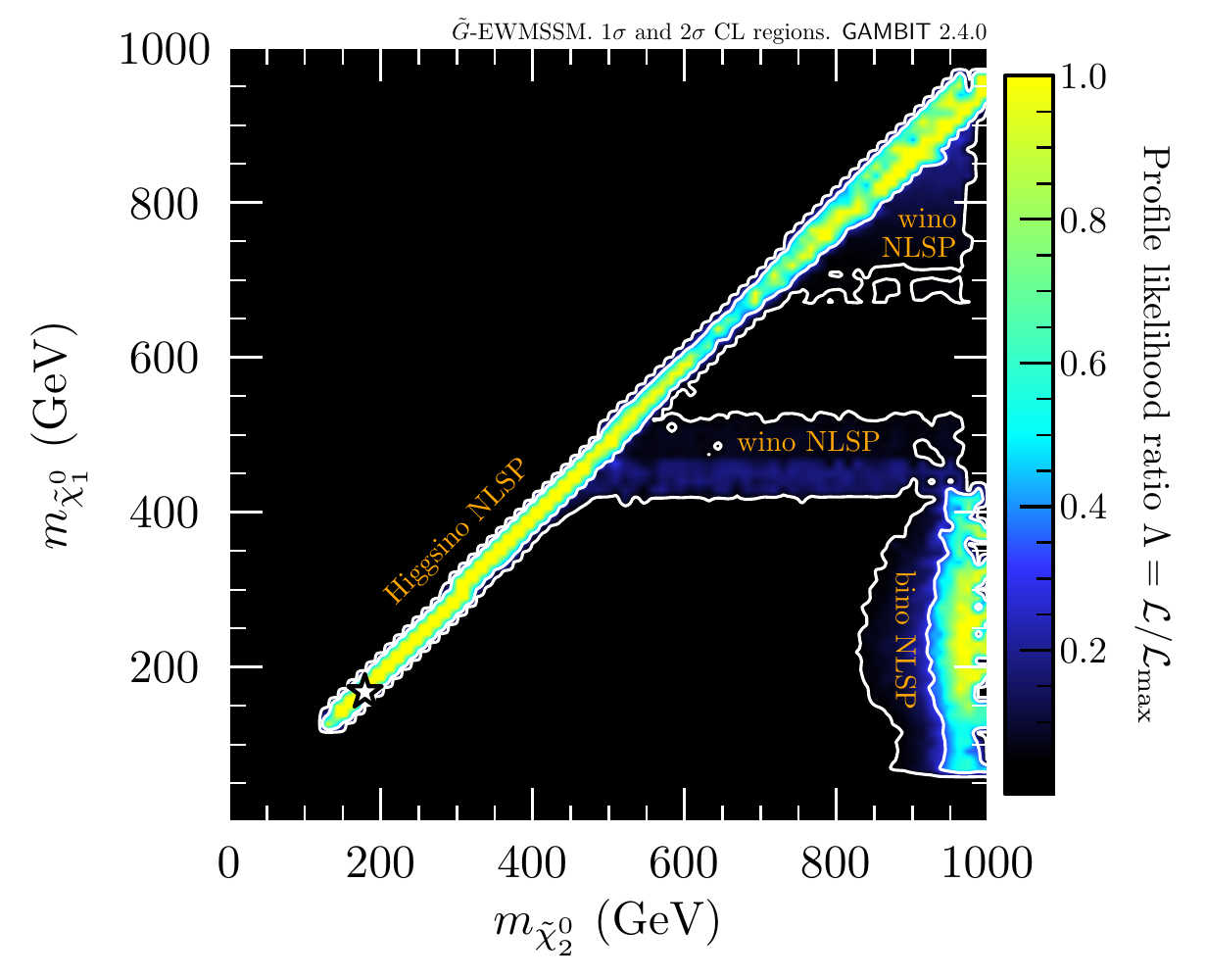}
  \includegraphics[height=0.8\columnwidth]{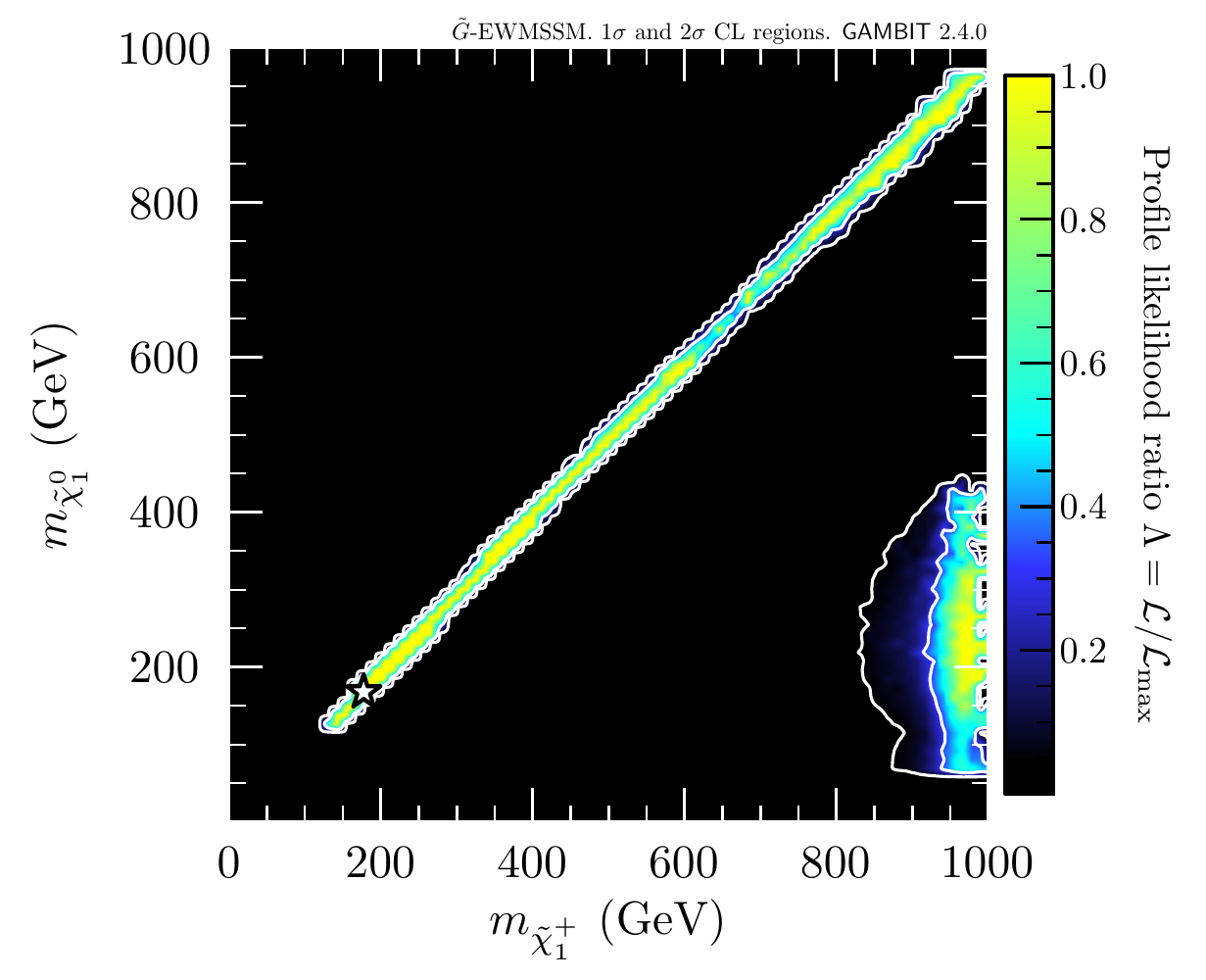}
  \caption{Capped profile likelihood in the $(m_{\tilde{\chi}_2^0},m_{\tilde{\chi}_1^0})$ plane (left) and in the $(m_{\tilde{\chi}_1^{\pm}},m_{\tilde{\chi}_1^{0}})$ plane (right). The white contour lines show the $1\sigma$ and $2\sigma$ confidence regions.
  }
  \label{fig:mass_planes_capped}
\end{figure*}
Assuming that these small data excesses are just background fluctuations rather than a true BSM signal, it is interesting to consider what electroweakino mass combinations the current combined data clearly exclude in the \GEWMSSM. We investigate this in Fig.~\ref{fig:mass_planes_capped} by showing profile likelihood plots where we use the capped likelihood, $\mathcal{L}_\text{searches}^\text{cap}$ (Eq.~\ref{eq:capped_lnlike}), as described in Sec.~\ref{sec:simulated_LHC_searches}.

\begin{figure*} 
  \centering
  \includegraphics[height=0.55\columnwidth]{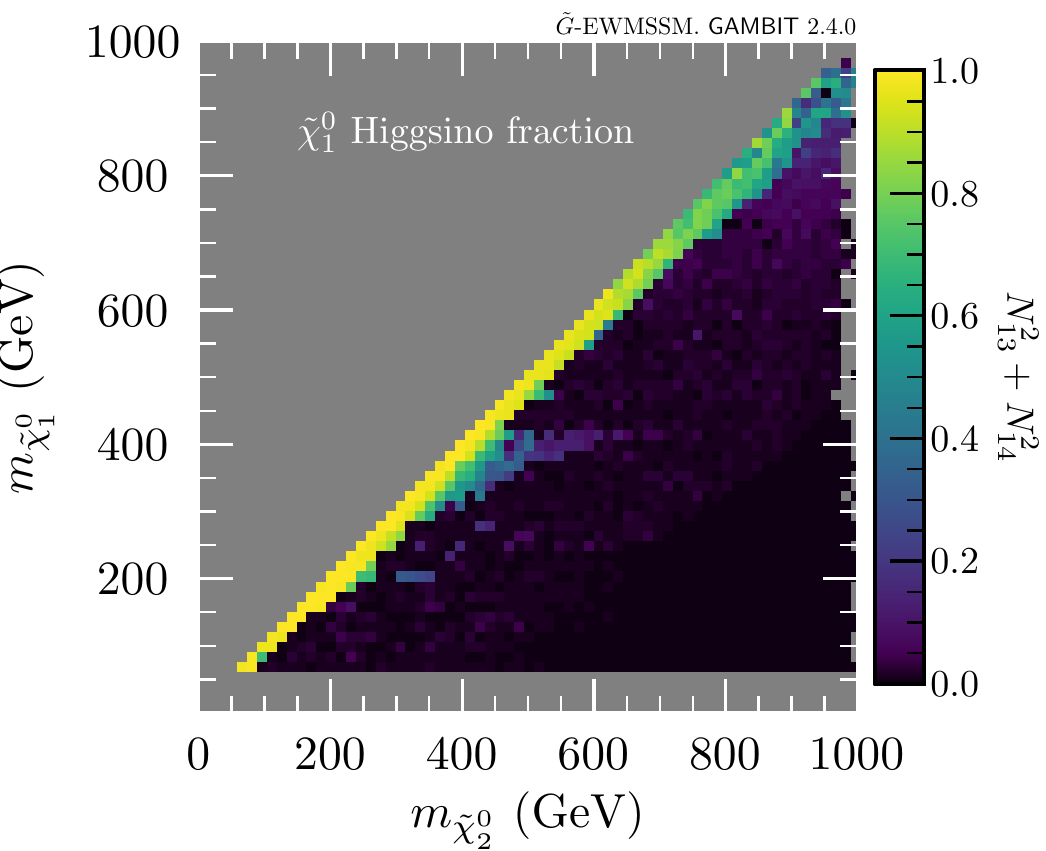}
  \includegraphics[height=0.55\columnwidth]{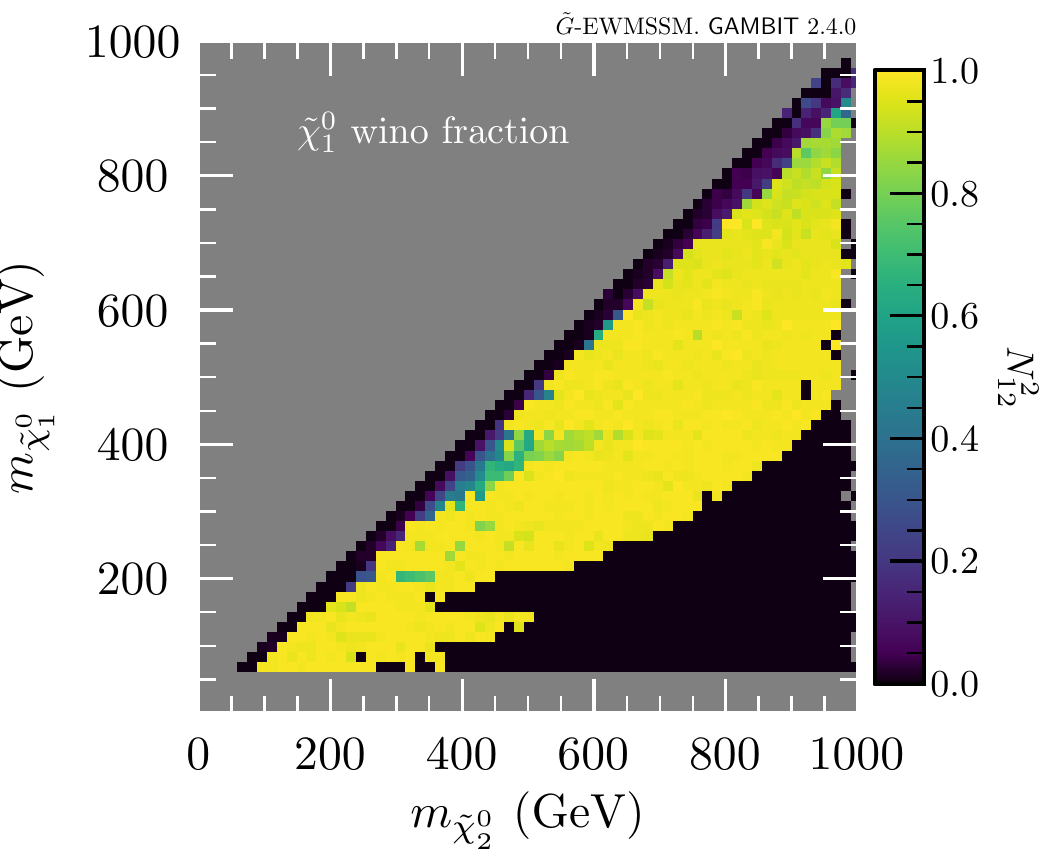}
  \includegraphics[height=0.55\columnwidth]{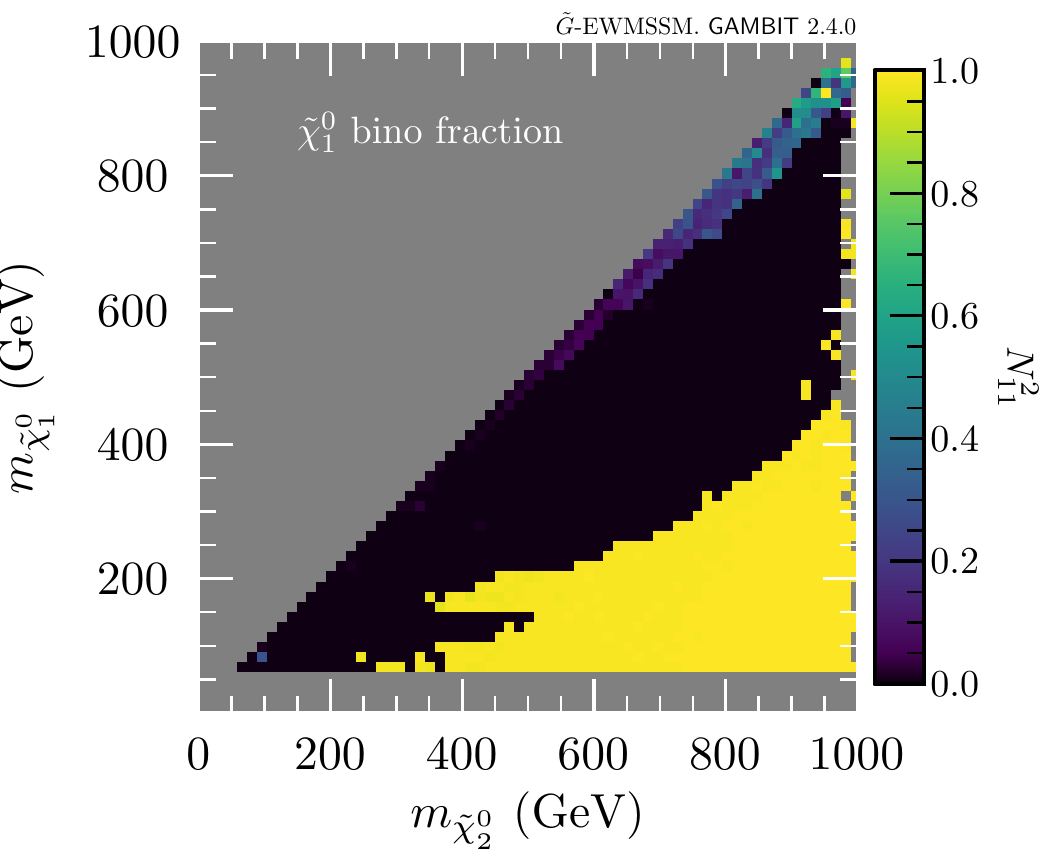}
  \caption{The Higgsino (left), wino (middle) and bino (right) fraction of the $\tilde{\chi}_1^0$, plotted across the profile-likelihood surface for the $(m_{\tilde{\chi}_2^0},m_{\tilde{\chi}_1^0})$ plane.}
  \label{fig:mass_planes_N1_mixture}
\end{figure*}

To understand the structures visible in Fig.~\ref{fig:mass_planes_capped}, we first consider Fig.~\ref{fig:mass_planes_N1_mixture}, where we show the Higgsino, wino and bino components of the lightest neutralino for the highest-likelihood point in each bin across the $(m_{\tilde{\chi}_2^0},m_{\tilde{\chi}_1^0})$ plane. This allows us to identify which of the three NLSP scenarios discussed in Sec.~\ref{sec:model} are preferred in different parts of the mass plane. We see clearly that the preferred scenarios along the diagonal are scenarios with a mostly Higgsino NLSP (left panel), as discussed above. Moving away from the diagonal, towards higher $m_{\tilde{\chi}_2^0}$, the best-fitting scenarios are wino NLSP scenarios (middle panel). We note that around $m_{\tilde{\chi}_1^0}, m_{\tilde{\chi}_2^0} \sim 400$\,GeV, the current collider data prefers a fairly even wino/Higgsino admixture for the $\tilde{\chi}_1^0$. Finally, at even higher $\tilde{\chi}_2^0$--$\tilde{\chi}_1^0$ mass splittings, the best possible fits are obtained for bino NLSP scenarios (right panel).\footnote{For $m_{\tilde{\chi}_2^0} \approx m_{\tilde{\chi}_1^0} \approx 1$\,TeV in Fig.~\ref{fig:mass_planes_N1_mixture}, all neutralino components contribute significantly to the composition of $\tilde{\chi}_1^0$. This is largely a consequence of our scan settings: Since we restrict our study to the parameter space that has \textit{all} electroweakino masses below $1\,\TeV$, having the lightest neutralino mass close to $1\,\TeV$ will correspond to parameter points with $|M_1| \sim M_2 \sim |\mu| \sim 1\,\TeV$.} 

We will in the following use the term \textit{profile-likelihood surface} to refer to the set of parameter samples that appear in figures like Fig.~\ref{fig:mass_planes_N1_mixture}, where for each bin in the given plane we visualise some property of the highest-likelihood parameter sample belonging to that bin. For the interpretation of these figures it is important to remember that apparent discontinuities, such as the boundaries between the yellow and black regions in Fig.~\ref{fig:mass_planes_N1_mixture}, typically result from the projection done by the profile likelihood procedure: two neighbouring bins in a mass plane can have their respective highest-likelihood points coming from very different parts of the four-dimensional \GEWMSSM parameter space. So for instance the black region in the right-hand panel of Fig.~\ref{fig:mass_planes_N1_mixture} does not imply that there are no parameter samples that predict the given $\tilde{\chi}_1^0$ and $\tilde{\chi}_2^0$ masses and a bino-dominated $\tilde{\chi}_1^0$, only that there for these mass predictions exist other parameter points that give a better fit to data and for which the $\tilde{\chi}_1^0$ is dominantly wino or Higgsino.

We can now go back and reconsider Fig.~\ref{fig:mass_planes_capped}. Along the diagonals of the two mass planes, we see the allowed scenarios with Higgsino-dominated $\tilde{\chi}_1^0$, $\tilde{\chi}_2^0$ and $\tilde{\chi}_1^\pm$. This region extends all the way up towards the edge of our scan range, corresponding to masses around $1\,\TeV$. In addition, there are three other non-excluded scenarios visible.

First, in the $(m_{\tilde{\chi}_2^0},m_{\tilde{\chi}_1^0})$ plane, we find an allowed horizontal region at around $m_{\tilde{\chi}_1^0} \approx 450\,\GeV$, with wino-dominated and mass degenerate $\tilde{\chi}_1^0$ and $\tilde{\chi}_1^\pm$. 
Second, in the region of $m_{\tilde{\chi}_1^0} \lesssim 450\,\GeV$ and $m_{\tilde{\chi}_2^0}, m_{\tilde{\chi}_1^\pm} \gtrsim 800\,\GeV$, we see solutions with a lonely, light, bino-dominated $\tilde{\chi}_1^0$. Lastly, in the $(m_{\tilde{\chi}_2^0},m_{\tilde{\chi}_1^0})$ plane around $m_{\tilde{\chi}_1^0} > 700\,\GeV$ and away from the diagonal, we see a region of solutions allowed at $2\sigma$, where again the $\tilde{\chi}_1^0$ and $\tilde{\chi}_1^\pm$ are mostly wino, though with non-negligible Higgsino components.

Before we explore these findings further, let us briefly compare them with the capped-likelihood results from our analysis of the EWMSSM~\cite{EWMSSM}. In \cite{EWMSSM} we found that essentially no combinations of $\tilde{\chi}_1^{\pm}$ and $\tilde{\chi}_1^{0}$ masses could be conclusively ruled out by the combination of LHC search results at the time of that study. The conclusion is markedly different in the present \GEWMSSM study, where only four distinct scenarios for electroweakinos below $1\,\TeV$ remain viable. There are several factors contributing to this result: (1) the overall stronger constraining power due to the now larger LHC data sets; (2) the diminishing of the data excesses that in \cite{EWMSSM} helped improve the fit for low-mass solutions in the EWMSSM; (3) the additional constraining power in the present study, coming from our inclusion of LHC measurements in addition to direct BSM searches; and (4) the distinctive \GEWMSSM collider signatures, in particular the photon signatures, that result in strong constraints on large parts of the \GEWMSSM parameter space.

\subsection{Impact of different searches}
\label{subsec:impact}

\begin{figure*} 
  \centering
  \includegraphics[height=0.8\columnwidth]{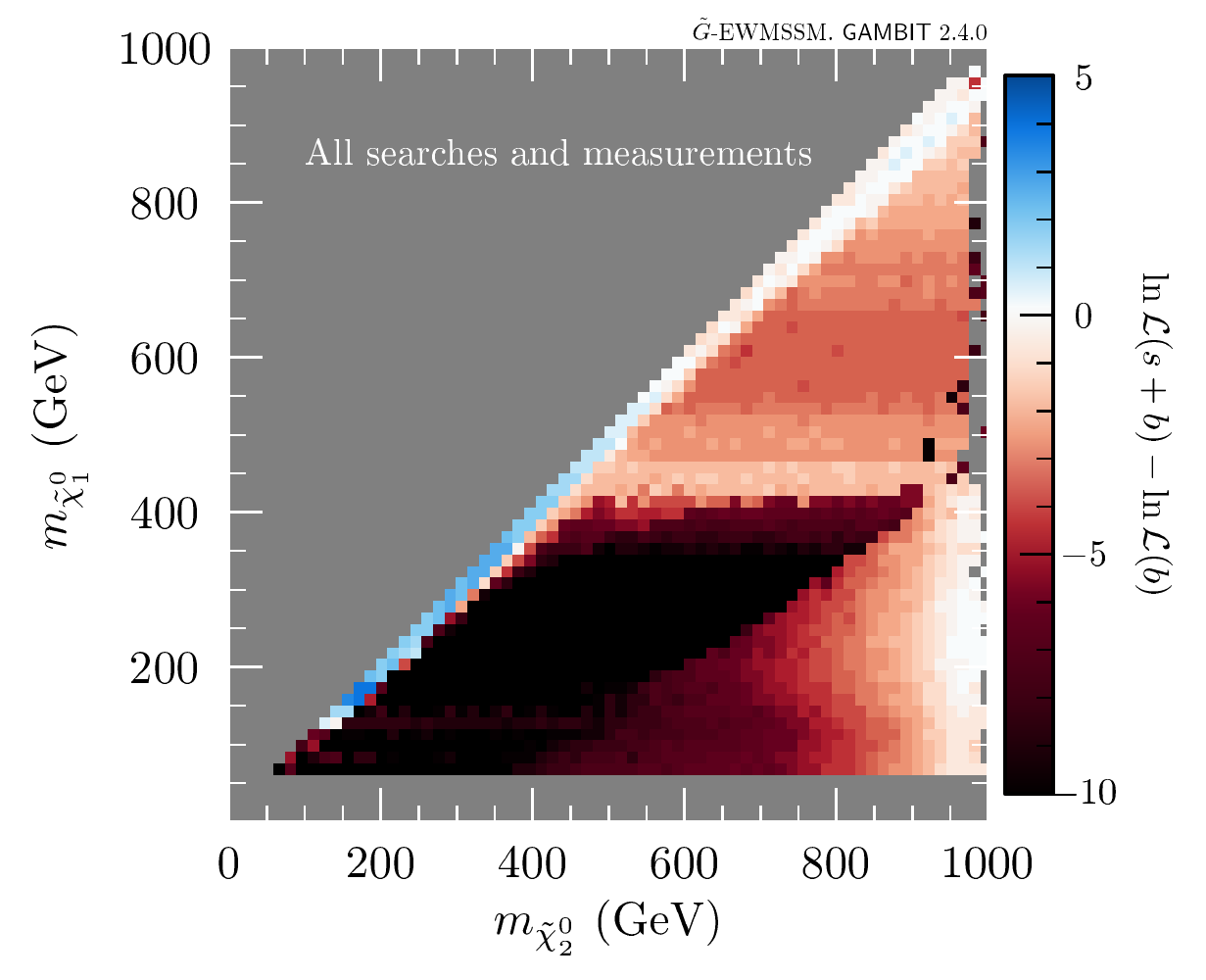}
  \includegraphics[height=0.8\columnwidth]{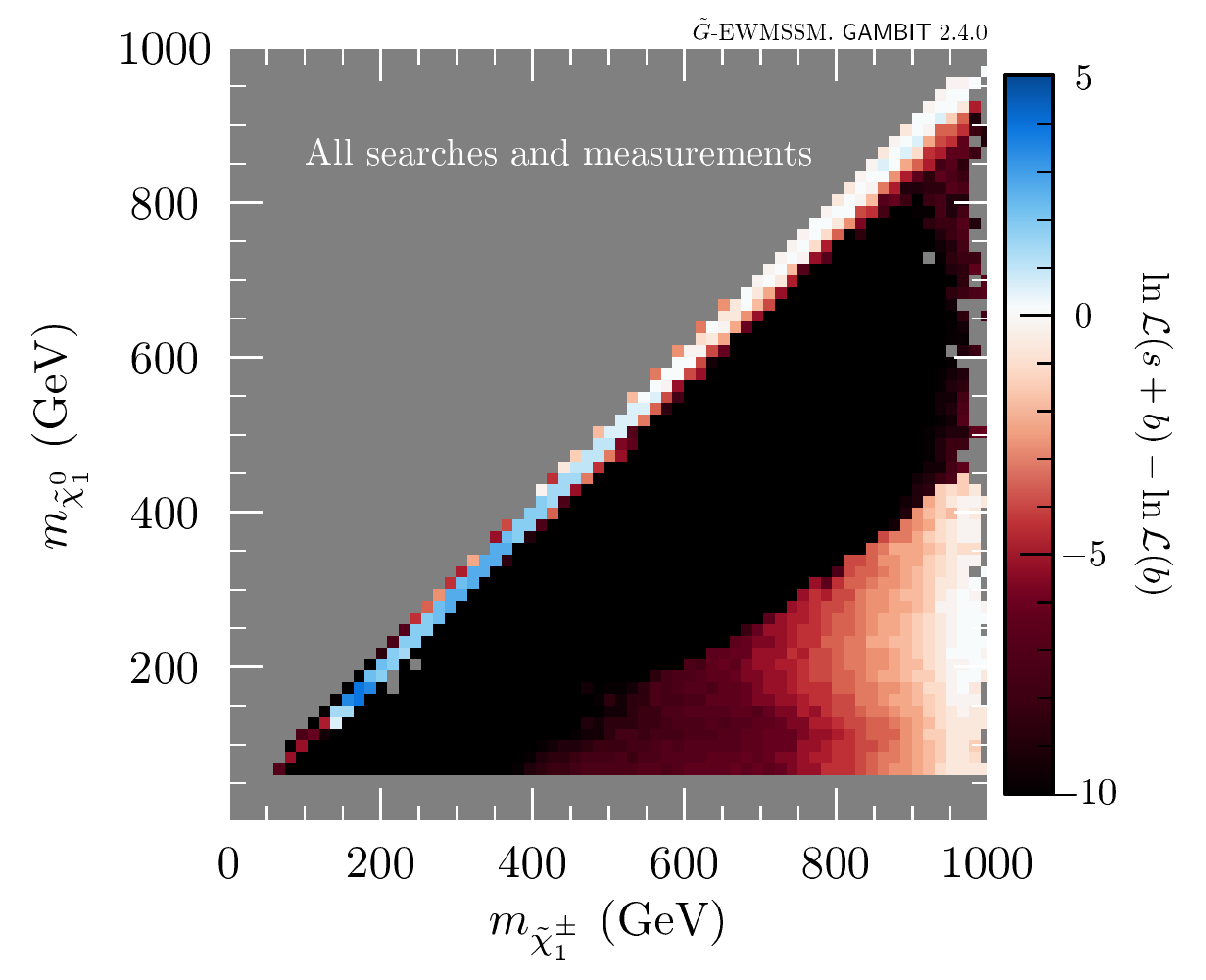}
  \caption{
  The total log-likelihood plotted across the profile-likelihood surface in the $(m_{\tilde{\chi}_2^0},m_{\tilde{\chi}_1^0})$ plane (left) and in the $(m_{\tilde{\chi}_1^{\pm}},m_{\tilde{\chi}_1^0})$ plane (right).}
  \label{fig:mass_planes_loglikes_all}
\end{figure*}

\begin{figure*} 
  \centering
  \includegraphics[height=0.8\columnwidth]{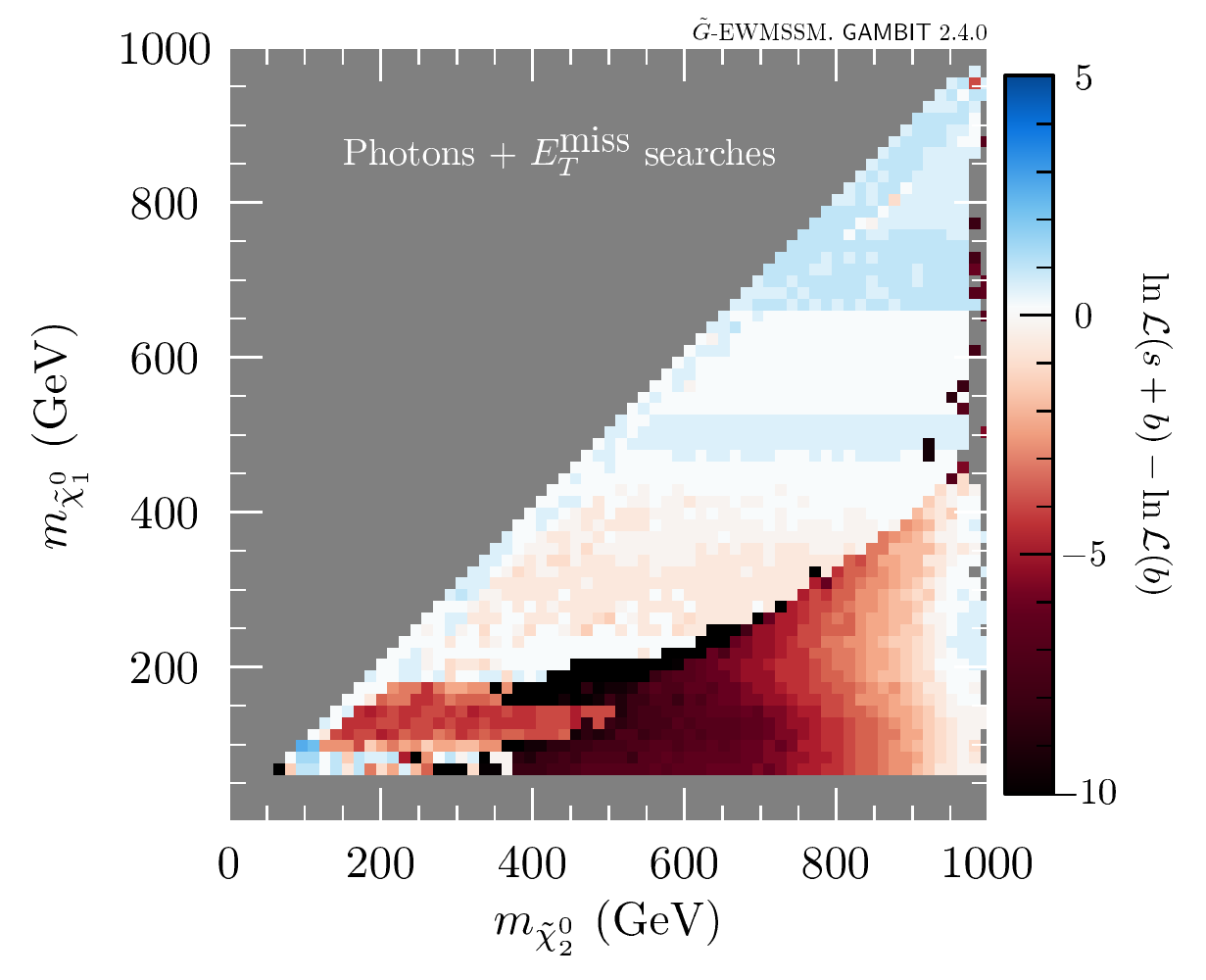}
  \includegraphics[height=0.8\columnwidth]{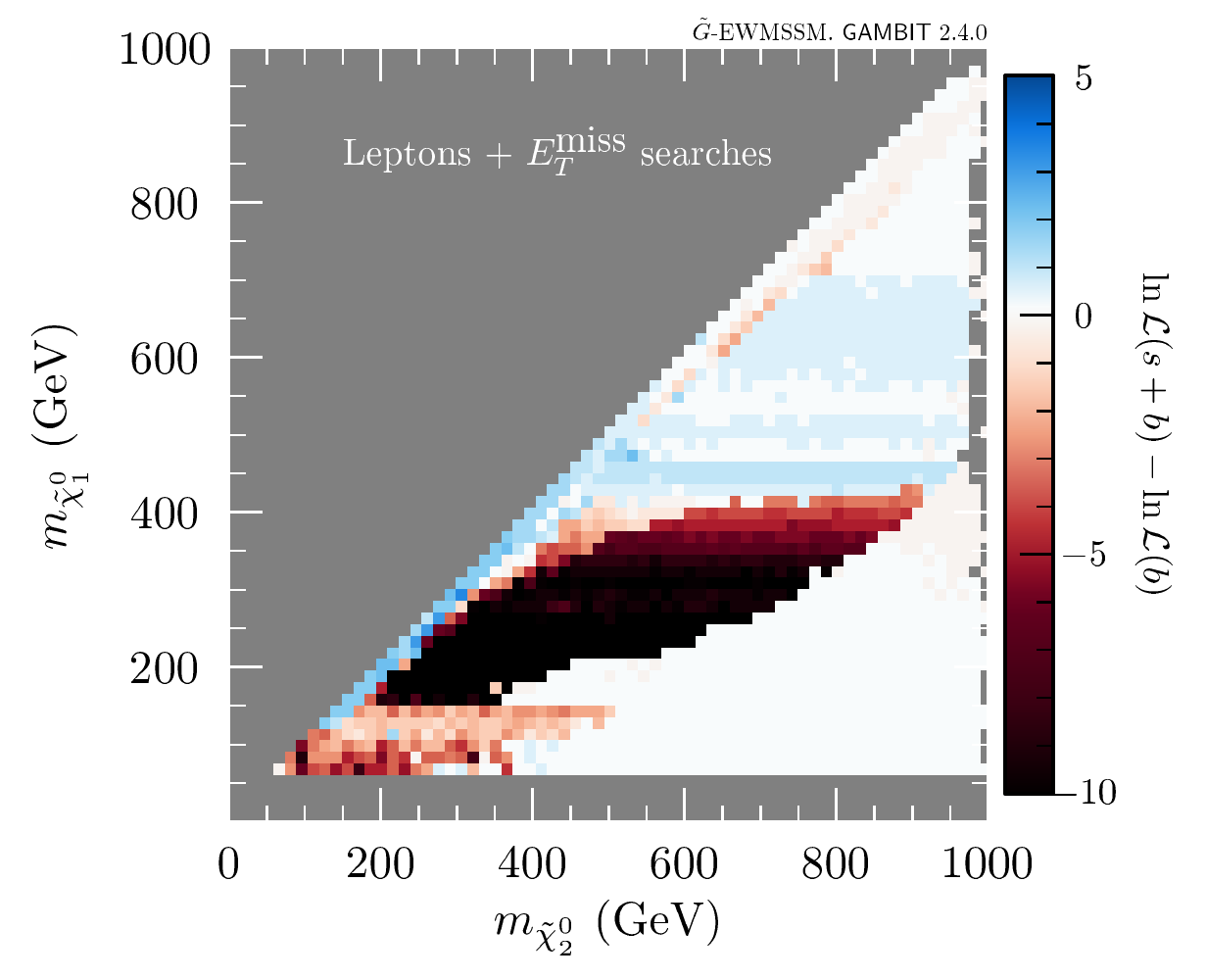}\\
  \includegraphics[height=0.8\columnwidth]{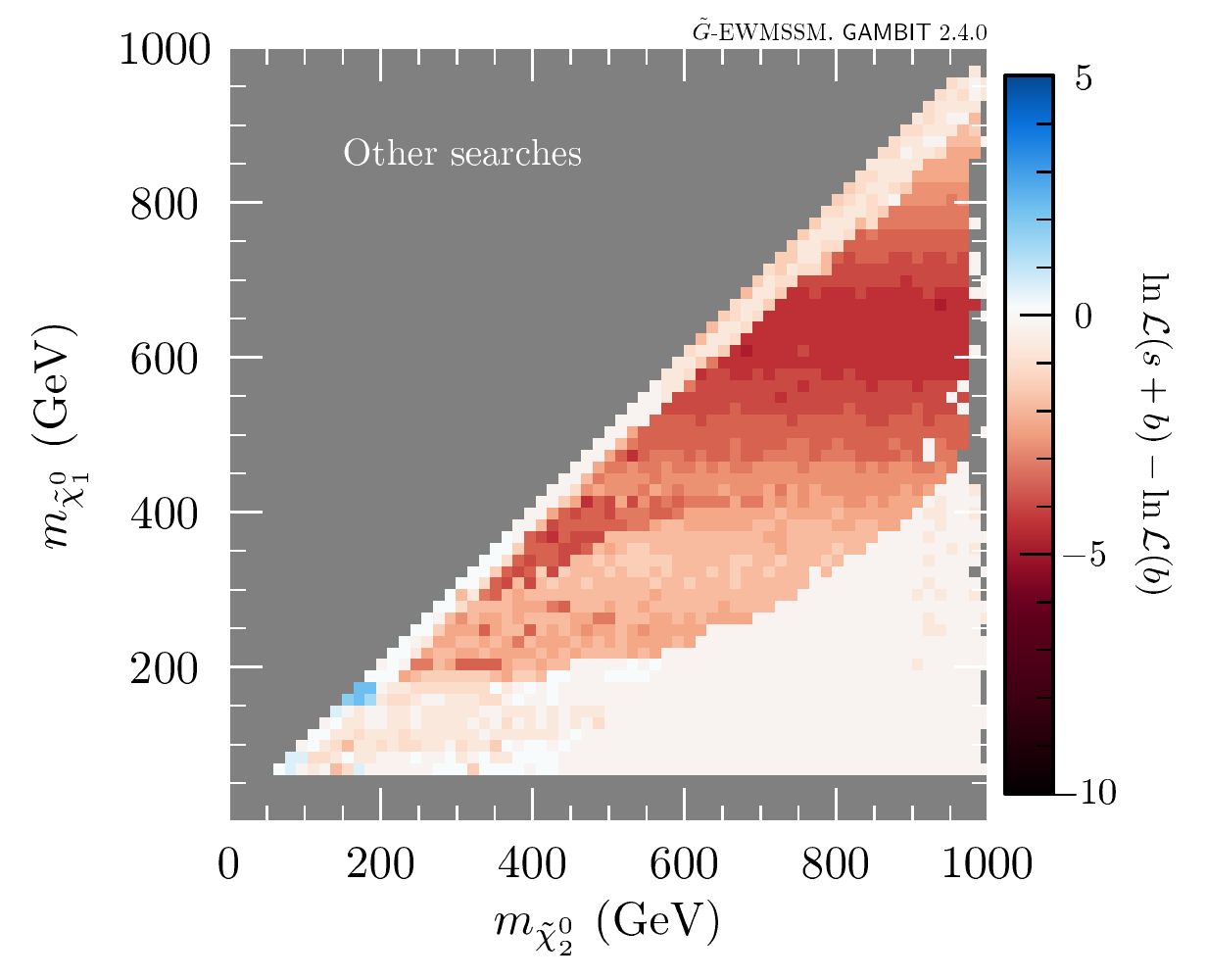}
  \includegraphics[height=0.8\columnwidth]{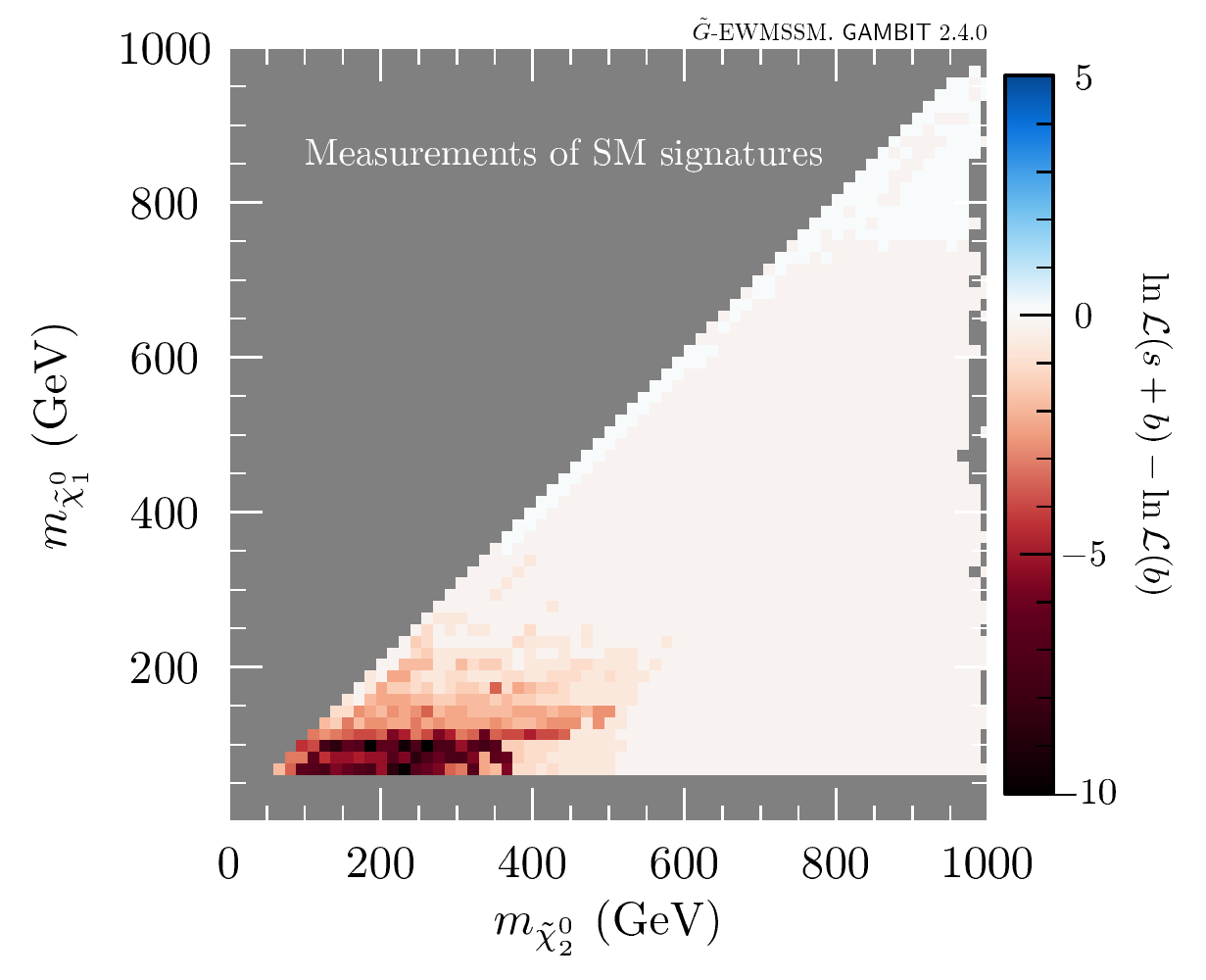}
  \caption{
  Log-likelihood contribution from various groups of LHC searches across the profile-likelihood surface for the $(m_{\tilde{\chi}_2^0},m_{\tilde{\chi}_1^0})$ plane.} 
  \label{fig:mass_planes_loglikes_groupedsearches}
\end{figure*}

\begin{figure*} 
  \centering
  \includegraphics[height=0.8\columnwidth]{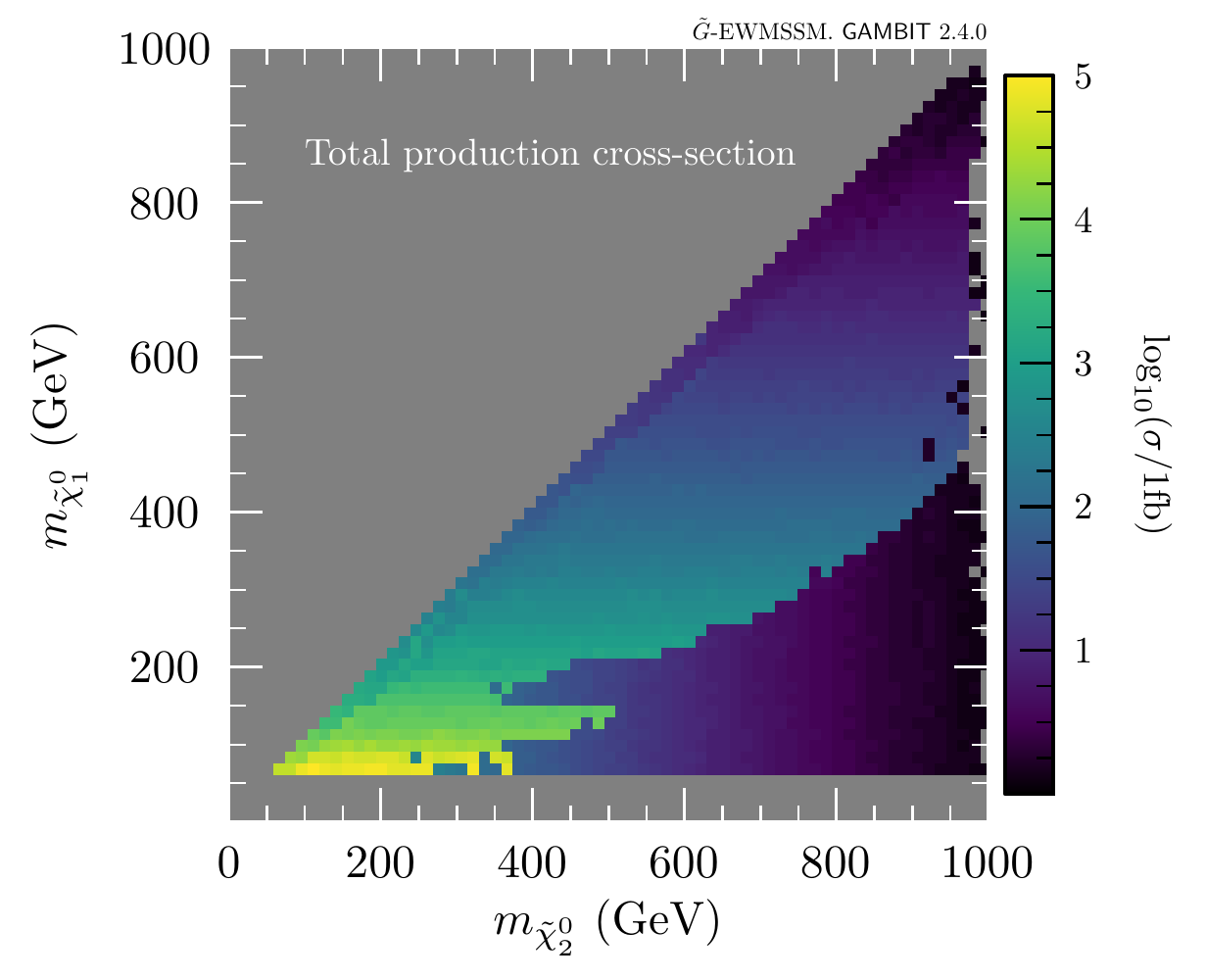}
  \includegraphics[height=0.8\columnwidth]{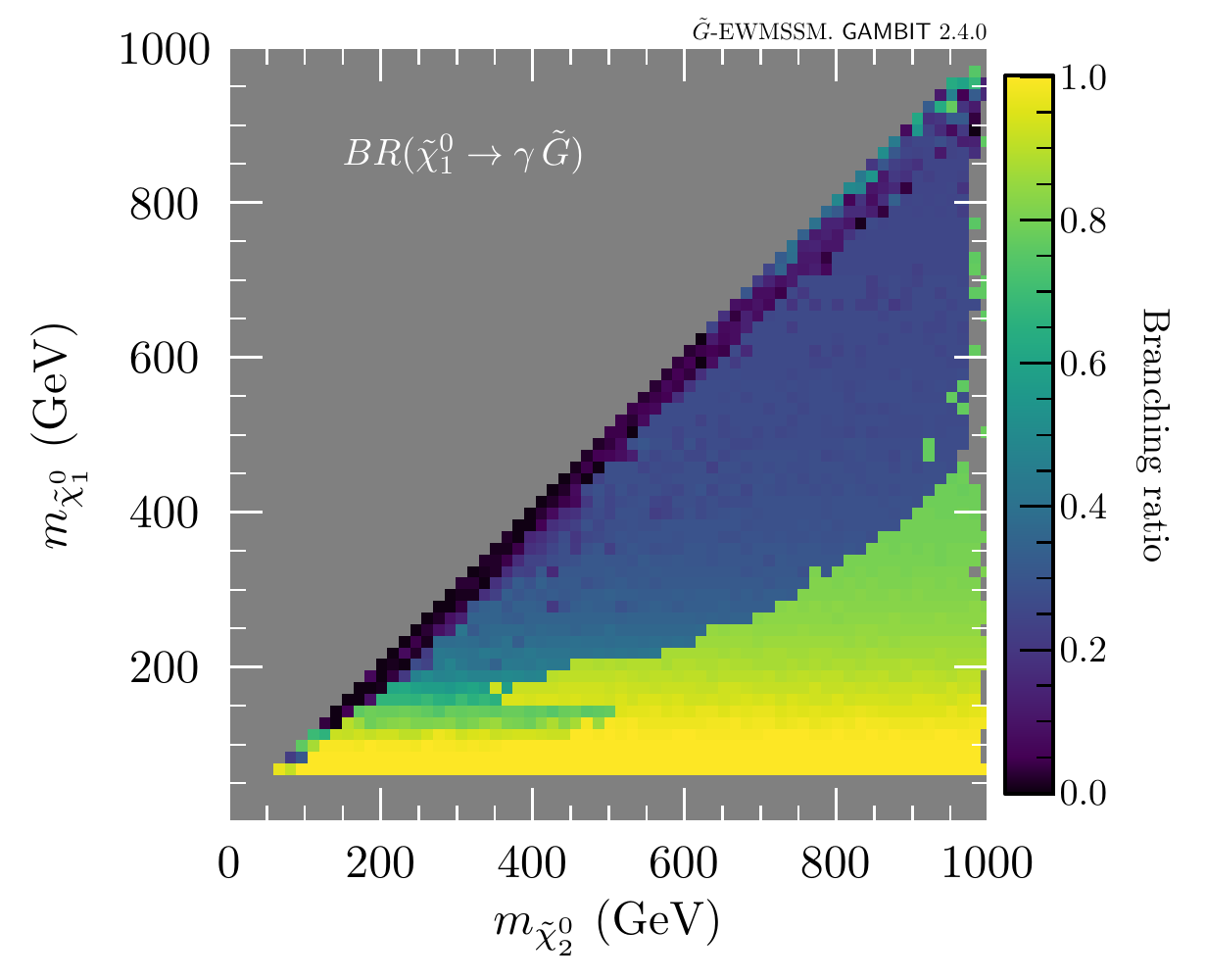}\\
  \includegraphics[height=0.8\columnwidth]{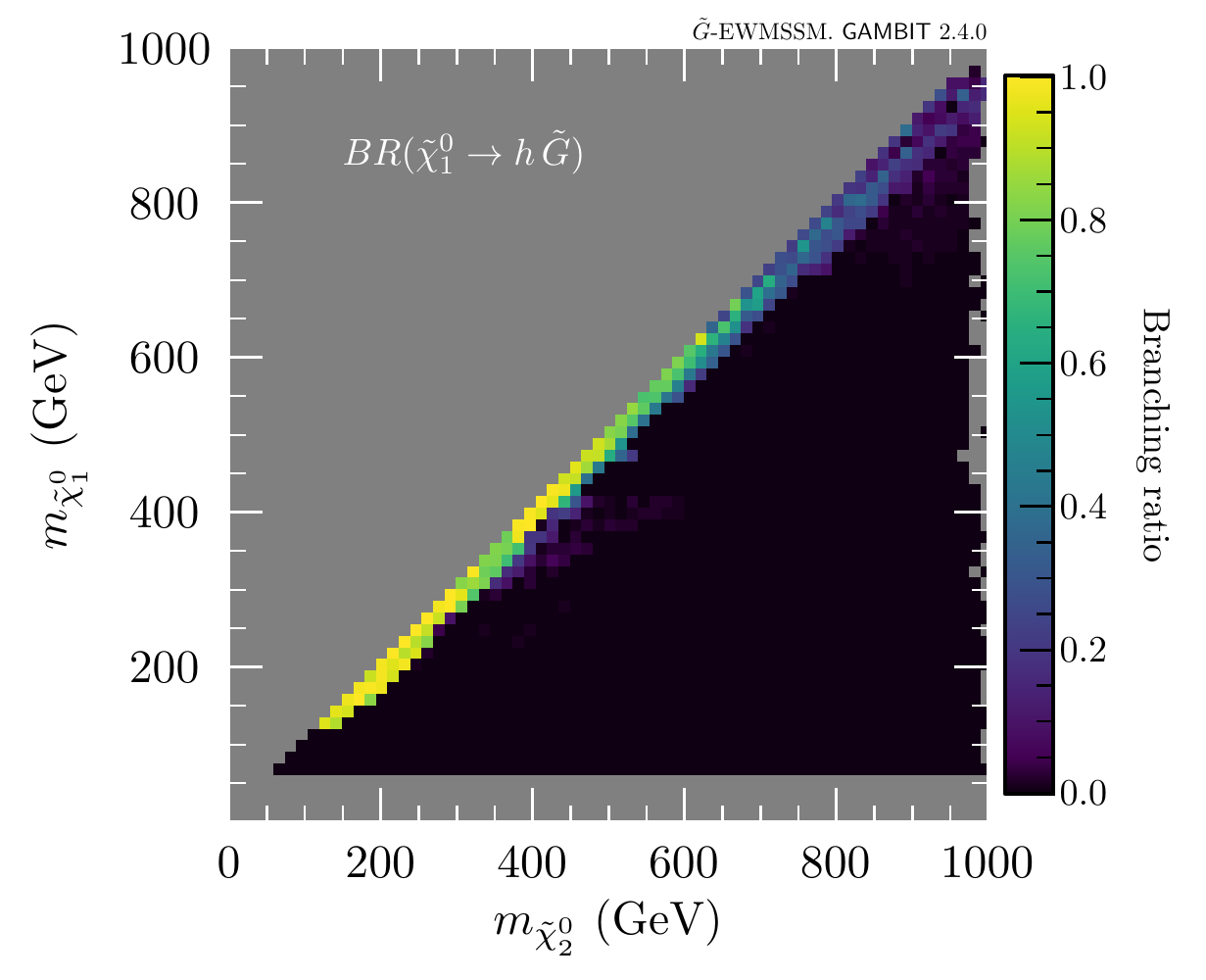}
  \includegraphics[height=0.8\columnwidth]{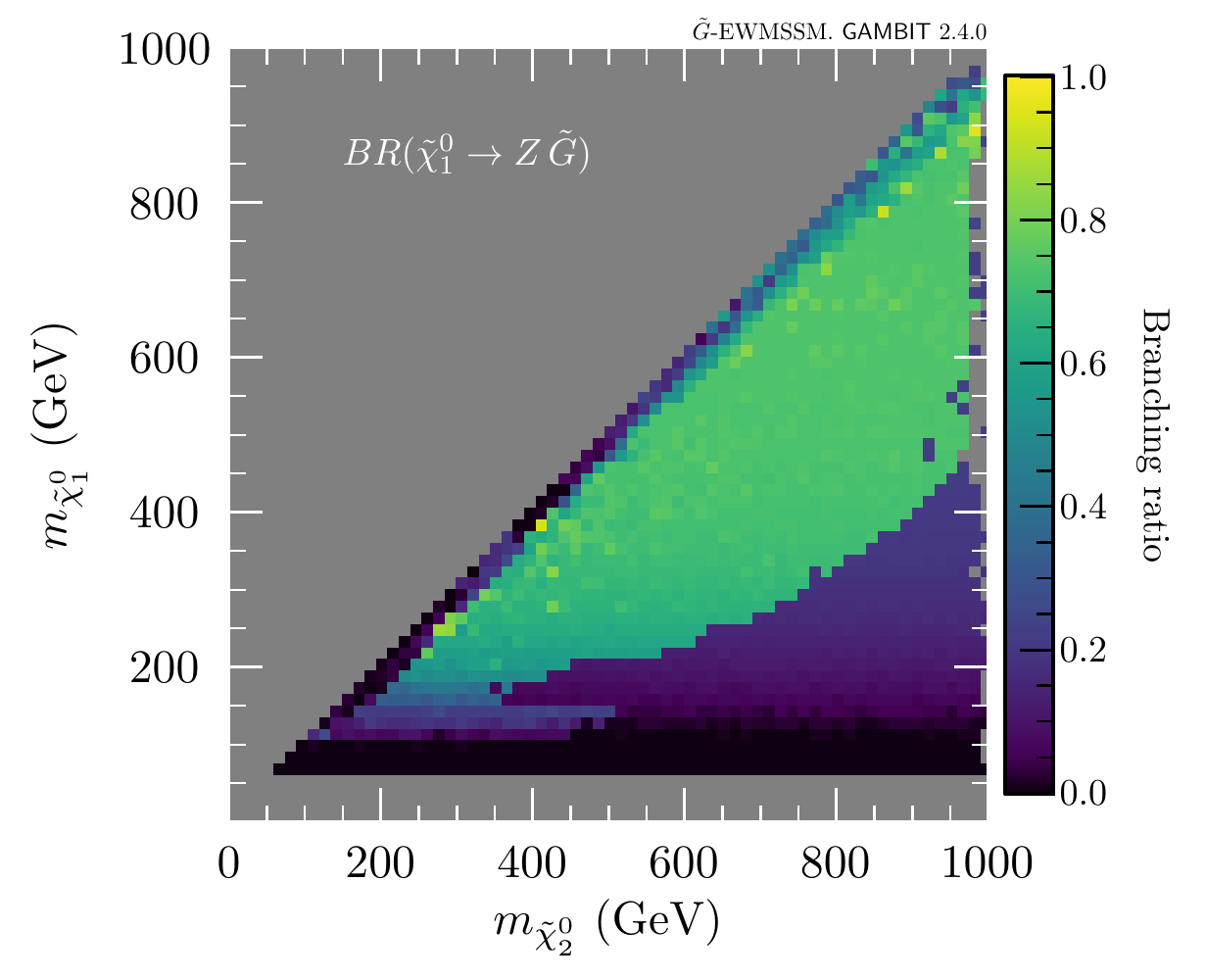}\\
  \includegraphics[height=0.8\columnwidth]{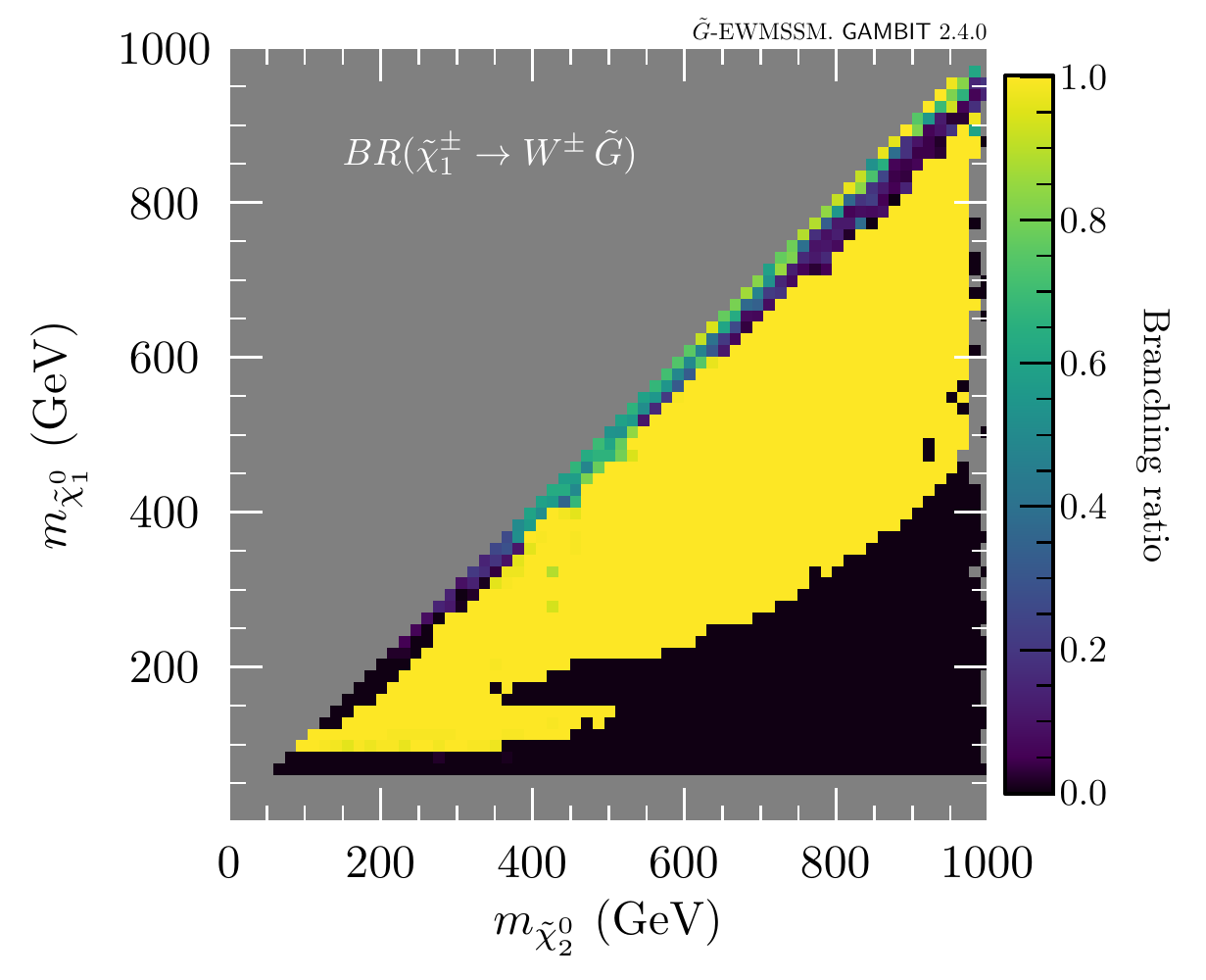}
  \includegraphics[height=0.8\columnwidth]{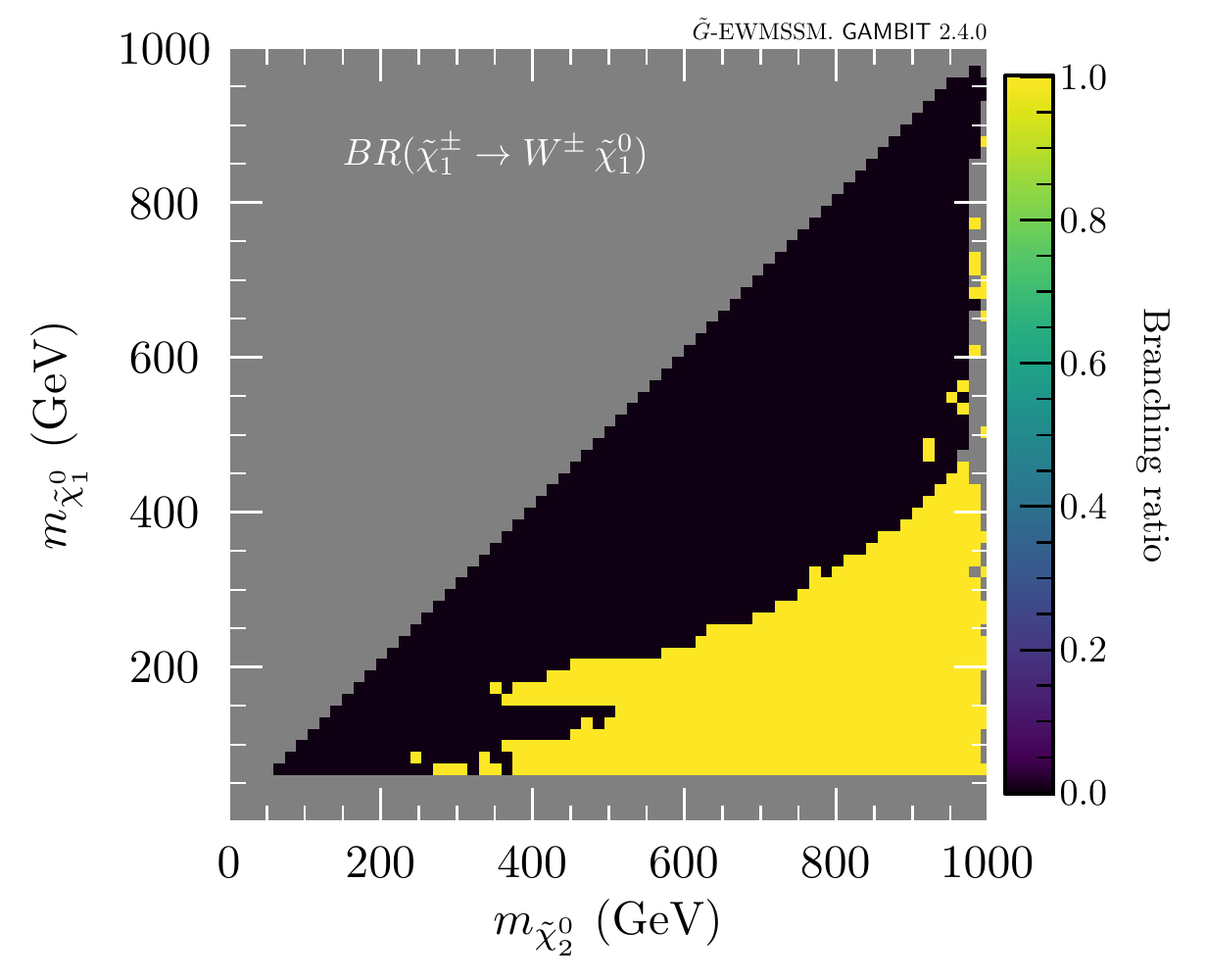}
  \caption{Total LHC production cross-section for electroweakinos, and selected branching ratios for the decays of $\tilde{\chi}_1^0$ and $\tilde{\chi}_1^\pm$, plotted across the profile-likelihood surface for the $(m_{\tilde{\chi}_2^0},m_{\tilde{\chi}_1^0})$ plane.}
  \label{fig:mass_planes_xsec_and_BRs}
\end{figure*}

To understand our results in greater detail, we will in the following discuss the contributions from the LHC searches and measurements that most strongly influence the fit result. To aid this discussion we consider Figs.\ ~\ref{fig:mass_planes_loglikes_all}, \ref{fig:mass_planes_loglikes_groupedsearches} and \ref{fig:mass_planes_xsec_and_BRs}: In Fig.~\ref{fig:mass_planes_loglikes_all} we show the total log-likelihood difference $\ln \mathcal{L}(\bm{s}) - \ln \mathcal{L}(\bm{s}=\bm{0})$. The various solutions in Figs.~\ref{fig:mass_planes_uncapped} and~\ref{fig:mass_planes_capped} are visible as regions of greater likelihood. In Fig.~\ref{fig:mass_planes_loglikes_groupedsearches} we consider the profile likelihood surface for the $(m_{\tilde{\chi}_2^0},m_{\tilde{\chi}_1^0})$ plane and break the total log-likelihood down into contributions from photon searches, lepton searches, other searches and measurements of SM-like final states. Finally, in the six panels of Fig.\ \ref{fig:mass_planes_xsec_and_BRs} we show the total electroweakino LHC production cross-section and a selection of relevant branching ratios across the $(m_{\tilde{\chi}_2^0},m_{\tilde{\chi}_1^0})$ profile likelihood surface.

The top-left panel of Fig.\ \ref{fig:mass_planes_loglikes_groupedsearches} shows that for the scenarios with a bino NLSP (see Fig.\ \ref{fig:mass_planes_N1_mixture}, right), the most constraining LHC analyses are the photons + $E_T^\textrm{miss}$ searches. This can be understood from the fact that for these scenarios the dominant $\tilde{\chi}_1^0$ decay mode is $\tilde{\chi}_1^0 \rightarrow \gamma \gravitino$ (Fig.\ \ref{fig:mass_planes_xsec_and_BRs}, top right), while the heavier wino- or Higgsino-dominated electroweakinos, which here dominate the production cross-section, decay via the $\tilde{\chi}_1^0$ rather than directly to a $\gravitino$ final state (Fig.\ \ref{fig:mass_planes_xsec_and_BRs}, bottom right). Towards larger masses for the heavier electroweakinos the production cross-section diminishes (Fig.\ \ref{fig:mass_planes_xsec_and_BRs}, top left) enough to leave an allowed region at $m_{\tilde{\chi}_1^0} \lesssim 450\,\GeV$ and $m_{\tilde{\chi}_2^0}, m_{\tilde{\chi}_1^\pm} \gtrsim 800\,\GeV$.

In the middle sector of the $(m_{\tilde{\chi}_2^0},m_{\tilde{\chi}_1^0})$ plane, where the highest-likelihood scenarios are wino NLSP scenarios (see Fig.\ \ref{fig:mass_planes_N1_mixture}, middle), the most important contributions to the profile likelihood surface come from the leptons + $E_T^\textrm{miss}$ searches (Fig.\ \ref{fig:mass_planes_loglikes_groupedsearches}, top right), and searches for jets + $E_T^\textrm{miss}$ final states, with or without leptons (Fig.\ \ref{fig:mass_planes_loglikes_groupedsearches}, bottom left). This is largely explained by the fact that the dominant decay modes of the now wino-dominated and near mass-degenerate $\tilde{\chi}_1^0$ and $\tilde{\chi}_1^\pm$ are $\tilde{\chi}_1^0 \rightarrow Z \gravitino$ and $\tilde{\chi}_1^\pm \rightarrow W^\pm \gravitino$, respectively (Fig.\ \ref{fig:mass_planes_xsec_and_BRs}, middle right and bottom left). Thus, $\tilde{\chi}_1^\pm \tilde{\chi}_1^0$ production will for these scenarios typically give rise to the same collider signatures as the commonly studied SUSY scenarios where wino-dominated $\tilde{\chi}_1^\pm \tilde{\chi}_2^0$ are produced and decay to final states with a stable, light $\tilde{\chi}_1^0$ through $\tilde{\chi}_2^0 \rightarrow Z \tilde{\chi}_1^0$ and $\tilde{\chi}_1^\pm \rightarrow W^\pm \tilde{\chi}_1^0$. However, while $\tilde{\chi}_1^\pm \tilde{\chi}_1^0$ is the most important production mode for these \GEWMSSM scenarios, relevant signal contributions can also arise from production of some of the heavier, Higgsino-dominated electroweakinos. Towards low $m_{\tilde{\chi}_1^0}$ ($m_{\tilde{\chi}_1^0} \lesssim 200\,\GeV$), phase space suppression of the $\tilde{\chi}_1^0 \rightarrow Z \gravitino$ decay makes $\tilde{\chi}_1^0 \rightarrow \gamma \gravitino$ the dominant decay mode for $\tilde{\chi}_1^0$ (Fig.\ \ref{fig:mass_planes_xsec_and_BRs}, top right). Here the photons + $E_T^\textrm{miss}$ searches contribute strongly to the total log-likelihood, as does the measurements of SM signatures, to be discussed in more detail below (Fig.\ \ref{fig:mass_planes_loglikes_groupedsearches}, top left and bottom right). At around $m_{\tilde{\chi}_1^0} \sim 450\,\GeV$, the reduction in the production cross-section with increasing mass (Fig.\ \ref{fig:mass_planes_xsec_and_BRs}, top left), combined with a balancing of the $\tilde{\chi}_1^0 \rightarrow \gamma \gravitino$ and $\tilde{\chi}_1^0 \rightarrow Z \gravitino$ branching ratios (Fig.\ \ref{fig:mass_planes_xsec_and_BRs}, top right and middle right) means that the combined constraining power of the searches is sufficiently weakened so that a horizontal band in the mass plane avoids exclusion at the $2\sigma$ level. This is also partly due to the model fitting some weak excesses in leptons + $E_T^\textrm{miss}$ and photons + $E_T^\textrm{miss}$ searches (light blue bands in Fig.\ \ref{fig:mass_planes_loglikes_groupedsearches}, top left and top right). However, towards even higher $m_{\tilde{\chi}_1^0}$, the ATLAS search for $E_T^\textrm{miss}$ + boosted bosons~\cite{ATLAS:2021yqv} gains sensitivity (Fig.\ \ref{fig:mass_planes_loglikes_groupedsearches}, bottom left) and the total likelihood therefore drops below the $2\sigma$ threshold for $m_{\tilde{\chi}_1^0}$ between $\sim 500\,\GeV$ and $\sim 700\,\GeV$.

As discussed above, the overall highest-likelihood scenarios are Higgsino NLSP scenarios, close to the diagonals of the $(m_{\tilde{\chi}_2^0},m_{\tilde{\chi}_1^0})$ and $(m_{\tilde{\chi}_1^\pm},m_{\tilde{\chi}_1^0})$ planes. Here the model obtains positive contributions to $\Delta \ln \mathcal{L}_\text{searches}$ from small excesses in leptons + $E_T^\textrm{miss}$ searches and the ATLAS $b$-jets + $E_T^\textrm{miss}$ search (Fig.\ \ref{fig:mass_planes_loglikes_groupedsearches}, top right and bottom left). Some examples of the balancing of different branching ratios that these scenarios exhibit, discussed in Sec.\ \ref{subsec:best_fit_scenarios}, can be seen in the middle left, middle right and bottom left panels of Fig.\ \ref{fig:mass_planes_xsec_and_BRs}.

\subsection{Impact of measurements}
\label{subsec:impact-measurements}

The present study is the first to include LHC measurements of SM signatures in a many-parameter BSM global fit. It is therefore interesting to explore what impact these likelihood contributions have on our results. The log-likelihood contribution $\Delta \ln \mathcal{L}_\text{meas}$ on the $(m_{\tilde{\chi}_2^0},m_{\tilde{\chi}_1^0})$ profile-likelihood surface is shown in the bottom-right panel of Fig.\ \ref{fig:mass_planes_loglikes_groupedsearches}. The contribution is significant in the regions with wino- or Higgsino-dominated $\tilde{\chi}_1^0$ with $m_{\tilde{\chi}_1^0} \lesssim 200\,\GeV$, where $\BR(\tilde{\chi}_1^0 \rightarrow \gamma \gravitino)$ is large. In particular, the SM signature measurements contribute to excluding low-mass scenarios where the constraints from leptons + $E_T^\textrm{miss}$ searches would otherwise have been largely balanced by positive log-likelihood contributions from the photons + $E_T^\textrm{miss}$ searches (Fig.\ \ref{fig:mass_planes_loglikes_groupedsearches}, top panels, $m_{\tilde{\chi}_1^0} \lesssim 100\,\GeV$).

\begin{figure*} 
  \centering
  \includegraphics[height=0.8\columnwidth]{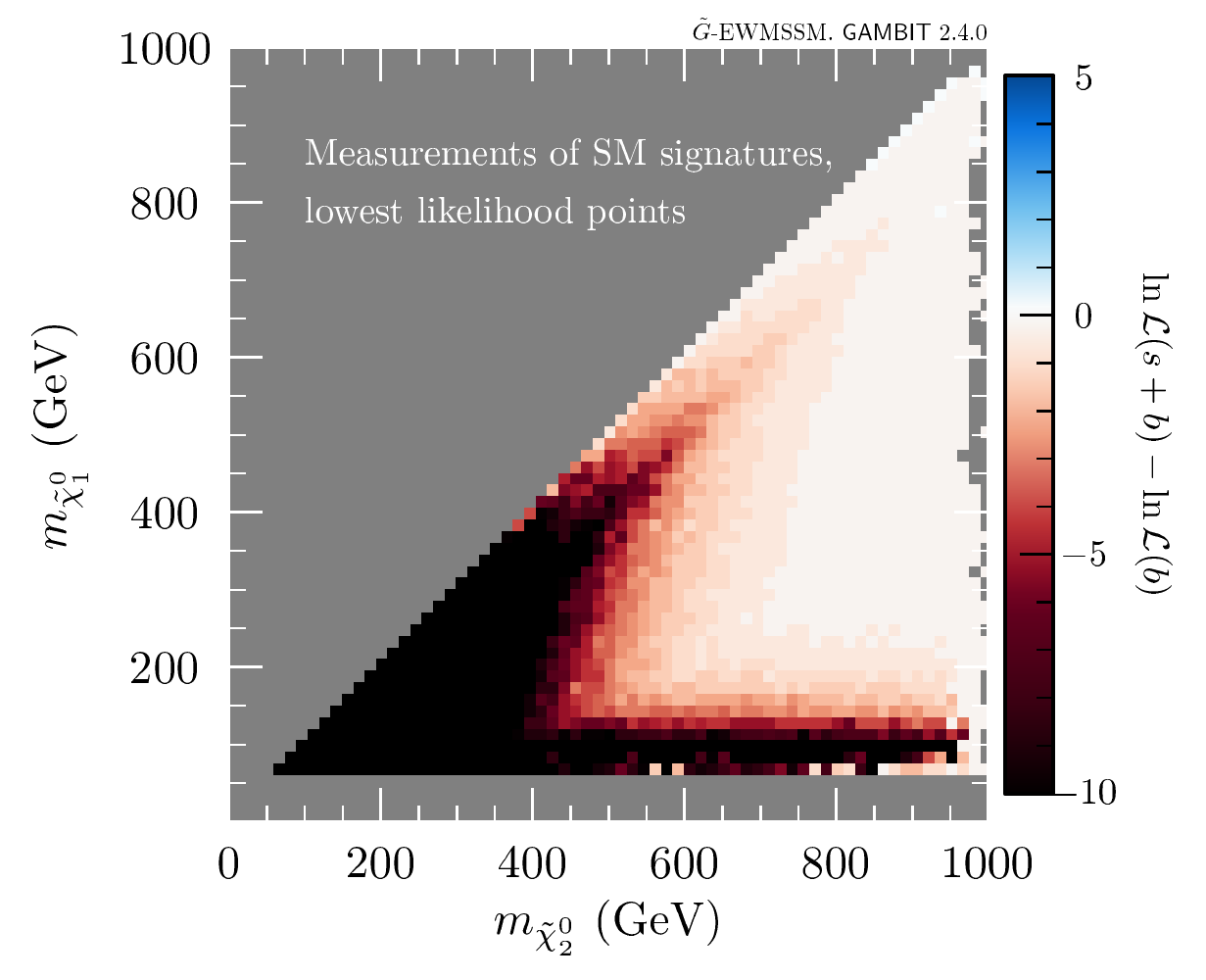}
  \includegraphics[height=0.8\columnwidth]{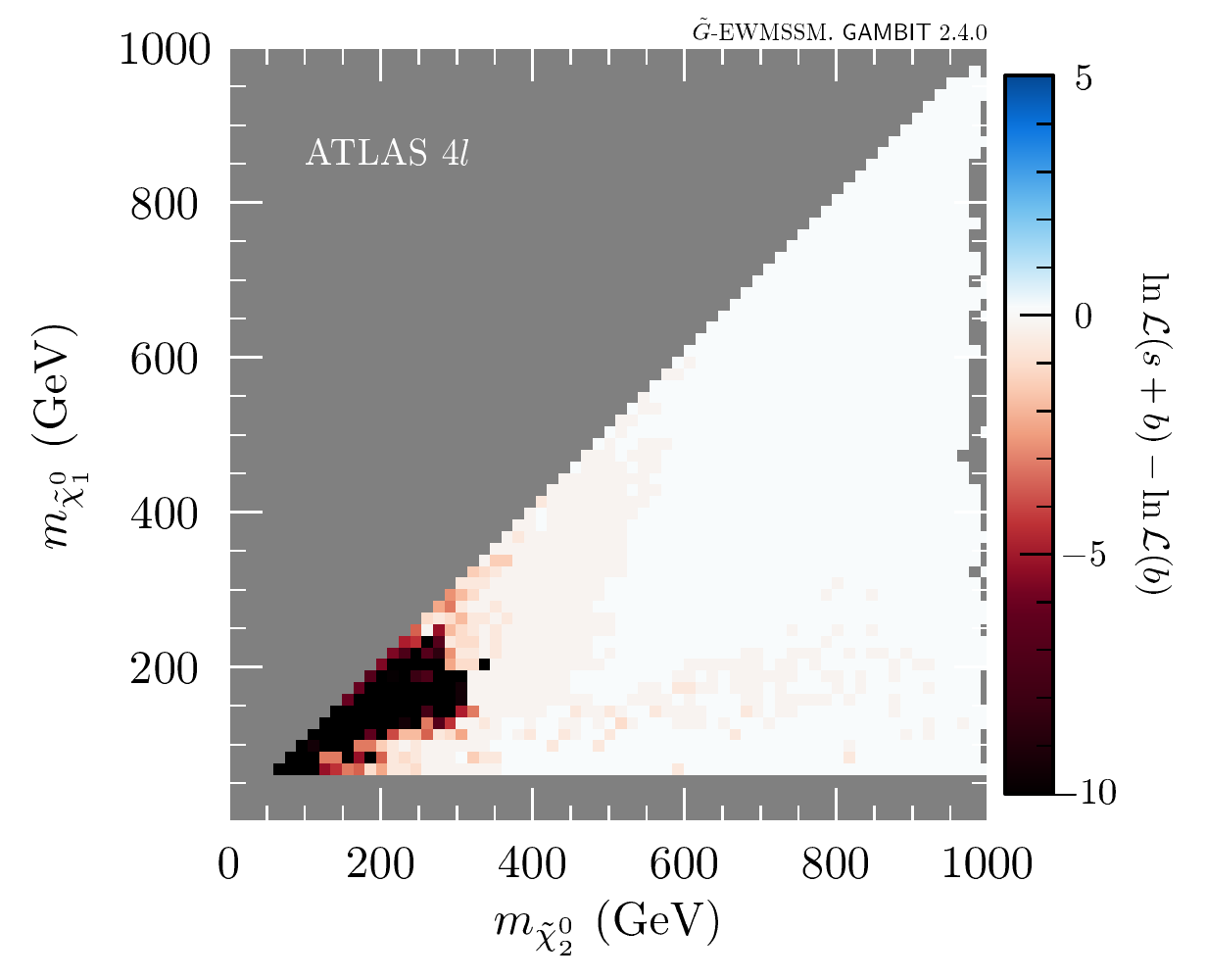}\\
  \includegraphics[height=0.8\columnwidth]{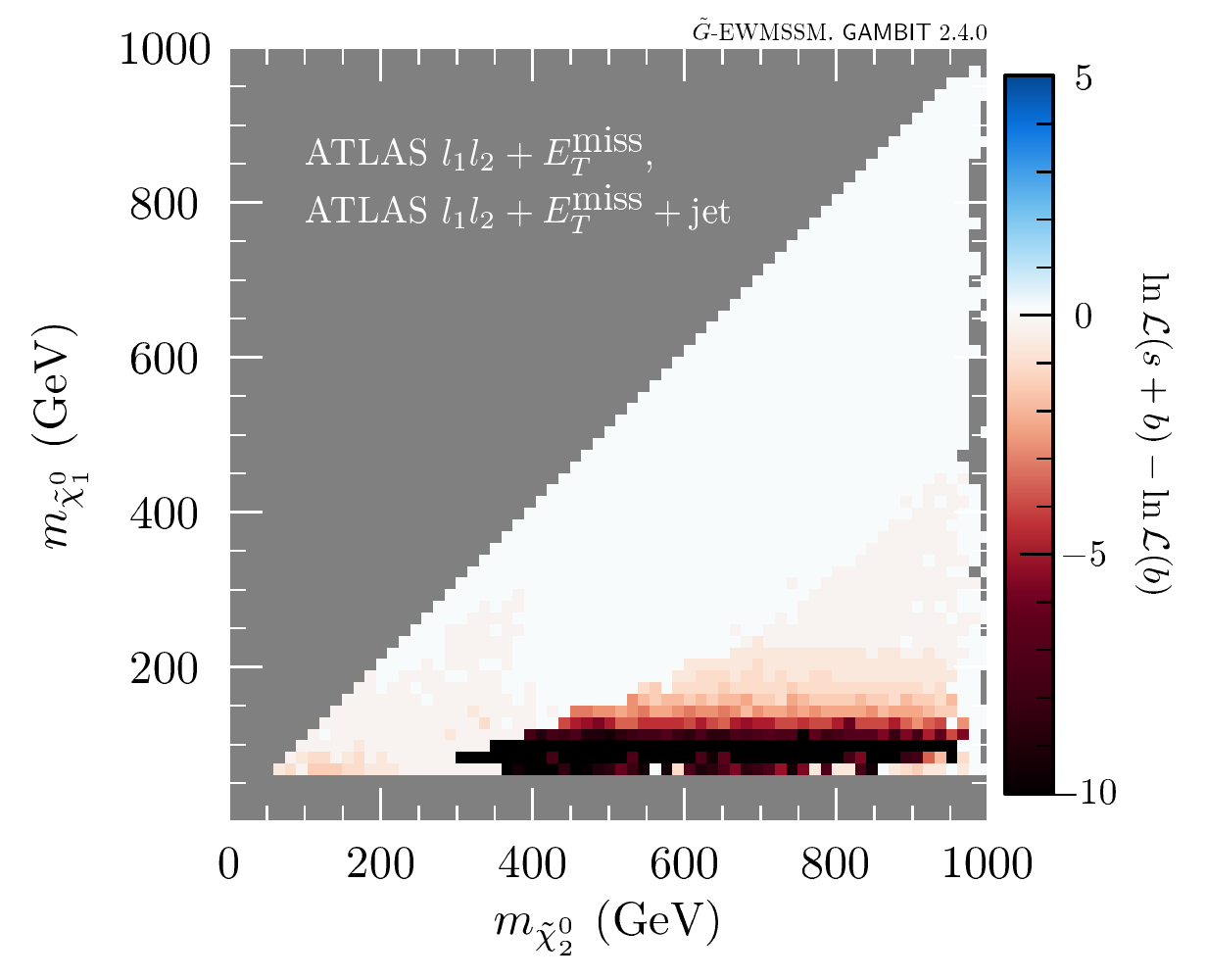}
  \includegraphics[height=0.8\columnwidth]{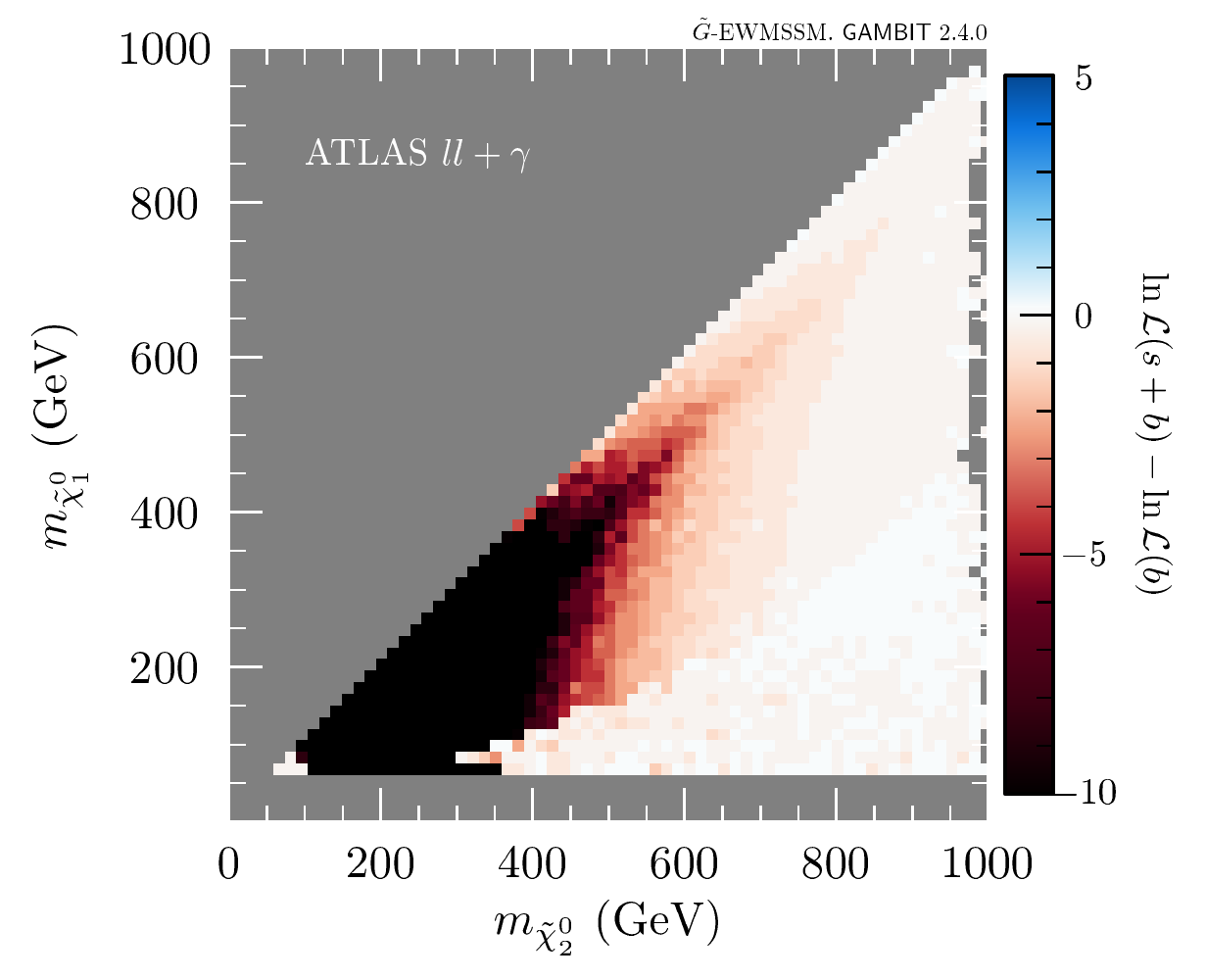}\\
  \caption{
  Log-likelihood contributions from different pools of LHC measurements, plotted across the $(m_{\tilde{\chi}_2^0},m_{\tilde{\chi}_1^0})$ plane for the scan points where the combined constraint from the LHC measurements is at its strongest.}
  \label{fig:mass_planes_loglikes_groupedmeasurements}
\end{figure*}

The $(m_{\tilde{\chi}_2^0},m_{\tilde{\chi}_1^0})$ profile likelihood surface discussed above is by definition made up of the overall \textit{least} constrained parameter sample within each $(m_{\tilde{\chi}_2^0},m_{\tilde{\chi}_1^0})$ bin. To get a more complete picture of the constraining power of the SM signature measurements, it is interesting to also look at $\Delta \ln \mathcal{L}_\text{meas}$ across the surface of parameter samples that are \textit{most strongly} constrained by this log-likelihood contribution. This is shown in the top-left panel of Fig.\ \ref{fig:mass_planes_loglikes_groupedmeasurements}. For the \GEWMSSM scenarios where the SM signature measurements have their largest sensitivity, they rule out scenarios that have both $m_{\tilde{\chi}_2^0}$ and $m_{\tilde{\chi}_1^0}$ below $\sim500\,\GeV$, and scenarios towards higher $m_{\tilde{\chi}_2^0}$ when $m_{\tilde{\chi}_1^0} \lesssim 150\,\GeV$. 
The three other panels in Fig.\ \ref{fig:mass_planes_loglikes_groupedmeasurements} show the individual log-likelihood contributions from the pools of measurements that contribute most strongly to the combined $\Delta \ln \mathcal{L}_\text{meas}$ in the upper-left panel: 
ATLAS measurements of the $pp \rightarrow ZZ \rightarrow 4l$ cross-section (top right) \cite{ATLAS:2021kog,ATLAS:2019qet,ATLAS:2017bcd}; 
ATLAS measurements of final states with two different flavour leptons and missing energy, with or without jets (bottom left) \cite{ATLAS:2021jgw,ATLAS:2019rob,ATLAS:2019ebv,ATLAS:2019hau}, where the dominant contribution is coming from the $pp \rightarrow W^+ W^-$ cross-section measurements in \cite{ATLAS:2021jgw,ATLAS:2019rob}; 
and an ATLAS measurement of the $pp \rightarrow Z(\rightarrow l^+ l^-) \gamma + X$ production cross-section \cite{ATLAS:2019gey} (bottom right).

\begin{figure*} 
  \centering
  \includegraphics[height=0.55\columnwidth]{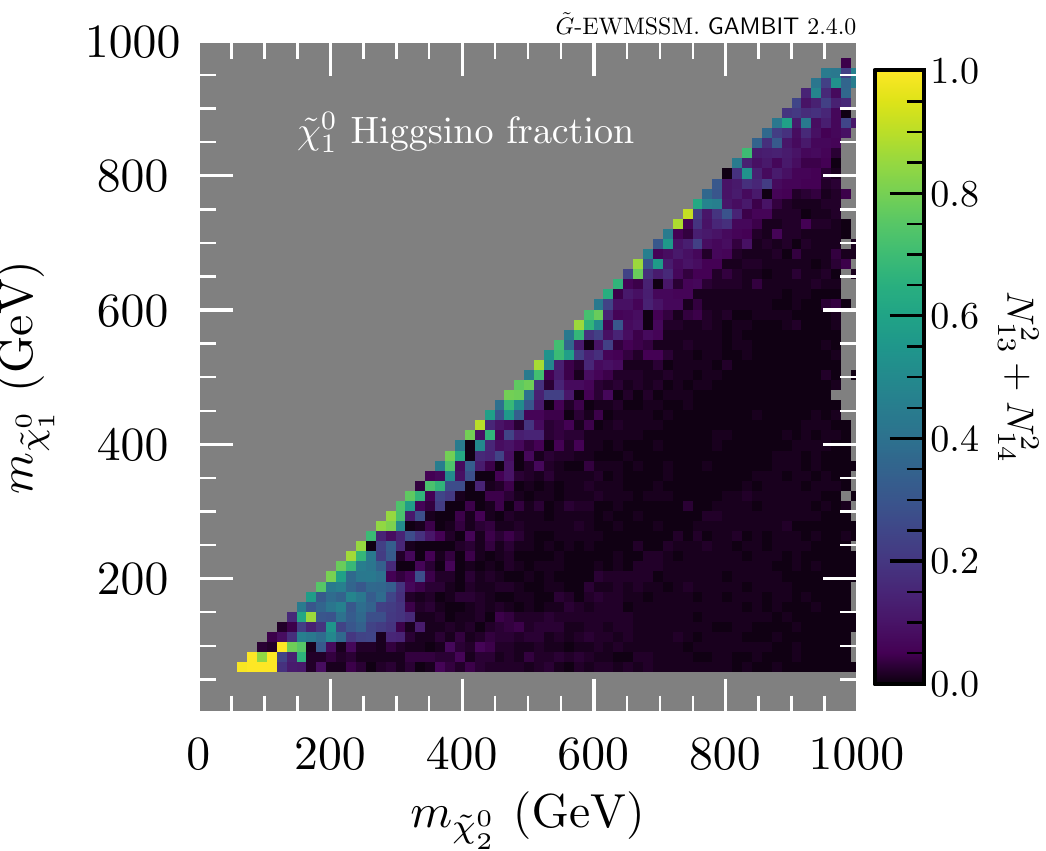}
  \includegraphics[height=0.55\columnwidth]{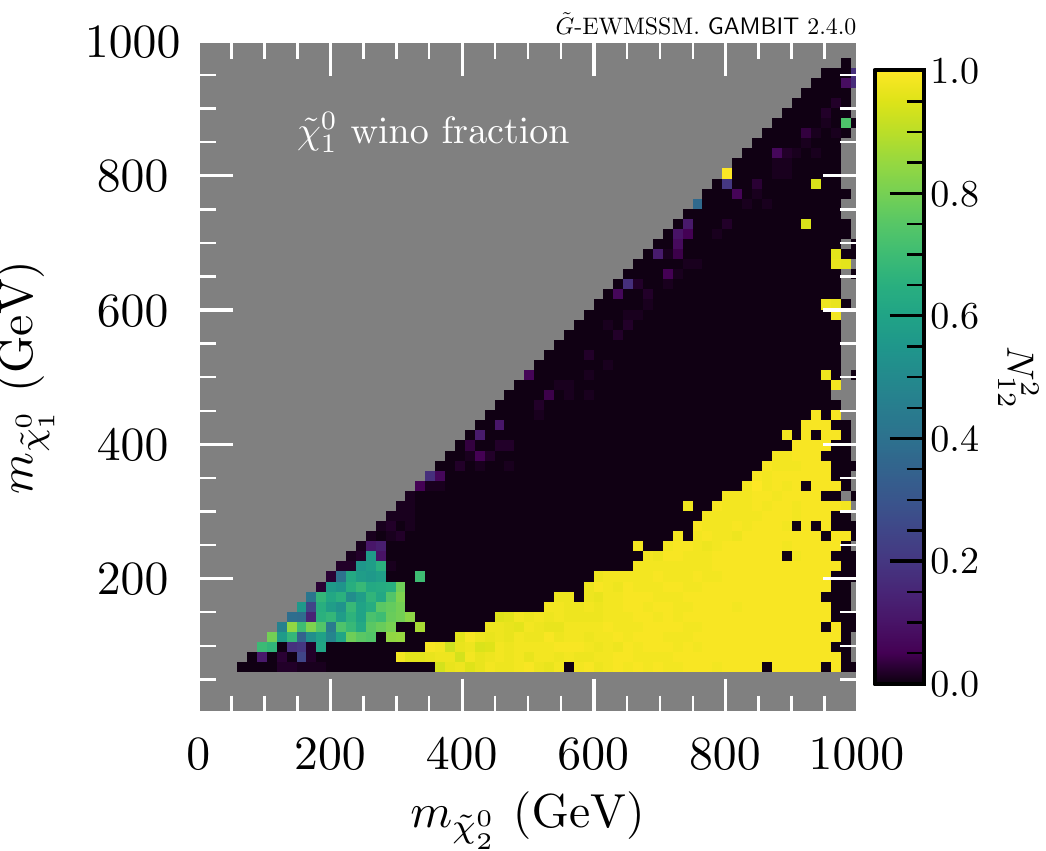}
  \includegraphics[height=0.55\columnwidth]{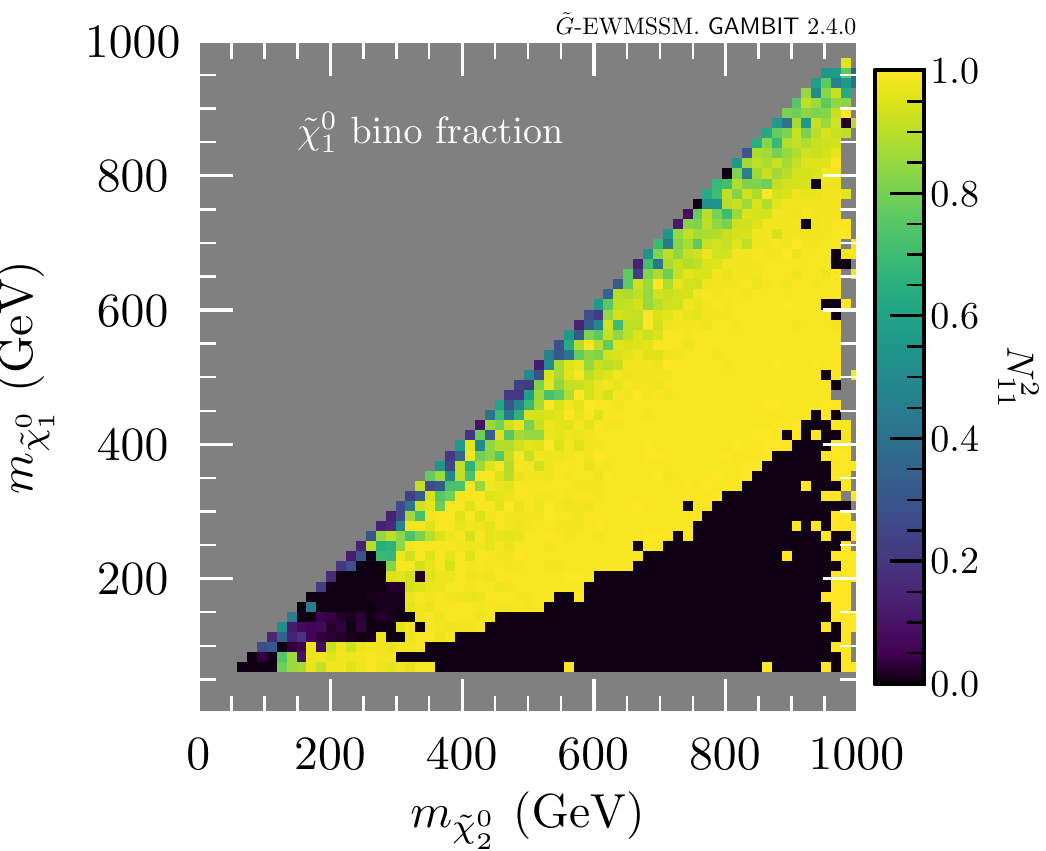}
  \caption{The Higgsino (left), wino (middle) and bino (right) fraction of the $\tilde{\chi}_1^0$, plotted across the $(m_{\tilde{\chi}_2^0},m_{\tilde{\chi}_1^0})$ plane for the scan points where the combined constraint from the LHC measurements is the largest.}
  \label{fig:mass_planes_N1_mixture_lowest_loglike}
\end{figure*}
In Fig.\ \ref{fig:mass_planes_N1_mixture_lowest_loglike} we show the $\tilde{\chi}_1^0$ composition for the parameter samples contributing to Fig.\ \ref{fig:mass_planes_loglikes_groupedmeasurements}. From Figs.\ \ref{fig:mass_planes_loglikes_groupedmeasurements} and \ref{fig:mass_planes_N1_mixture_lowest_loglike} we see that the $ZZ$ cross-section measurements most strongly constrain low-mass scenarios where the $\tilde{\chi}_1^0$ is dominantly Higgsino or a wino-Higgsino mixture. These \GEWMSSM scenarios combine a large total electroweakino production cross-section,\footnote{A balanced wino-Higgsino mixture for a low-mass $\tilde{\chi}_1^0$ implies that $M_2$ for these points typically is within $\sim100\,\GeV$ of $|\mu|$. This means that at least four of the five heavier electroweakino states will have masses not too much larger than $m_{\tilde{\chi}_1^0}$.} with significant branching ratios for some of the decays $\tilde{\chi}_{i}^0 \rightarrow Z \gravitino$ and/or $\tilde{\chi}_{i}^0 \rightarrow Z \tilde{\chi}_{j}^0$.
The measurements of $W^+ W^-$ production cross-sections exclude low-mass scenarios with wino-dominated $\tilde{\chi}_1^0$. Here the strongest $W^+W^-$ signal contribution comes from the production of pairs of light, wino-dominated $\tilde{\chi}_1^\pm$, which decay as $\tilde{\chi}_1^\pm \rightarrow W^\pm \gravitino$.
Finally, the $Z(\rightarrow l^+ l^-) \gamma + X$ cross-section measurement constrains scenarios with bino-dominated $\tilde{\chi}_1^0$. These scenarios typically have a large $\BR(\tilde{\chi}_1^0 \rightarrow \gamma \gravitino)$ and a non-negligible $\BR(\tilde{\chi}_1^0 \rightarrow Z \gravitino)$, such that production of any pair of electroweakinos that decay to $\tilde{\chi}_1^0$'s can result in signal contributions to the measured cross-section. 

\begin{figure*} 
  \centering
  \includegraphics[width=0.8\textwidth]{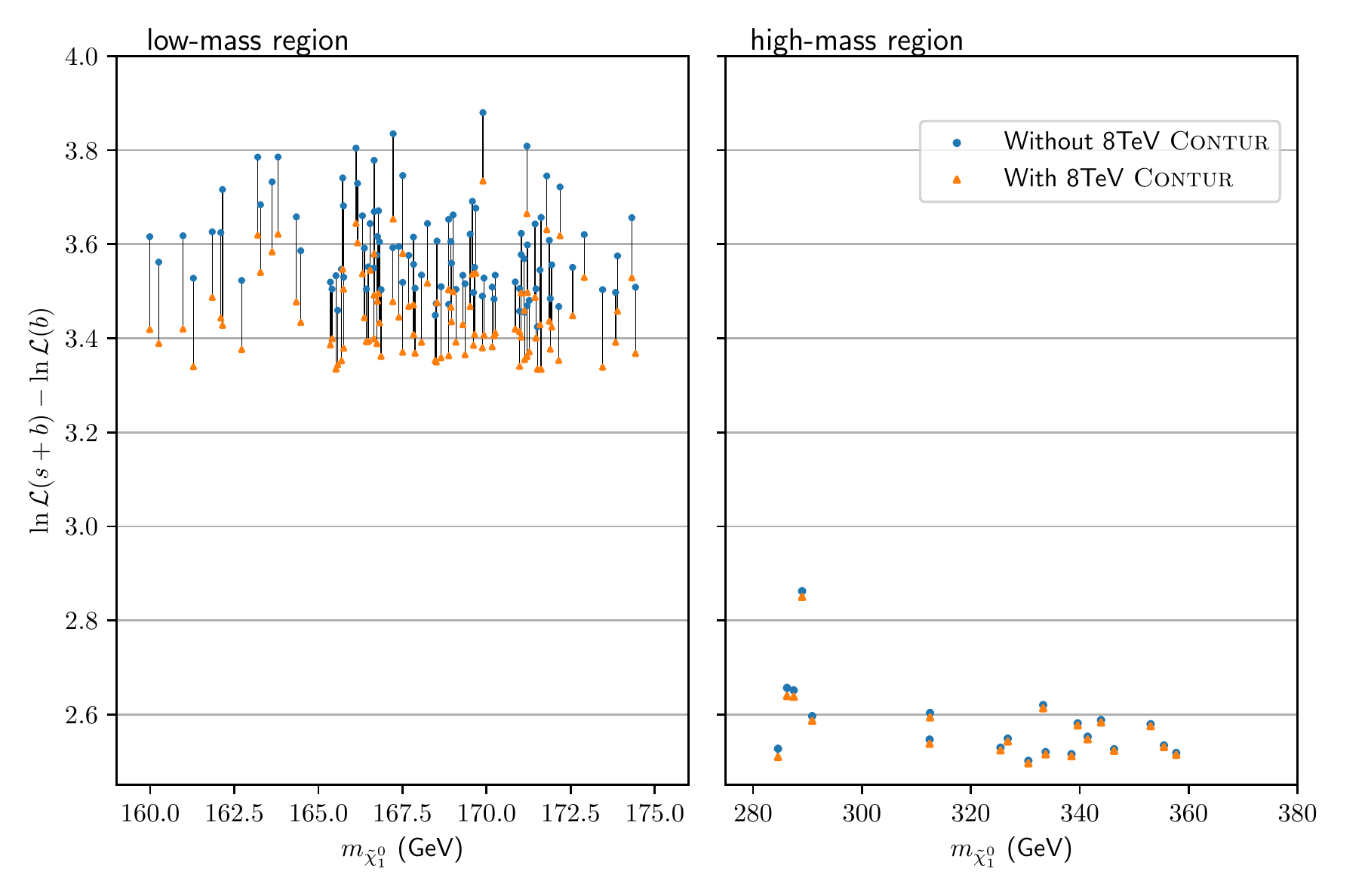}
  \caption{The log-likelihood impact from including $8\,\TeV$ LHC measurements of SM signatures, shown for the highest-likelihood scan points in the mass region around the best-fit point (left) and a higher-mass region (right).}
  \label{fig:8TeV_impact}
\end{figure*}
Since the best-fit region predicts light Higgsinos, at masses around $170\,\GeV$, the LHC searches and measurements performed at $\sqrt{s} = 8\,\TeV$ can also be relevant. A full investigation of the impact of $8\,\TeV$ results is beyond the scope of this study, as it would effectively double the computational cost of our parameter scans. However, to gauge the possible impact, we generate \num{100000} events at $8\,\TeV$ for each of our 100 highest-likelihood parameter points. We pass the events through \rivet and \contur to compute a log-likelihood contribution from the collection of $\sqrt{s} = 8\,\TeV$ measurements in \rivet. The result of this is illustrated in Fig.\ \ref{fig:8TeV_impact}, where we show the change in the total log-likelihood for each point when the contribution from $8\,\TeV$ measurements is added. In the left-hand panel we show the points close to the best-fit point at $m_{\tilde{\chi}_1^0} \sim 170\,\GeV$. Of our 100 highest-likelihood points, some also belong to the higher-mass region, at $m_{\tilde{\chi}_1^0} \gtrsim 280\,\GeV$, shown in the right-hand panel. For the best-fit points in the low-mass region, including the $8\,\TeV$ measurements reduces the total log-likelihood by around $0.2$ units. As expected, there is a smaller impact on points in the higher-mass region.

\subsection{Scenarios with a chargino NLSP}

\begin{figure*} 
  \centering
  \includegraphics[height=0.8\columnwidth]{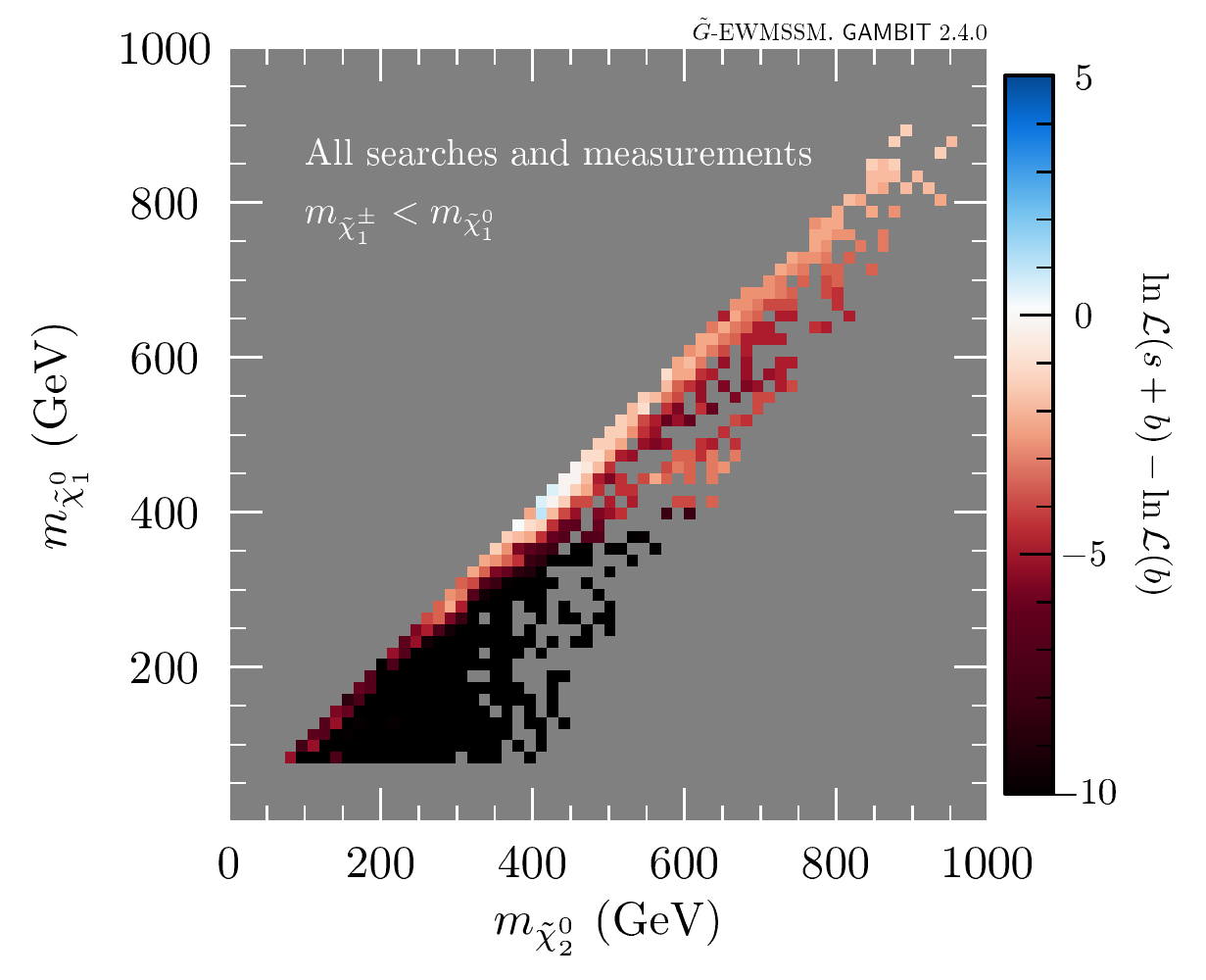}
  \includegraphics[height=0.8\columnwidth]{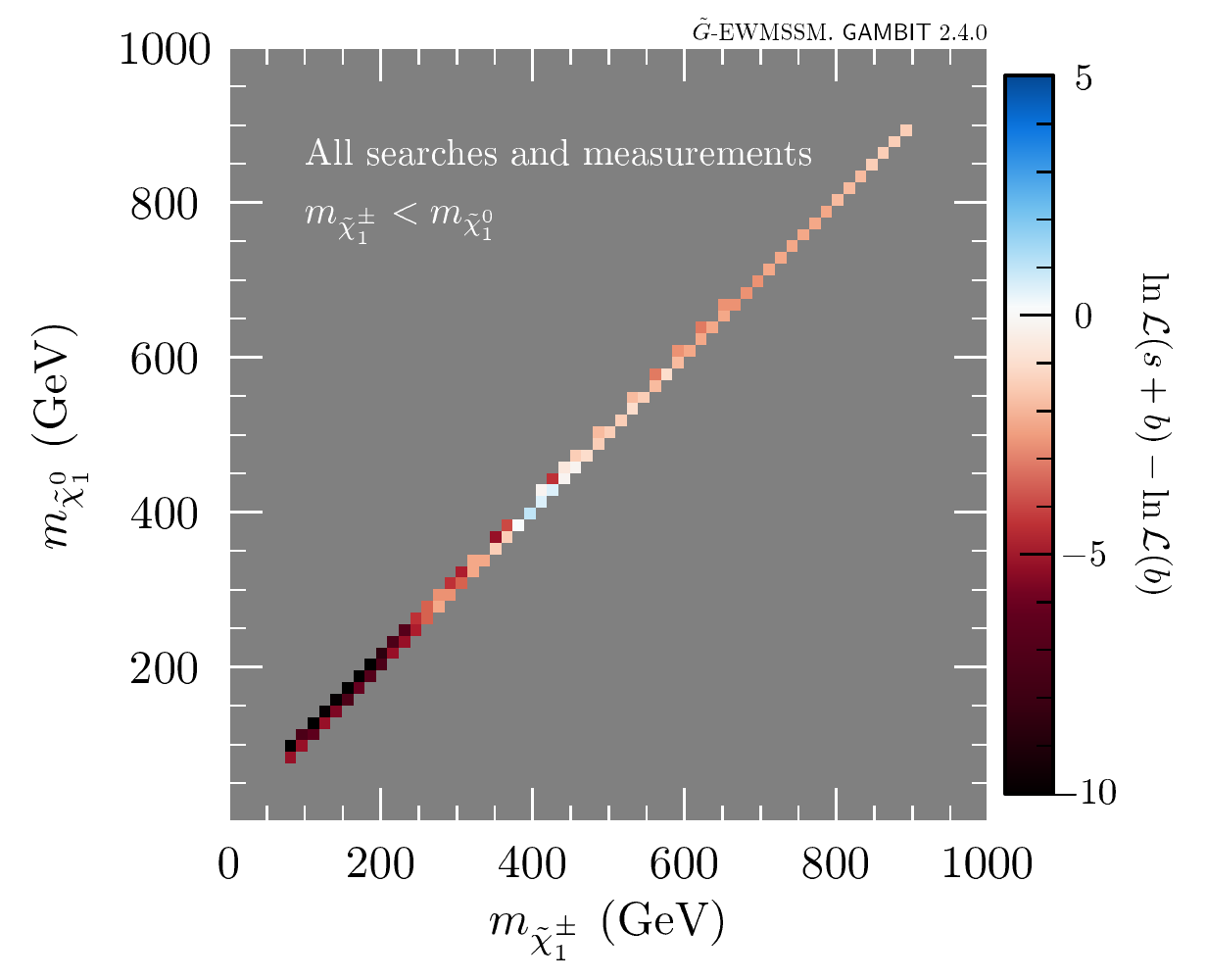}
  \caption{
  The total log-likelihood plotted across the profile-likelihood surface for the subset of points with $m_{\tilde{\chi}_1^\pm} < m_{\tilde{\chi}_1^0}$, shown in the $(m_{\tilde{\chi}_2^0},m_{\tilde{\chi}_1^0})$ plane (left) and in the $(m_{\tilde{\chi}_1^{\pm}},m_{\tilde{\chi}_1^0})$ plane (right).}
  \label{fig:mass_planes_loglikes_all_C1NLSP}
\end{figure*}

In contrast with the EWMSSM, the \GEWMSSM admits the possibility of a chargino as the lightest electroweakino. Such a scenario was highlighted in Fig.~\ref{fig:neutralinoBR} where the gray band signals a sudden drop in branching ratio due to $m_{\tilde{\chi}_1^{\pm}} < m_{\tilde{\chi}_1^{0}}$. While rare for MSSM-like electroweakino mass matrices, and featuring small mass differences, our scan identified still-viable parameter regions with $m_{\tilde{\chi}_1^{\pm}} < m_{\tilde{\chi}_1^{0}}$, shown in  Fig.~\ref{fig:mass_planes_loglikes_all_C1NLSP}. 

We find that in these cases, the points with the highest likelihoods have Higgsino-like electroweakinos, with only small splittings for the $\tilde{\chi}_1^{\pm}$, $\tilde{\chi}_1^0$ and $\tilde{\chi}_2^0$, with masses preferred to be in the region of $400$--$500\,\GeV$. Here, the decay mode for $\tilde{\chi}_1^{\pm}$ is always $\tilde{\chi}_1^{\pm}\rightarrow W \gravitino$. Hence, the detectable signal for $\tilde{\chi}_1^{\pm}\tilde{\chi}_1^{\pm}$ pair production is two on-shell $W$ bosons and some missing energy from the gravitinos. For the $\tilde{\chi}_1^{0}$, the dominant decay modes are $\tilde{\chi}_1^{0}\rightarrow Z \gravitino$ and $\tilde{\chi}_1^{0}\rightarrow h \gravitino$ due to the dominant Higgsino component. The detectable signal for $\tilde{\chi}_1^{0}\tilde{\chi}_1^{\pm}$ production would then be on-shell $WZ$ or $Wh$ plus missing energy from the gravitinos. Finally, $\tilde{\chi}_2^0$ decays to soft SM fermions and the $\tilde{\chi}_1^0$ or $\tilde{\chi}_1^\pm$. Thus, the production of $\tilde{\chi}_1^{0}\tilde{\chi}_2^{0}$ and $\tilde{\chi}_1^{\pm}\tilde{\chi}_2^{0}$ will in effect enhance the cross sections for $\tilde{\chi}_1^{0}\tilde{\chi}_1^{\pm}$ and $\tilde{\chi}_1^{\pm}\tilde{\chi}_1^{\pm}$ production.

\section{Conclusions}
\label{sec:conclusions}
In this study we have investigated the current viability of the $\tilde G$-EWMSSM, the simplest realisation of a light supersymmetric electroweak sector together with a nearly massless gravitino LSP. We have confronted the $\tilde G$-EWMSSM with a comprehensive selection of the relevant Run 2 searches at the LHC, relevant past searches at LEP, and, we have, for the first time in a global fit, used a broad set of SM measurements at the LHC to constrain the model by building a new interface between \gambit and \contur.

Our best-fit region for the model is where $|\mu|<|M_1|,M_2$, and is characterised phenomenologically by a trio of relatively light degenerate Higgsinos in the mass range of $140$--$500\,\GeV$, with a best fit point around $170\,\GeV$. Due to the collective effect of small excesses over multiple ATLAS and CMS searches we find closed $2\sigma$ contours in the parameter space, but we emphasise that this is a model-specific best-fit region and does not constitute a measure of goodness-of-fit.

Our main result is that the bulk of the $\tilde G$-EWMSSM parameter space with electroweakino masses below $1\,\TeV$ is excluded by collider searches and measurements. The four exceptions, classified according to the nature of the lightest electroweakinos, are:
\begin{itemize}
\item[\it i)] degenerate Higgsinos from $140\,\GeV$ and up, 
\item[\it ii)] a region of degenerate winos around $400$--$500\,\GeV$ allowed at the $2\sigma$ level, 
\item[\it iii)] degenerate winos above $700\,\GeV$, and
\item[\it iv)] a `lonely' bino from $62\,\GeV$ and up, decoupled from heavier Higgsinos and winos lying above $800\,\GeV$. 
\end{itemize}

For Run 3 of the LHC the degenerate Higgsino region, {\it i)}, will be challenging to test fully.
Drawing from the lessons learnt in this study, the measurement of SM multi-lepton signatures will continue to be important to exclude parameter space at the low-mass end of the region. Potential improvements to searches sensitive to the important $\tilde{\chi}_1^{0}\to h \tilde G$ decay (see Fig.~\ref{fig:mass_planes_xsec_and_BRs}, middle left), will also improve the reach.
However, fully excluding this still very viable region will need future $e^+e^-$ or muon colliders operating at high enough centre-of-mass energies.

On the other hand, the surviving wino band, {\it ii)}, with masses around $450\,\GeV$ seems to be fully excludable with the slightly higher Run 3 centre-of-mass energy and more data, in particular since its survival is already marginal. 
For the same reason it should also be possible to push the remaining wino region, {\it iii)}, to somewhat higher masses with higher cross sections and more data. 

For the lonely bino region, {\it iv)}, the search for pair production of light binos decaying to photons is also hampered by low production cross sections. However, we expect some impact here with increasing statistics in Run 3 and beyond to the High-Luminosity LHC, in particular on the parts of parameter space where there is bino production through the decay of heavier electroweakinos, which could realistically be pushed out beyond 1 TeV. 

We emphasise the still open interesting possibility of a reverse mass hierarchy of charginos and neutralinos, with $m_{\tilde{\chi}_1^{\pm}} < m_{\tilde{\chi}_1^{0}}$, with distinct signal predictions for LHC Run 3 searches. Although the base production cross section is not so high given their Higgsino nature, the preferred region of this scenario should be within reach of Run 3 statistics and the slightly higher centre-of-mass energy, when considering all final states $WW$, $WZ$ and $Wh$.

We make all our generated parameter samples available from \Zenodo for further study \cite{Zenodo_gravitino}.

\begin{acknowledgements}
We thank our colleagues in the GAMBIT Community for helpful discussions and comments. For computing resources, we thank PRACE for awarding us access to Marconi at CINECA and Joliot-Curie at CEA. Computing resources were also provided by Sigma2, the National Infrastructure for HPC in Norway, under project NN9284K.
AF was supported by the National Natural Science Foundation of China (NNSFC) Research Fund for International Excellent Young Scientists grant 1950410509. ABu and JB were supported by the UK Science and Technology Facilities Council (STFC) Consolidated Grant programme awards ST/S000887/1 and ST/S000666/1 respectively, ABe by STFC grant ST/T00679X/1, and TP by the STFC Doctoral Training Programme. The work of CB was supported by the Australian Research Council Discovery Project grant DP210101636. AK and AR were supported by the Research Council of Norway FRIPRO grant 323985. TEG was funded by the Deutsche Forschungsgemeinschaft (DFG) through the Emmy Noether Grant No.~KA 4662/1-1. The work of VA was supported by the European Union Framework Programme for Research and Innovation Horizon 2020 (2014--2021) under the Marie Sklodowska-Curie Grant Agreement No.~765710. YZ was supported by the NNSFC under grant No.~12105248 and 12047503 (Peng-Huan-Wu Theoretical Physics Innovation Center). MJW is supported by the ARC Centre of Excellence for Dark Matter Particle Physics (CE200100008). We made use of \textsf{pippi v2.2} \cite{pippi} for this work.
\end{acknowledgements}

\begin{appendices}
\renewcommand \thetocsection {\Alph{section}}


\section{Profile likelihood maps for the input parameters} \label{app:input_params}
\begin{figure*} 
  \centering
  \includegraphics[height=0.55\columnwidth]{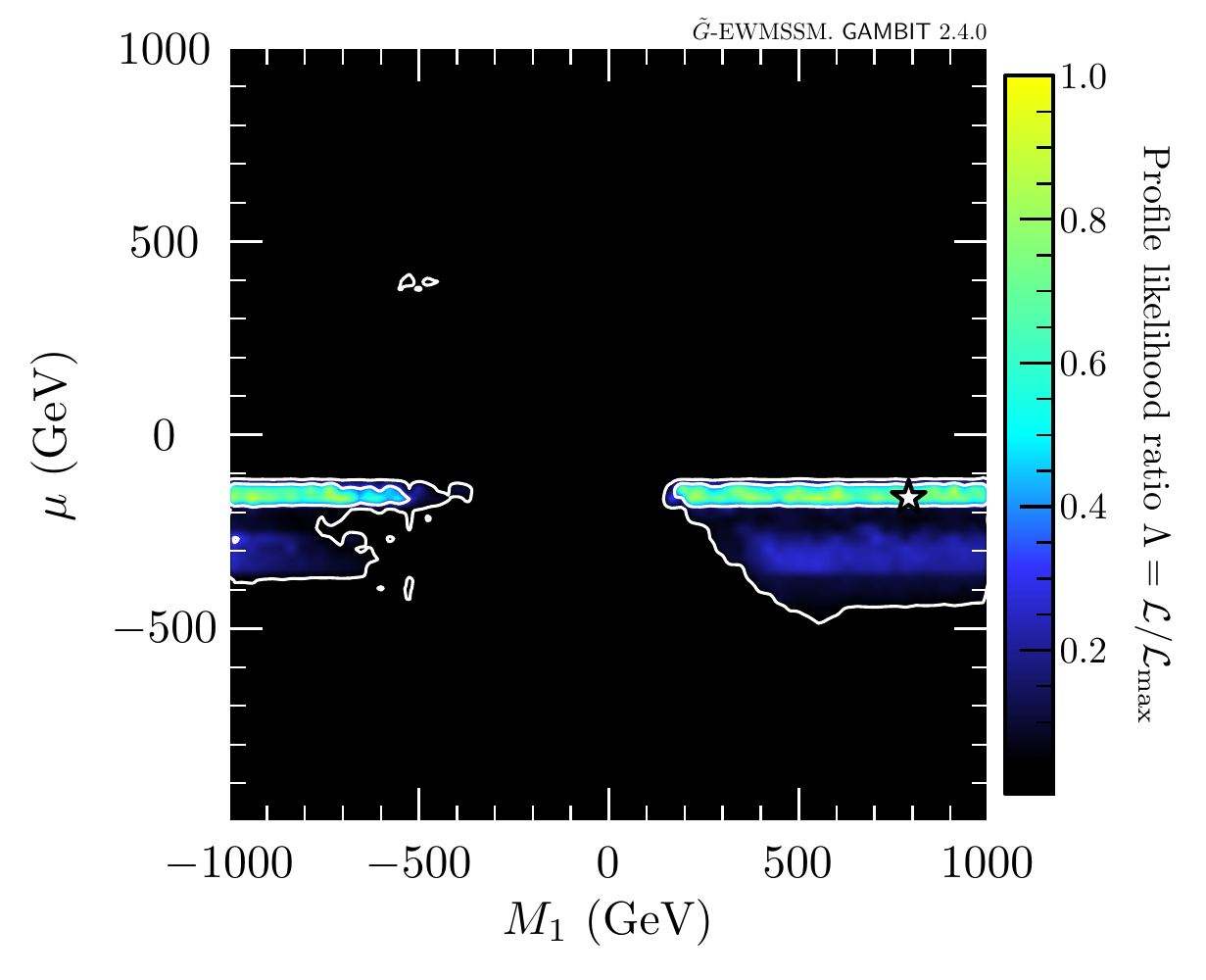}
  \includegraphics[height=0.55\columnwidth]{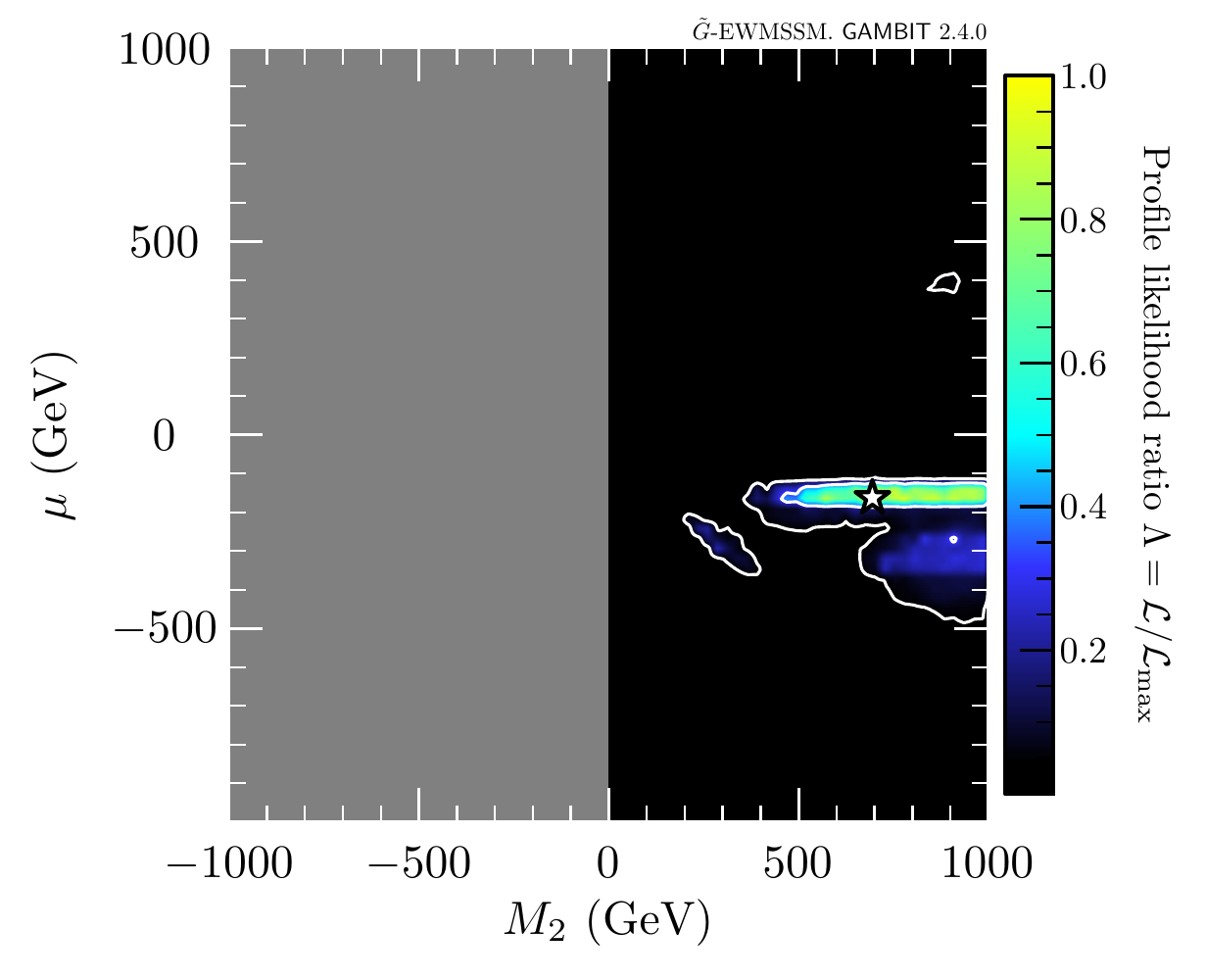}
  \includegraphics[height=0.55\columnwidth]{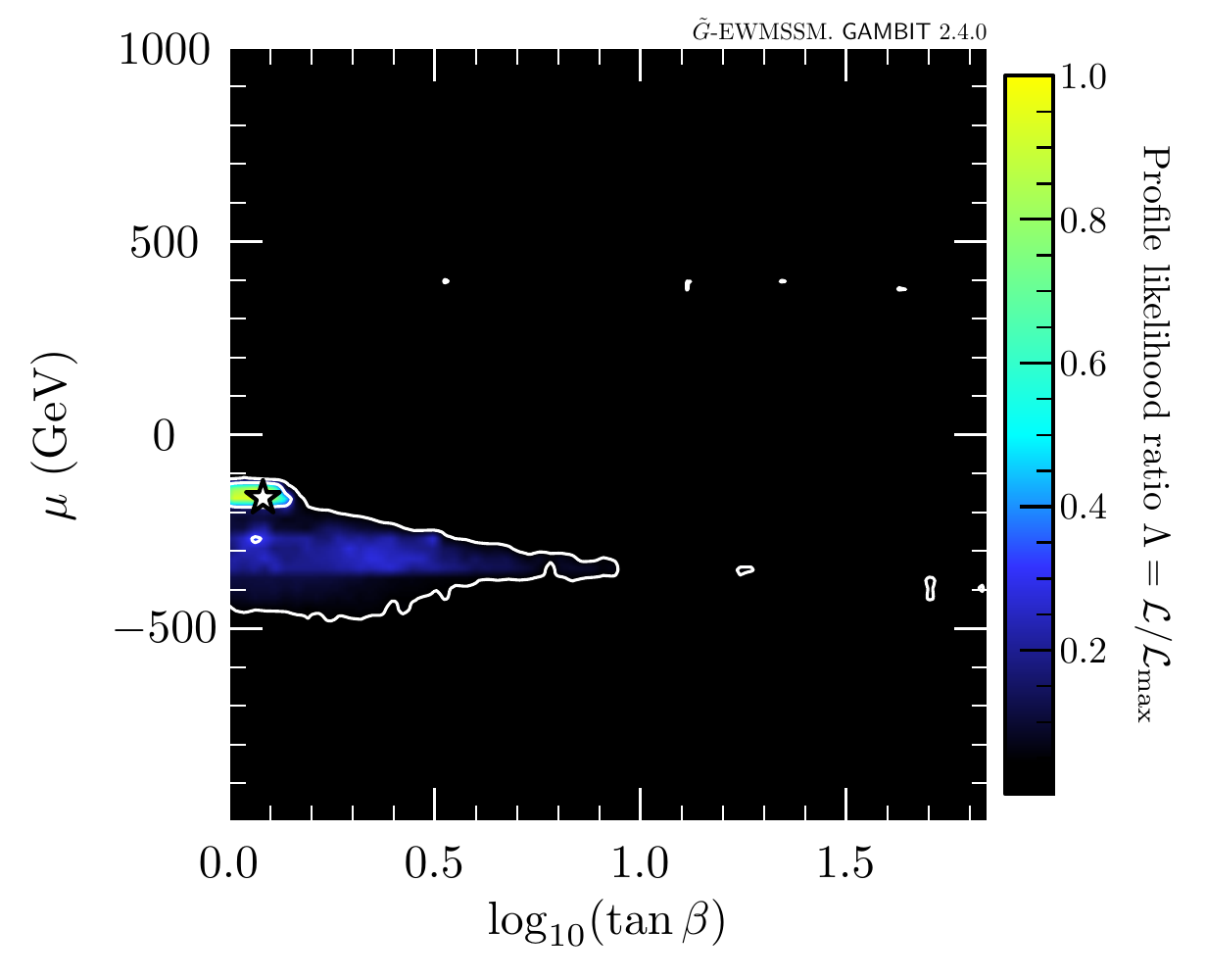} \\
  \includegraphics[height=0.55\columnwidth]{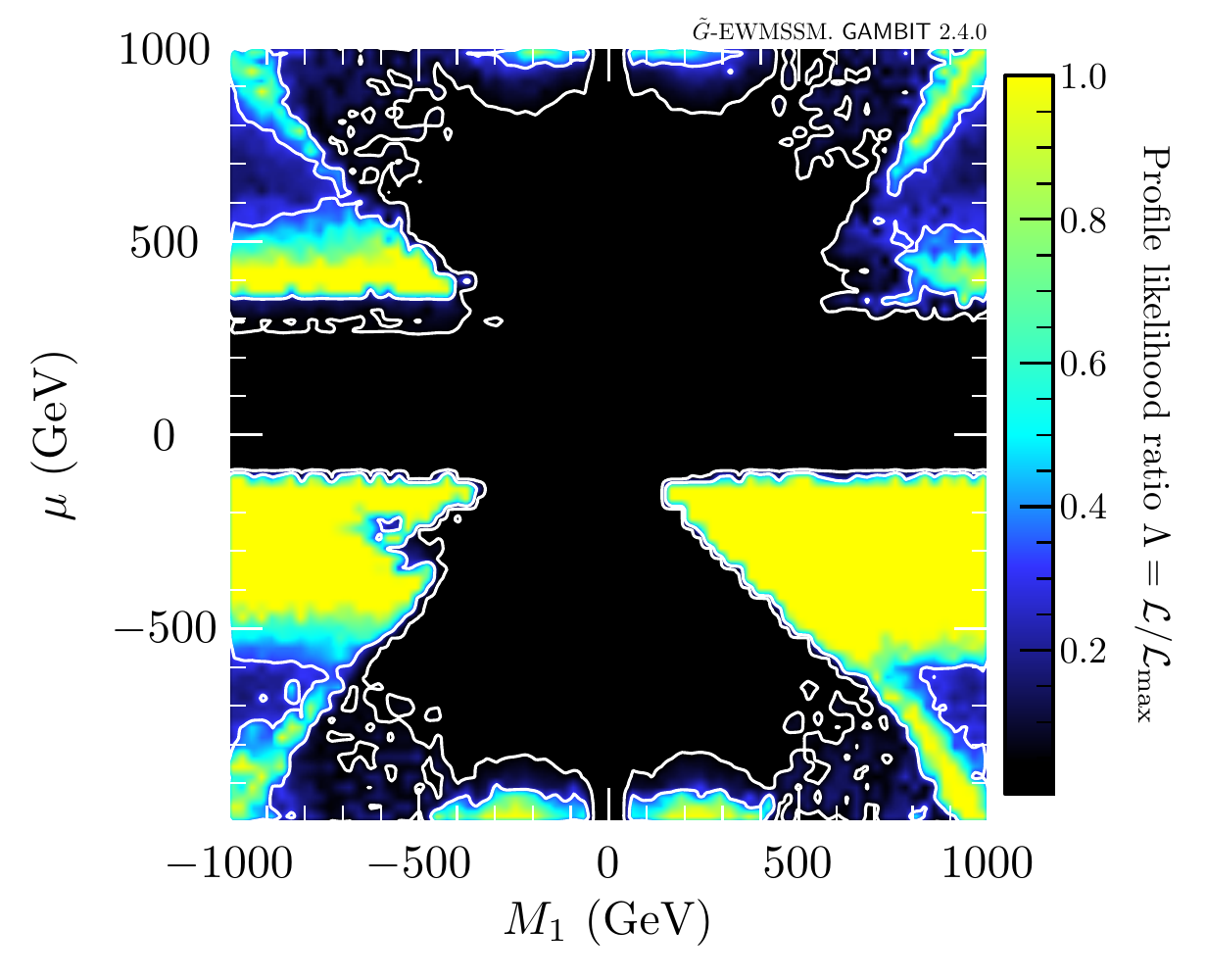}
  \includegraphics[height=0.55\columnwidth]{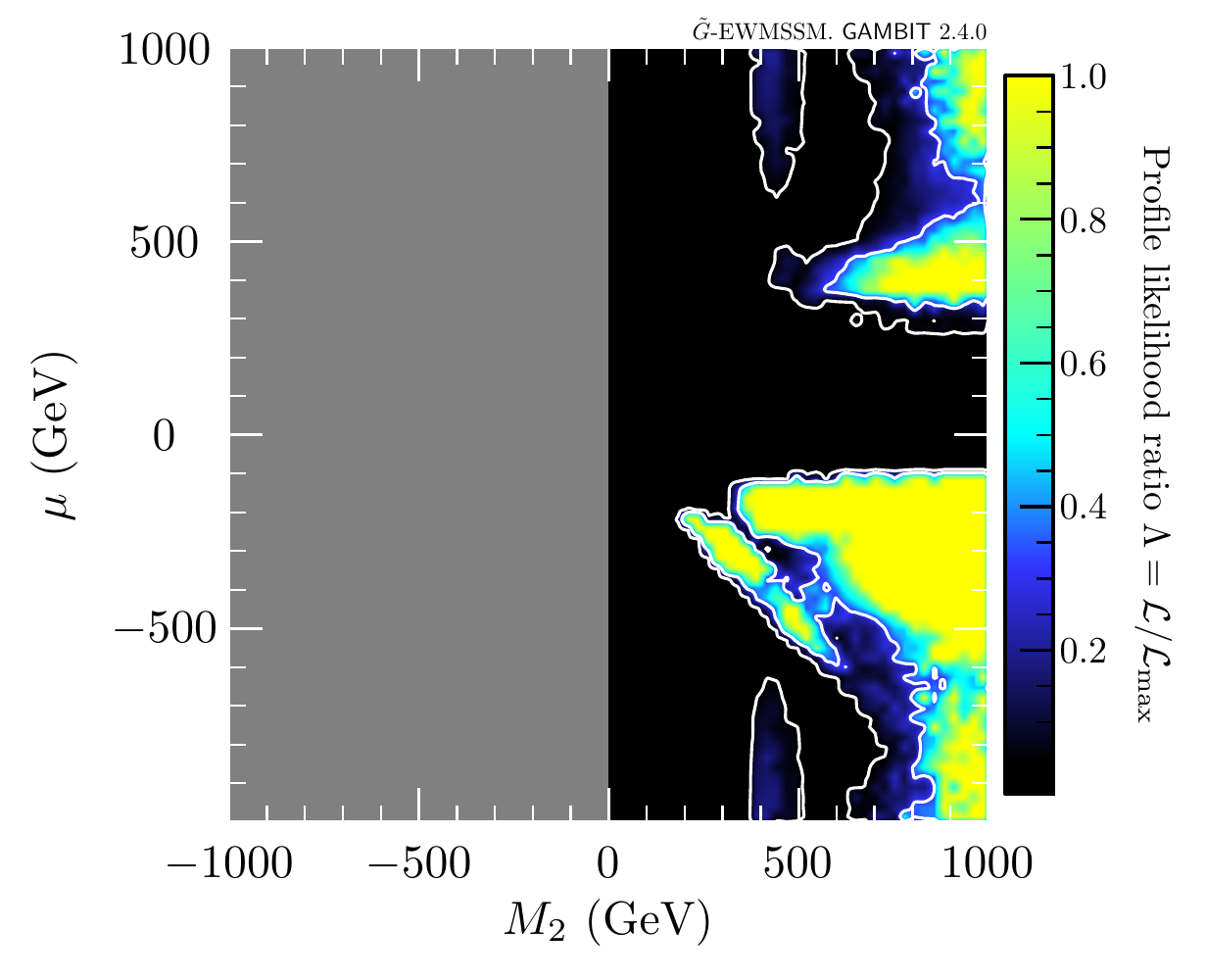}
  \includegraphics[height=0.55\columnwidth]{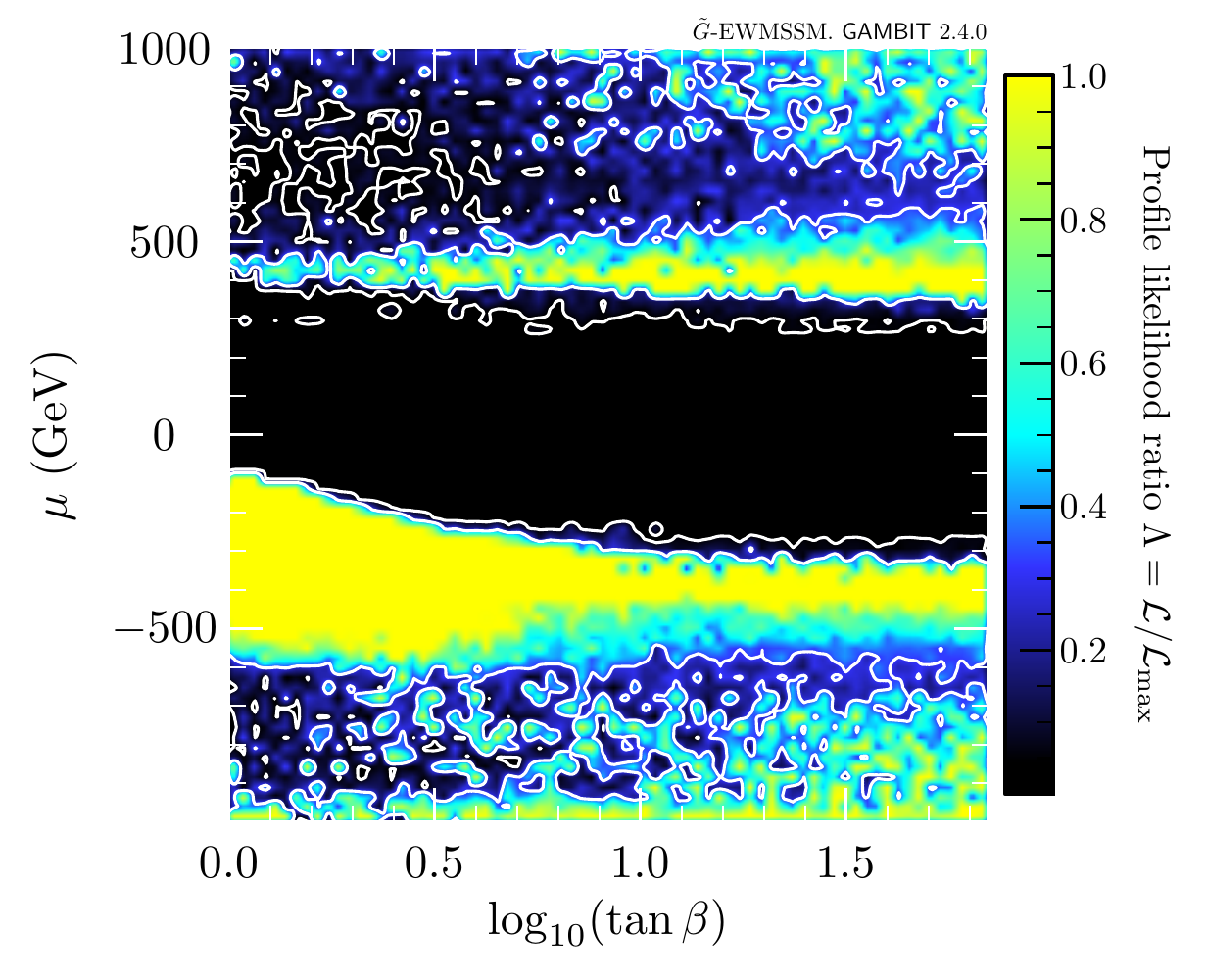}
  \caption{Profile likelihood maps for the $(M_1,\mu)$, $(M_2,\mu)$ and $(\tan\beta,\mu)$ parameter planes, using the full likelihood (top row) or the capped likelihood (bottom row). The white contour lines show the $1\sigma$ and $2\sigma$ confidence regions, and the white star in the top-row panels mark the point of highest likelihood. The region with $M_2 < 0$ (grey) in the middle panels is included only to ensure equal aspect ratio for all three mass parameters.}
  \label{fig:parameter_planes}
\end{figure*}
In Fig.~\ref{fig:parameter_planes} we show profile likelihood results for three different planes of the \GEWMSSM parameters. The panels in the top row show likelihood maps using the full likelihood, i.e.\ corresponding to the results in Fig.~\ref{fig:mass_planes_uncapped}. In the bottom row we show results for the same parameter planes using the capped likelihood (see Sec.~\ref{sec:simulated_LHC_searches}), corresponding to the results in Fig.~\ref{fig:mass_planes_capped}. As discussed in Sec.~\ref{subsec:best_fit_scenarios}, the highest-likelihood solutions are found for $|\mu| < |M_1|, M_2$, $\mu < 0$ and $\tan\beta$ close to 1. When $|\mu|$ is larger than $|M_1|$ or $M_2$, the likelihood is only very weakly dependent on $|\mu|$ and $\tan\beta$. This explains the patchiness of the capped profile likelihood in the bottom-right panel, since the set of high-likelihood scan samples (which pick out the required $M_1$ or $M_2$ values) is spread out across large regions in the $(\tan\beta,\mu)$ plane.

\section{LHC searches} \label{app:LHC_searches}

Below we give a short description of each 13\,TeV LHC search we include in our study, and point out which signal regions our simulation includes. A list of all the searches, along with the corresponding labels used in \colliderbit, is given in Table~\ref{tab:lhclike}. \\

\begin{table*}[h!]
  \centering
  \begin{tabular}{lll}
   \toprule
    Search label & Luminosity & Source \\
    \midrule

    \textsf{ATLAS\_2BoostedBosons} & 139\,\invfb & ATLAS hadronic chargino/neutralino search~\cite{ATLAS:2021yqv} \\

    \textsf{ATLAS\_0lep} & 139\,\invfb & ATLAS 0-lepton search~\cite{ATLAS:2020syg} \\

    \textsf{ATLAS\_0lep\_stop} & 36\,\invfb & ATLAS 0-lepton stop search~\cite{ATLAS:2017drc}\\
    
    \textsf{ATLAS\_1lep\_stop} & 36\,\invfb & ATLAS 1-lepton stop search~\cite{ATLAS:2017eoo}\\
    
    \textsf{ATLAS\_2lep\_stop} & 139\,\invfb & ATLAS 2-lepton stop search~\cite{ATLAS:2021hza}\\
    
    \textsf{ATLAS\_2OSlep\_Z} & 139\,\invfb & ATLAS stop search with Z/H final states~\cite{ATLAS:2020aci}\\

    \textsf{ATLAS\_2OSlep\_chargino} & 139\,\invfb & ATLAS 2-lepton chargino search~\cite{ATLAS:2019lff}\\

    \textsf{ATLAS\_2b} & 36\,\invfb & ATLAS 2-$b$-jet stop/sbottom search~\cite{ATLAS:2017avc}\\

    \textsf{ATLAS\_3b} & 24\,\invfb & ATLAS 3-$b$-jet Higgsino search~\cite{ATLAS:2018tti}\\
    
    \textsf{ATLAS\_3lep} & 139\,\invfb & ATLAS 3-lepton chargino/neutralino search~\cite{ATLAS:2021moa}\\

    \textsf{ATLAS\_4lep} & 139\,\invfb & ATLAS 4-lepton search~\cite{ATLAS:2021yyr}\\    

    \textsf{ATLAS\_MultiLep\_strong} & 139\,\invfb & ATLAS leptons + jets search~\cite{ATLAS:2019fag}\\
    
    \textsf{ATLAS\_PhotonGGM\_1photon} & 139\,\invfb & ATLAS 1-photon GGM search~\cite{ATLAS:2021ijy}\\

    \textsf{ATLAS\_PhotonGGM\_2photon} & 36\,\invfb & ATLAS 2-photon GGM search~\cite{ATLAS:2018nud}\\

    \textsf{ATLAS\_Z\_photon} & 80\,\invfb & ATLAS Z + photon search~\cite{ATLAS:2018vzq}\\

    \textsf{CMS\_0lep} & 137\,\invfb & CMS 0-lepton search~\cite{CMS:2019zmd}\\
    
    \textsf{CMS\_1lep\_bb} & 36\,\invfb & CMS 1-lepton + $b$-jets chargino/neutralino search~\cite{CMS:2017kyj}\\

    \textsf{CMS\_1lep\_stop} & 36\,\invfb & CMS 1-lepton stop search~\cite{CMS:2017gbz}\\

    \textsf{CMS\_2lep\_stop} & 36\,\invfb & CMS 2-lepton stop search~\cite{CMS:2017jrd}\\

    \textsf{CMS\_2lep\_soft} & 36\,\invfb & CMS 2 soft lepton search~\cite{CMS:2018kag} \\

    \textsf{CMS\_2OSlep} & 137\,\invfb & CMS 2-lepton search~\cite{CMS:2020bfa} \\ 

    \textsf{CMS\_2OSlep\_chargino\_stop} & 36\,\invfb & CMS 2-lepton chargino/stop search~\cite{CMS:2018xqw}\\

    \textsf{CMS\_2SSlep\_stop} & 137\,\invfb & CMS 2 same-sign lepton stop search~\cite{CMS:2020cpy}\\

    \textsf{CMS\_MultiLep} & 137\,\invfb & CMS multilepton chargino/neutralino search~\cite{CMS:2021cox-fix}\\

    \textsf{CMS\_photon} & 36\,\invfb & CMS 1-photon GMSB search~\cite{CMS:2017brl}\\

    \textsf{CMS\_2photon} & 36\,\invfb & CMS 2-photon GMSB search~\cite{CMS:2019vzo}\\

    \textsf{CMS\_1photon\_1lepton} & 36\,\invfb & CMS 1-photon + 1-lepton GMSB search~\cite{CMS:2018fon}\\

    \bottomrule
  \end{tabular}
  \caption{The different ATLAS and CMS searches we simulate for our LHC likelihood, with associated short-hand labels.}
  \label{tab:lhclike}
\end{table*}

\noindent {\bf The ATLAS search for electroweak production of charginos and neutralinos in final states with two boosted, hadronically-decaying bosons and missing transverse momentum~\cite{ATLAS:2021yqv}}: This search (\textsf{ATLAS\_2BoostedBosons}) targets the pair production of electroweakinos, where each of them is assumed to decay into the LSP and an on-shell $W$, $Z$ or SM Higgs boson. The mass difference between the produced electroweakinos and the LSP is assumed to be at least 400 GeV. The analysis is optimised on three different scenarios: 1) a baseline MSSM scenario where the produced electroweakinos and the LSPs can be either binos, winos or Higgsinos, 2) a general gauge mediation-inspired scenario in which the LSP is a gravitino and the heavier particles are Higgsinos and 3) a scenario with an axino LSP, where the heavier particles are assumed to be Higgsinos. Various simplified models are considered in each case. The analysis is performed in two fully-hadronic final states: the $qqqq$ final state arising from $W$/$Z$ bosons each decaying to light-flavour quarks/antiquarks, and the $bbqq$ final state which arises from a $Z$ or Higgs boson decaying to $b\bar{b}$ and a $W$ or $Z$ boson decaying to light-flavour quarks/antiquarks. The analysis uses events with at least two large-$R$ jets, and counts the $b$-multiplicity of each of these jets using a $b$-tagged track jet procedure. Boosted boson tagging algorithms are then defined to identify various SM boson decays in the two leading large-$R$ jets: $W_{qq}(Z_{qq})$-tagging targets $W(Z)\rightarrow qq$, whilst $Z_{bb}(h_{bb})$-tagging targets $Z(h)\rightarrow bb$. $V_{qq}$-tagging is used to denote the logical OR of $W_{qq}$- and $Z_{qq}$-tagging. Signal regions are then defined using the multiciplities of the different boson tags $n(W_{qq})$, $n(Z_{qq})$, $n(V_{qq})$, $n(Z_{bb})$ and $n(h_{bb})$. Additional background rejection is provided by selections such as a veto on $b$-jets that do not originate from the boosted boson candidates, lower bounds on the effective mass $m_{\text{eff}}$ (defined as the scalar sum of the $p_T$ of the two leading large-$R$ jets and $E_{\text{T}}^{\text{miss}}$), lower bounds on $E_{\text{T}}^{\text{miss}}$, cuts on an event shape variable, and a lower bound on the stransverse mass $m_{\text{T2}}$ constructed from the two leading large-$R$ jets. Our implementation of this search includes the signal regions \textsf{4Q-WW}, \textsf{4Q-WZ}, \textsf{4Q-ZZ} and \textsf{4Q-VV}. Due to difficulties with reproducing the $b$-tagging for small radius track jets we do not include the signal regions that rely on this.\\

\noindent {\bf The ATLAS search for gluino and squark production in final states with jets and missing transverse momentum~\cite{ATLAS:2020syg}:} This is the flagship ATLAS supersymmetry search for squarks and gluinos (\textsf{ATLAS\_0lep}), targeting events with multiple jets and significant missing transverse momentum. Although it is optimised on models of squark and gluino production, similar final states can be produced by electroweakino production with subsequent cascade decay processes that produce hadronically-decaying gauge bosons. We implement the optimised single-bin signal regions that are designed to present the ATLAS results in a model-independent way (\textsf{2j-1600}, \textsf{2j-2200}, \textsf{2j-2800}, \textsf{4j-1000}, \textsf{4j-2200}, \textsf{4j-3400}, \textsf{5j-1600}, \textsf{6j-1000}, \textsf{6j-2200} and \textsf{6j-3400}). The signal region selections include requirements on the multiplicity and transverse momenta of the jets in each event, the angular separation between the jets and the missing transverse momentum vector, the aplanarity, $E_{\text{T}}^{\text{miss}}/\sqrt{H_{\text{T}}}$ and $m_{\text{eff}}$. \\

\noindent {\bf The ATLAS search for top squarks in the jets plus missing transverse momentum final state~\cite{ATLAS:2017drc}:} This search (\textsf{ATLAS\_0lep\_stop}) seeks to uncover evidence of stop production in final states with four or more jets plus missing transverse momentum. Five sets of signal region are defined in the analysis, targeting different stop simplified models, with a range of different included sparticles and sparticle mass differences. The six \textsf{SRA} and \textsf{SRB} regions employ top-mass reconstruction to increase sensitivity to models in which the stop produces a top quark, which makes them less relevant for the scenario considered in this paper. The five \textsf{SRC} regions use recursive jigsaw variables to target regions with a small $\tilde{t}_1-\tilde{\chi}^0_1$ mass difference, the details of which are highly-dependent on the treatment of initial state radiation in the Monte Carlo generator used to model LHC events. We do not include these \textsf{SRC} regions due to known deficiencies of the Pythia initial state radiation model in this region. The two \textsf{SRD} regions are optimised for direct top squark production where both top squarks decay via $\tilde{t}\rightarrow b\tilde{\chi}^\pm_1$. At least five jets are required, two of which must be $b$-tagged, and further requirements are placed on the jet transverse momenta and the scalar sum of the transverse momenta of the two jets with the highest $b$-tag weights. Finally, the \textsf{SRE} signal region is designed for models with highly boosted top quarks. Requirements on the jet mass of reclustered fat jets are used, alongside requirements on the main discriminating variables $H_T$, $E_T^\text{miss}$ and $E_T^\text{miss}/\sqrt{H_T}$. \\

\noindent {\bf The ATLAS search for top squarks in final states with one lepton, jets plus missing transverse momentum~\cite{ATLAS:2017eoo}:} This search (\textsf{ATLAS\_1lep\_stop}) is optimised on simplified models of stop production with decays that produce one lepton (through a real or off-shell leptonically-decaying $W$ boson), and also on a dark matter model with a spin-0 mediator produced in association with two top quarks. All signal region are required to have exactly one signal lepton, and 2, 3 or 4 jets. Five regions labelled \textsf{tN} are optimised for the decay pattern $\tilde{t}\rightarrow t \tilde{\chi}^0_1$, using selections on variables such as the $am_{\text{T2}}$ variable, the transverse mass $m_T$ formed from the lepton and missing transverse momentum, the $H^{\text{miss}}_{\text{T,sig}}$\footnote{$H^{\text{miss}}_{\text{T,sig}}$ is defined as $H^{\text{miss}}_{\text{T,sig}}=\frac{|\vec{H}_\text{T}^{\text{miss}}|-M}{\sigma_{|\vec{H}_\text{T}^{\text{miss}}|}}$, where $\vec{H}_\text{T}^{\text{miss}}$ is the negative vectorial sum of the momenta of the signal jets and the lepton, $M=100$~GeV is an offset parameter, and the denominator is computed from the per-event jet energy uncertainties.} and the mass of a reconstructed hadronic top quark. Note that we do not include three signal regions that use a boosted decision tree in the definition of the signal region, since this is very difficult to reproduce outside of the ATLAS collaboration. Two additional signal regions, \textsf{bWN} and \textsf{bffN}, are dedicated to the three-body ($\tilde{t}\rightarrow bW\tilde{\chi}^\pm_1$) and four-body ($\tilde{t}\rightarrow bff'\tilde{\chi}^\pm_1$) decay searches. Six signal regions target various $\tilde{t}\rightarrow b\tilde{\chi}^\pm_1$ scenarios: three are optimised on a simplified model that assumes $m_{\tilde{\chi}^\pm_1}=2m_{\tilde{\chi}^0_1}$
(labels \textsf{bC2x\_diag}, \textsf{bC2x\_med}, \textsf{bCbv}), and three are designed to search for the case of a Higgsino LSP, in which the $\tilde{\chi}^\pm_1$, $\tilde{\chi}^0_2$ and $\tilde{\chi}^0_1$ are close in mass (labels \textsf{bCsoft\_diag}, \textsf{bCsoft\_med}, \textsf{bCsoft\_high}). In the latter case, the signature is characterised by low-momentum leptons or jets from highly off-shell $W$ or $Z$ bosons, and the analysis benefits from a dedicated soft lepton reconstruction. Finally, three extra signal regions (\textsf{DM\_low\_loose}, \textsf{DM\_low}, \textsf{DM\_high}) are optimised on the dark matter mediator model, with the analysis using similar variables to the regions targeting the decay  $\tilde{t}\rightarrow t \tilde{\chi}^0_1$. \\

\noindent {\bf The ATLAS search for top squarks in final states with two opposite-charge leptons and missing transverse momentum~\cite{ATLAS:2021hza}:} This search (\textsf{ATLAS\_2lep\_stop}) is optimised on similar models of direct stop production to the 0 lepton and 1 lepton searches. Events are required to have exactly two light leptons (electrons or muons) of opposite charge, with an invariant mass outside of the $Z$ boson mass window in the case of same flavour leptons. A series of discriminating variables are constructed from the missing transverse momentum and $p_T$ values of the leading leptons and jets, with other useful variables including a variant of $m_{\text{T2}}$ and the super-razor variables first defined in Ref.~\cite{Buckley:2013kua}. Various signal regions are optimised for 2-body, 3-body and 4-body stop decays. For the case of 2-body decays, the ATLAS analysis also defines a set of seven inclusive signal regions (labelled \textsf{SR2bInc}) intended to provide less model-specific sensitivity. Our implementation of the search uses this set of inclusive signal regions. \\

\noindent {\bf The ATLAS search for top squarks in events with a Higgs or $Z$ boson~\cite{ATLAS:2020aci}:} This search (\textsf{ATLAS\_2OSlep\_Z}) is optimised on various simplified models of top squark production in which a top squark decays to produce a Higgs or $Z$ boson. Top squark decays involving $Z$ bosons are targeted using a 3-lepton selection, with at least one same-flavour-opposite-sign pair (SFOS) whose invariant mass is consistent with $Z$ boson mass. Further selections are placed on the transverse momenta of the three leading leptons, the jet multiplicity, the $b$-jet multiplicity, the transverse momenta of the leading jet and $b$-jet, the missing transverse energy, a variant of $m_{\text{T2}}$ and the transverse momentum of the SFOS pair. Events containing Higgs bosons are targeted using a 1-lepton event selection, with further selections placed on the jet and $b$-jet multiplicity, the transverse mass formed from the lepton and the missing transverse momentum, and the missing transverse energy significance. In addition, a Higgs tagger built from a neural network is used to identify Higgs boson candidates, and events must contain at least one of them. Due to the difficulty of reproducing this Higgs tagging with sufficient accuracy, our implementation of this search covers only the 3-lepton final states (labels \textsf{SRZ1A}, \textsf{SRZ1B}, \textsf{SRZ2A}, \textsf{SRZ2B}). \\

\noindent {\bf The ATLAS search for charginos and sleptons in final states with two leptons and missing transverse momentum~\cite{ATLAS:2019lff}:} This search (\textsf{ATLAS\_2OSlep\_chargino}) is optimised on simplified models of slepton and chargino production, targeting chargino pair production with decays to lightest neutralinos and $W$ bosons, chargino cascade decays through sleptons to lightest neutralinos, and the direct production of slepton pairs. Events are required to have exactly two opposite-charge light leptons with an invariant mass greater than 100 GeV. Selected events must also have no $b$-tagged jets, and large values of $E_T^{\text{miss}}$ and  $E_T^{\text{miss}}$ significance. Further discrimination comes from the use of the $m_{\text{T2}}$ variable. Four sets of signal regions are defined (labels \textsf{SR-SF-0J}, \textsf{SR-SF-1J}, \textsf{SR-DF-0J}, \textsf{SR-DF-1J}) based on whether the leptons have the same or a different flavour, and whether the events have 0 or 1 non-$b$-tagged jets. From this, the ATLAS analysis defines a total of 16 inclusive signal regions, intended for more model-independent sensitivity, and a set of 36 signal regions with fine-grained binning in $m_{\text{T2}}$, to maximise sensitivity to the simplified model studied by ATLAS. In our study we use the inclusive signal regions. \\

\noindent {\bf The ATLAS search for bottom and top squarks in final states with two $b$-tagged jets and missing transverse momentum~\cite{ATLAS:2017avc}:} This search (\textsf{ATLAS\_2b}) is optimised on various simplified models of stop and sbottom production, targeting final states with 2 $b$-tagged jets, large missing transverse momentum and either zero leptons or one lepton. A long list of discriminating variables is used, including the minimum $\Delta\Phi$ between any of the leading jets and the missing transverse momentum vector, $H_{\text{T}}$ (defined as the scalar sum of the $p_T$ values of a subset of the jets in the event), $m_{\text{eff}}$, ratios of the missing transverse energy with $m_{\text{eff}}$ and $\sqrt{H_{\text{T}}}$, the contranverse mass, $am_{\text{T2}}$ and others. Three zero lepton signal regions and three one lepton signal regions are defined. Due to challenges in reproducing the cuts based on $am_{\text{T2}}$ our study only uses the zero lepton signal regions (labels \textsf{0L\_SRA350}, \textsf{0L\_SRA450}, \textsf{0L\_SRA550}, \textsf{0L\_SRB}, \textsf{0L\_SRC}). \\

\noindent {\bf The ATLAS search for Higgsinos in final states with at least three $b$-tagged jets~\cite{ATLAS:2018tti}:} This search (\textsf{ATLAS\_3b}) targets Higgsino production and decay in gauge-mediated supersymmetry scenarios, in which each Higgsino is assumed to decay to a Higgs boson and a gravitino. Two complementary analyses, targeting high- and low-mass signals, are performed. For the high-mass analysis, events with at least three $b$-tagged jets are selected, and jet pairs are assigned to two Higgs candidates. For the low-mass analysis, events with four $b$-jets are analysed by grouping the jets into Higgs candidates. Selections are placed on a number of kinematic variables including $m_{\text{eff}}$, $E_{\text{T}}^{\text{miss}}$, the mass of the Higgs boson candidates, angular variables and the minimum transverse mass formed with the missing transverse momentum vector and any of the leading four jets. Due to some overlaps between the signal regions for the low-mass and high-mass analyses, our analysis only uses the low-mass signal regions. From this analysis we have implemented all the 46 signal regions optimised for exclusion. \\

\noindent {\bf The ATLAS search for chargino-neutralino pair production in final states with three leptons and missing transverse momentum~\cite{ATLAS:2021moa}:} This search (\textsf{ATLAS\_3lep}) is optimised on two scenarios of electroweakino production. In the first, a $\tilde{\chi}_1^\pm$ and $\tilde{\chi}_2^0$ are produced (both wino-dominated), with subsequent decay to a bino-dominated $\tilde{\chi}_1^0$. In the second, the $\tilde{\chi}_1^\pm$, $\tilde{\chi}_2^0$ and $\tilde{\chi}_1^0$ are pure Higgsino states, and are therefore typically more mass degenerate (although an arbitrary mass hierarchy is assigned in order to define a parameter plane in which to optimise the analysis). The analysis has three dedicated selections to cover different mass regimes and assumptions, including an on-shell $WZ$ selection, an off-shell $WZ$ selection and a $Wh$ selection. All consider final states with exactly three leptons, possible ISR jets and $E_{\text{T}}^{\text{miss}}$. Events with at least one SFOS pair are divided into three bins of the SFOS pair invariant mass, $m_{ll}$, covering the regions below, on and above the $Z$ mass. Each $m_{ll}$ bin is further divided into $E_{\text{T}}^{miss}$ and $m_{\text{T}}$ bins, where the transverse mass $m_{T}$ is defined using the lepton that is not in the SFOS pair (and which can therefore be assumed to arise from a $W$ boson decay). Events are further separated by their jet multiplicity, and by two different variants of $H_\text{T}$, defined as the scalar sum of the transverse momenta of jets or leptons depending on the definition. Signal regions for events with a different-flavour-opposite-sign (DFOS) lepton pair are defined separately, using selections on the jet multiplicity, $E_{\text{T}}^{\text{miss}}$ significance, transverse momentum for the third-leading lepton, and the $\Delta R$ between the DFOS leptons and the same-flavour-same-sign lepton that is nearest in $\phi$. Our implementation includes 39 of the 41 signal regions targeting on-shell $WZ$ or $Wh$ production, leaving out the two regions \textsf{SR-Wh-DFOS-1} and \textsf{SR-Wh-DFOS-2} for which some cuts rely on object resolution variables that are not available in our fast event simulation framework. \\

\noindent {\bf The ATLAS search for gluino, electroweakino or slepton production in final states with four or more leptons~\cite{ATLAS:2021yyr}:} This search (\textsf{ATLAS\_4lep}) is optimised on different R-parity violating and R-parity conserving SUSY scenarios that can produce lepton-rich final states. The simplified model used for the R-parity conserving scenarios is a model with Higgsino production and a gravitino LSP, thus highly relevant for our study. Events are required to have four or more leptons (electrons, muons or hadronically-decaying taus). The signal regions for four-lepton events are classified by whether the events have four light leptons and zero taus, three light leptons and one tau, or two light leptons and two taus. Further selections are placed on the number of $b$-tagged jets, the presence or absence of a $Z$ boson, $E_T^{\text{miss}}$ and $m_{\text{eff}}$. A further signal region is also defined, requiring at least five light leptons, subject to no kinematic requirements. Our implementation includes all the zero-tau signal regions, i.e.\ the five-lepton region (label \textsf{SR5L}) and the seven regions with four light leptons (the regions with the \textsf{SR0} label). In particular, this includes the the two signal regions \textsf{SR0-ZZ-loose-bveto} and \textsf{SR0-ZZ-tight-bveto} designed to target the Higgsino plus gravitino scenario. \\

\noindent {\bf The ATLAS search for squarks and gluinos in final states with same-sign leptons and jets~\cite{ATLAS:2019fag}:} This search (\textsf{ATLAS\_MultiLep\_strong}) is optimised on simplified models of gluino and squark production, covering the case of both R-parity conversation and R-parity violation. Events are required to have two same-sign leptons and may contain additional leptons. In the case of the R-parity conserving search, large missing transverse momentum is required. Five signal regions are defined using the number of leptons and their relative electric charges, the number of jets and the number of $b$-tagged jets. Key kinematic variables used include the effective mass $m_{\text{eff}}$, $E_T^{\text{miss}}$ and its ratio to $m_{\text{eff}}$ and the invariant mass of same-sign electron pairs (which reduces contamination from $Z\rightarrow e^+ e^-$ decays where the charge of one electron is mismeasured). We include all five signal regions in our analysis. \\

\noindent {\bf The ATLAS search for gauge-mediated supersymmetry in final states with photons, jets and missing transverse momentum~\cite{ATLAS:2021ijy}:} This search (\textsf{ATLAS\_PhotonGGM\_1photon}) is optimised on a simplified model in which pair-produced gluinos decay to neutralinos, which in turn decay to a gravitino, at least one photon and jets. Three signal regions are defined which target the cases of large, medium and small mass differences between the gluino and neutralino, and all of them veto leptons in the selected events. Further selections are placed on the transverse momentum of the leading photon, the jet multiplicity, the angular separations of the jet and photon momenta with the missing transverse energy vector, $E_{\text{T}}^{\text{miss}}$, $H_{\text{T}}$ and a variable called $R_{\text{T}}^4$, defined as the ratio of the scalar sum of the $p_{\text{T}}$ for the four leading jets, and the scalar sum of the $p_{\text{T}}$ for all signal jets in the event. Our implementation includes all three signal regions. \\

\noindent {\bf The ATLAS search for photonic signatures from gauge-mediated supersymmetry models~\cite{ATLAS:2018nud}:} This search (\textsf{ATLAS\_PhotonGGM\_2photon}) is optimised on models of both strong and electroweak sparticle production, and targets final states with either a single photon and multiple jets, or two photons, plus significant missing transverse momentum in both cases. Discriminating variables include $m_{\text{eff}}$, $E_{\text{T}}^{\text{miss}}$, a variant of $H_{\text{T}}$, the angular separation between photons and the missing transverse momentum vector and $R_{\text{T}}^4$ (the scalar sum of the transverse momenta of the four leading jets divided by the scalar sum of the transverse momenta of all jets in the event). Our implementation contains all signal regions from the paper, but in our analysis we only use the two-photon signal regions (labels \textsf{SRaa\_SL}, \textsf{SRaa\_SH}, \textsf{SRaa\_WL}, \textsf{SRaa\_WH}), since the one-photon signal regions largely overlap with (and are superseded by) the signal regions in \textsf{ATLAS\_PhotonGGM\_1photon} above. \\

\noindent {\bf The ATLAS search for exotic decays of the Higgs boson to at least one photon and missing transverse momentum~\cite{ATLAS:2018vzq}:} This search (\textsf{ATLAS\_Z\_photon}) targets exotic decays of the Higgs boson to, for example, a gravitino and a lightest neutralino, with the neutralino subsequently decaying to a gravitino and a photon, This generates a final state with a single photon plus missing transverse energy, and one can reduce SM backgrounds by looking for events with a Higgs boson produced in association with a $Z$ boson. One can also generate final states with two photons if the Higgs boson decays to a pair of neutralinos that subsequently decay. Events are selected in the analysis if they have at least one photon, moderate $E_{\text{T}}^{\text{miss}}$ and two opposite-sign electrons or muons with an invariant mass within 10 GeV of the $Z$ boson mass (and no additional leptons). A $\gamma E_{\text{T}}^{\text{miss}}$ system is defined by performing the vector sum of the photon momentum or momenta and the missing momentum vector in the transverse plane. The search relies on two discriminating variables that quantify the angular separation of the two-lepton and $\gamma E_{\text{T}}^{\text{miss}}$ systems and the $p_T$ asymmetry of the two systems, and the analysis has only one signal region. \\

\noindent {\bf The CMS search for gluino and squark production in final states with multiple jets and missing transverse momentum~\cite{CMS:2019zmd}:} This is the flagship CMS search for gluino and squark production (\textsf{CMS\_0lep}), and is the CMS equivalent of the ATLAS search presented in Ref.~\cite{ATLAS:2020syg}. Search regions are defined in a four-dimensional space of variables given by the total number of jets, the number of $b$-tagged jets, the scalar sum of the jet $p_{\text{T}}$ values ($H_{\text{T}}$), and the magnitude of the vector $p_{\text{T}}$ sum of the jets ($H_{\text{T}^{\text{miss}}}$). In total, there are 174 exclusive signal region bins. For our analysis, we implement 12 aggregate search bins which are presented in an appendix of the paper, and which are constructed from the original search bins after taking correlations into account. \\

\noindent {\bf The CMS search for chargino and neutralino production in the WH final state~\cite{CMS:2017kyj}:} This search (\textsf{CMS\_1lep\_bb}) is optimised on simplified models of chargino and neutralino production, with subsequent decays to a lightest neutralino and either a $W$ or Higgs boson. Events are required to have an electron or muon, two $b$-tagged jets with an invariant mass close to the Higgs boson mass and significant missing transverse momentum. Discriminating variables include the transverse mass of the lepton-neutrino system, the contranverse mass, and the $E_{\text{T}}^{\text{miss}}$. The paper defines two signal regions, distinguished by the selection on $E_{\text{T}}^{\text{miss}}$. In our analysis we use both signal regions, and account for the correlated background uncertainties using the correlation coefficient provided by the CMS collaboration. \\

\noindent {\bf The CMS search for stop production in final states with one lepton~\cite{CMS:2017gbz}:} This search (\textsf{CMS\_1lep\_stop}) is optimised on simplified models of stop production, and targets events with a single isolated electron or muon, jets and large missing transverse momentum. Two sets of signal regions are defined; one for a large range of $\tilde{t}_1-\tilde{\chi}_1^0$ mass splittings, and one for compressed spectra. For the first set of 27 signal regions, events are selected based on the number of jets, the $E_{\text{T}}^{\text{miss}}$, the invariant mass of the lepton and the closest $b$-tagged jet and a special ``topness'' variable. For the second set of 4 signal regions, at least five jets are required, and with the highest $p_{\text{T}}$ jet must not be $b$-tagged since it is expected to arise from initial state radiation. Further selections are placed on the angular separations of the missing transverse momentum and the lepton/jets, plus the $p_{\text{T}}$ of the lepton. Our analysis makes use of a smaller set of six aggregated signal regions that are provided in an appendix of the paper. \\

\noindent  {\bf The CMS search for stop production and dark matter in final states with two opposite-charge leptons~\cite{CMS:2017jrd}:} This search (\textsf{CMS\_2lep\_stop}) is optimised on models of stop production, and on dark matter models with a scalar or pseudo-scalar mediator in which the mediator is produced in association with a pair of top quarks. Events are selected if they have exactly two leptons with opposite charge and, in the case of a same-flavour lepton pair, the invariant mass of the lepton pair must not be close to the $Z$ mass. Events must also have at least two jets, at least one $b$-tagged jet and a moderate amount of missing transverse momentum. Signal regions are defined in bins of three variables: two variants of $m_{\text{T2}}$, and $E_{\text{T}}^{\text{miss}}$, giving 13 signal regions, which are further split into 26 regions by separating events with same-flavour and different-flavour lepton pair. Our analysis includes all 26 signal regions, plus makes use of the covariance matrix provided by the CMS collaboration to account for the correlated background uncertainties. \\

\noindent {\bf The CMS search for charginos and neutralinos in final states with two low-momentum opposite-charge leptons~\cite{CMS:2018kag}:} This search (\textsf{CMS\_2lep\_soft}) is optimised on simplified models of chargino and neutralino production where the mass difference between the mass-degenerate $\tilde{\chi}_2^0$ and $\tilde{\chi}_1^\pm$ and the  $\tilde{\chi}_1^0$ is small, such that decays proceed via off-shell $W$ and $Z$ bosons. A separate series of signal regions targets stop production. Selected events in both cases must contain two opposite-charge leptons (of either the same or different flavour) with a low transverse momentum, moderate $E_{\text{T}}^{\text{miss}}$ and at least one jet. No $b$-tagged jets must be present and further selections are applied to variables such as the invariant mass and transverse momentum of the dilepton pair, $E_{\text{T}}^{\text{miss}}/H_{\text{T}}$, $H_{\text{T}}$ and the transverse masses form from the leptons and the missing transverse momentum. For the electroweakino search, signal regions are defined in bins of $E_{\text{T}}^{\text{miss}}$ and the dilepton invariant mass. For the stop search, signal regions are defined in bins of $E_{\text{T}}^{\text{miss}}$ and the transverse momentum of the leptons. We implement all of the signal regions in our analysis, and treat the correlated background uncertainties using the covariance matrices provided by CMS. \\

\noindent {\bf The CMS search for supersymmetry in final states with two opposite-sign same-flavour leptons and missing transverse energy~\cite{CMS:2020bfa}:} This search (\textsf{CMS\_2OSlep}) is optimised on various simplified models of gluino, squark, slepton and electroweakino production, and targets three potential signatures: 1) an excess of events with a lepton pair, whose invariant mass is consistent with the $Z$ boson mass, 2) a kinematic edge in the invariant mass distribution of the lepton pair and 3) non-resonant production of two leptons. A set of strong production signal regions is defined using selections on the jet and $b$-jet multiplicities, $H_{\text{T}}$, a variant of $m_{\text{T2}}$ and $E_{\text{T}}^{\text{miss}}$, with the signal regions binned in the latter of these variables. On-$Z$ electroweak production signal regions are defined using selections on the jet and $b$-jet multiplicities, the dijet invariant mass (sometimes defined using the $b$-jets), two different variants of $m_{\text{T2}}$ and $E_{\text{T}}^{\text{miss}}$. In addition, a set of boosted signal regions is defined by requiring that there is a large radius jet with $p_T>200$~GeV, consistent with a hadronically-decaying gauge boson. For the dilepton edge search, a first approach is based on a fit to the dilepton invariant mass using events that pass selections on $m_{\text{T2}}$ and $E_{\text{T}}^{\text{miss}}$. A second approach uses counts in various bins of $E_{\text{T}}^{\text{miss}}$, after applying other selections on $m_{\text{T2}}$ and the jet and $b$-jet multiplicities. An additional requirement is placed on a novel variable that characterises how ``$t\bar{t}$-like'' the events are. Finally, a set of slepton search regions are defined using selections on the jet and $b$-jet multiplicities, $m_{\text{T2}}$, $E_{\text{T}}^{\text{miss}}$ and the ratio of the transverse momenta of the sub-leading lepton with the leading jet. Our implementation includes all signal regions except the edge fit regions, due to difficulty of implementing these outside of the CMS collaboration. We use the covariance matrices provided by CMS to account for correlated background uncertainties. \\

\noindent {\bf The CMS search for charginos and stops in final states with two opposite-charge leptons~\cite{CMS:2018xqw}:} This search (\textsf{CMS\_2OSlep\_chargino\_stop}) is optimised on various simplified models of chargino and stop production and decay. Events are selected if they contain two opposite-charge electrons or muons, plus missing transverse momentum. For events with a same-flavour lepton pair, the invariant mass of the dilepton pair must not be close to the $Z$ mass. For the chargino search, signal regions are defined in bins of the $E_{\text{T}}^{\text{miss}}$, number of $b$-tagged jets, number of jets, same-flavour or different-flavour status of the leptons and $m_{\text{T2}}$. For the stop search, an extra requirement is added on the number of ``ISR jets'', defined as jets with $p_{\text{T}}>$150~GeV and no $b$-tag. We implement all of the chargino and stop regions, and make use of the covariance matrices provided by the CMS collaboration. \\

\noindent {\bf The CMS search for beyond-Standard Model physics in final states with jets and two same-sign or at least three charged leptons~\cite{CMS:2020cpy}:} This search (\textsf{CMS\_2SSlep\_stop}) is optimised on various simplified models of gluino production and decay, including both R-parity conserving and violating processes. Six exclusive categories of events are defined using preliminary selections on the lepton multiplicity and charge, plus $E_{\text{T}}^{\text{miss}}$. For events with two leptons, the leptons must have the same sign. For events with at least three charged leptons, separate categories are defined for the cases where there either is or is not an opposite-sign same-flavour pair with an invariant mass consistent with the $Z$ boson mass. The main search is performed using a very large number of binned regions, and we instead implement the set of 17 inclusive signal regions (labels \textsf{ISR1}--\textsf{ISR17}) that are designed for easier interpretation of the results.  These are defined in bins of the minimum transverse mass formed from either of the leptons and the missing transverse momentum, $E_{\text{T}}^{\text{miss}}$, the jet and $b$-tagged jet multiplicities and $H_{\text{T}}$.\\

\noindent {\bf The CMS search for electroweak production of charginos and neutralinos in final states with three or four leptons, up to two hadronically-decaying $\tau$ leptons or two same-sign light leptons~\cite{CMS:2021cox-fix}: }This search (\textsf{CMS\_MultiLep}) is optimised on a broad range of simplified models of chargino and neutralino production, including $\tilde{\chi_2}^0\tilde{\chi_1}^\pm$ production with decays to lightest neutralinos via intermediate sleptons or with Higgs, $W$ or $Z$ bosons, and $\tilde{\chi_1}^0\tilde{\chi_1}^0$ production with decays to gravitinos and $Z$ or Higgs bosons. A number of event categories are defined in the analysis, based on the lepton multiplicity, whether there is a same-sign lepton pair or same-flavour-opposite-sign lepton pair, and the hadronically-decaying $\tau$ lepton multiplicity. The analysis is performed using a large number of binned signal regions for each category. The key variables used are various variants of $m_{\text{T2}}$, the transverse momentum of the dilepton system in 2 lepton events, $E_{\text{T}}^{\text{miss}}$, the dilepton invariant mass of same-flavour-opposite-sign lepton pairs, the transverse mass formed from the trilepton system, dilepton system or a single lepton and the missing transverse energy vector, $H_{\text{T}}$, the $\Delta R$ separation between leptons, and the invariant mass formed from a light lepton and a hadronically-decaying $\tau$ lepton. Our analysis includes the signal regions labelled \textsf{SS01}--\textsf{SS20} (two same-sign leptons), \textsf{A01}--\textsf{A64} (three leptons, one SFOS pair), \textsf{B01}--\textsf{B03} (three leptons, no SFOS pair), \textsf{G01}--\textsf{G05} (four leptons, two SFOS pairs) and \textsf{H01}--\textsf{H03} (four leptons, one or zero SFOS pairs). \\

\noindent {\bf The CMS search for gauge-mediated supersymmetry in events with at least one photon and missing transverse momentum~\cite{CMS:2017brl}:} This search (\textsf{CMS\_photon}) is optimised on various simplified models of gauge-mediated supersymmetry breaking, in which neutralinos and charginos decay to produce gravitinos and photons, as well as $Z$, $W$ and Higgs bosons. Events are required to contain at least one high-energy photon plus large missing transverse momentum. The signal regions feature common selections on the transverse mass formed from the missing transverse momentum and the photon, and the $E_{\text{T}}^{\text{miss}}$. After these selections, four signal regions are defined as bins in $S_{\text{T}}^\gamma$, the scalar sum of $E_{\text{T}}^{\text{miss}}$ and the $p_{\text{T}}$ of all photons in the event. We implement all four signal regions in our analysis. \\

\noindent {\bf The CMS search for gauge-mediated supersymmetry in events with two photons and missing transverse momentum~\cite{CMS:2019vzo}:} This search (\textsf{CMS\_2photon}) is optimised on simplified models of squark and gluino production, with cascade decays terminating in lightest neutralinos that always decay to a photon and a gravitino. Events are required to have two photons and large missing transverse momentum, and to satisfy various selections on their electromagnetic activity. Six signal regions are defined by selecting different ranges of $E_{\text{T}}^{\text{miss}}$, and all of these are implemented in our analysis. \\

\noindent {\bf The CMS search for gauge-mediated supersymmetry in events with a photon, an electron or muon and large missing transverse momentum~\cite{CMS:2018fon}:} This search (\textsf{CMS\_1photon\_1lepton}) is optimised on simplified models of gauge-mediated supersymmetry breaking that include gluino and quark production, plus direct production of neutralinos and charginos. Events are required to have at least one photon and at least one electron or muon. In the case of more than one light lepton, the lepton with the highest transverse momentum is used in the analysis. Selections are placed on the transverse mass formed from the lepton plus missing transverse momentum, and the $E_{\text{T}}^{\text{miss}}$, and multiple signal regions are defined in bins of $E_{\text{T}}^{\text{miss}}$, the transverse momentum of the photon and $H_{\text{T}}$ (defined as the scalar sum of the jet $p_{\text{T}}$ values in the event). The CMS search defines 18 signal regions per lepton channel (i.e.\ electron or muon), and we implement all regions in our analysis.

\section{Code extensions} \label{app:codeextensions}

In this Appendix we describe the extensions to the \gambit framework introduced for this study, and how to use them. See Refs.\ \cite{gambit,GUM} for a general introduction to the \gambit framework and Ref.\ \cite{ColliderBit} for an introduction to the \colliderbit module.

\subsection{\gambit models with a light gravitino}

The first modification is the addition of a new family of models \bfsf{MSSMXatY\_mG}, mirroring the existing family of MSSM models \bfsf{MSSMXatY} in \gambit, supplemented with a new parameter \texttt{mG} which codifies the mass of a light gravitino. As in the existing \bfsf{MSSMXatY} models, \bfsf{X} refers to the number of parameters and \bfsf{Y} to the scale at which the parameters are defined (which itself could be a parameter in \bfsf{MSSMXatQ} models) (e.g.\ \bfsf{MSSM30atMGUT} has 30 parameters defined at the GUT scale). Models with alternate parametrisation are labelled with increasing alphabetical letters after the number of parameters (e.g.\ \bfsf{MSSM10batQ}) or in the specific case of reparametrisation with $m_A$ and $\mu$ instead of $m_{H_u}$ and $m_{H_d}$, models are labelled with the suffix \bfsf{\_mA} (e.g.\ \bfsf{MSSM30atMGUT\_mA}. The specific model used in the study is the \bfsf{MSSM11atQ\_mA\_mG}, defined as

\begin{description}
 \item[\bfsf{MSSM11atQ\_mA\_mG}: \label{MSSM11atQ_mA-mG}] \term{M1, M2, mu, TanBeta, Ad_3,} \term{Ae_3, Au_3, M3, Qin, mA, mG, ml2, mq2}

 An MSSM parametrisation with 11 parameters plus a gravitino mass \texttt{mG}, of which we vary only four in this study: \texttt{TanBeta}, \texttt{M1}, \texttt{M2} and \texttt{mu}. We fix the other parameters to \texttt{mG} = 1\,eV, \texttt{Ad\_3} = \texttt{Ae\_3} = \texttt{Au\_3} = 0, \texttt{M3} = \texttt{mA} = 5\,TeV and \texttt{ml2} = \texttt{mq2} = (3\,TeV)$^2$, to decouple all superpartners other than the electroweakinos and the gravtino from the low energy phenomenology. The input parameters are defined at a scale \texttt{Qin} = 3\,TeV. We choose to use a model parametrised with \texttt{mA} and \texttt{mu} instead of \texttt{mHu2} and \texttt{mHd2}, as $\mu$ controls the Higgsino masses.

\end{description}

\subsection{Additions to \decaybit}

\begin{table*}[tp]
\centering
 \caption{Capabilities and module functions added to \decaybit for this study. The }
 \scriptsize
 \makebox[\linewidth]{
 \begin{tabular}{p{5.4cm} p{4.8cm} p{6cm} }
  \toprule
  \textbf{Capability}  & \textbf{Function} (\textbf{type}) & \textbf{Dependencies [type] / \newline Backend reqs [type (args)]} \\ \midrule
   \cpp{neutralino\_}\py{i}\cpp{\_decay\_rates\_gravitino} &\cpp{neutralino\_}\py{i}\cpp{_decays\_gravitino} (\cpp{DecayTable::Entry}) & {\cpp {MSSM\_spectrum [Spectrum]} \newline \cpp{Z\_decay\_rates [DecayTable::Entry]}}  \\
   \midrule
   \cpp{chargino\_}\py{j}\cpp{\_decay\_rates\_gravitino} &\cpp{chargino\_}\py{j}\cpp{\_decays\_gravitino} (\cpp{DecayTable::Entry}) & \cpp {MSSM\_spectrum [Spectrum]} \newline \cpp{W\_plus\_decay\_rates [DecayTable::Entry]}  \\
   \midrule
   \cpp{neutralino\_}\py{i}\cpp{\_decay\_rates} &\cpp{neutralino\_}\py{i}\cpp{_decays\_all} (\cpp{DecayTable::Entry}) & \cpp{neutralino\_}\py{i}\cpp{\_decay\_rates\_gravitino [DecayTable::Entry]} \newline \cpp{neutralino\_}\py{i}\cpp{\_decay\_rates\_SH [DecayTable::Entry]}  \\
   \midrule
   \cpp{chargino\_}\py{j}\cpp{\_decay\_rates} &\cpp{chargino\_}\py{j}\cpp{\_decays\_all} (\cpp{DecayTable::Entry}) & \cpp{chargino\_}\py{j}\cpp{\_decay\_rates\_gravitino  [DecayTable::Entry]} \newline \cpp{chargino\_}\py{j}\cpp{\_decay\_rates\_SH  [DecayTable::Entry]} \\
   \midrule
  \end{tabular}
 }
 \label{tab:decays}
\end{table*}

The \gambit module \decaybit takes care of the computation of the decay widths of the various BSM particles~\cite{SDPBit}. As detailed on Section \ref{sec:model}, the main channel for the production of gravitinos is through the decays of the light electroweakinos. Therefore, for the purpose of this study we have implemented new capabilities and module functions for the computation of the decays of neutralinos and charginos to gravitinos, which can be seen in Table~\ref{tab:decays}. The capabilities \cpp{neutralino\_}\py{i}\cpp{\_decay\_rates\_gravitino}, where \py{i} runs through the neutralino eigenstates, i.e.\ \py{i}$=1,\dots,4$, compute the decay of each of the neutralinos to a gravitino and a $\gamma$, $h$ or $Z$, following eqs.~\eqref{chitogammaG}-\eqref{chitohG}. The capabilities \cpp{chargino\_}\py{j}\cpp{\_decay\_rates\_gravitino}, where \py{j} runs through the chargino eigenstates, i.e.\ \py{j}$=1,2$, compute the decay of each of the charginos to a gravitino and a (possibly off-shell) $W$-boson. 

In order to combine the newly added decays to gravitinos with the pre-existing decay channels of neutralinos and charginos, we implemented a new set of module functions that provide the capabilities \cpp{neutralino\_}\py{i}\cpp{\_decay\_rates} and \cpp{chargino\_}\py{j}\cpp{\_decay\_rates}, respectively. These module functions, also seen in Table~\ref{tab:decays} are called \cpp{neutralino\_}\py{i}\cpp{\_decays\_all} and \cpp{chargino\_}\py{j}\cpp{\_decays\_all}, and simply combine the decay tables computed by \texttt{SUSY-HIT}, via the capabilities \cpp{neutralino\_}\py{i}\cpp{\_decay\_rates\_SH} and \cpp{chargino\_}\py{j}\cpp{\_decay\_rates\_SH}, with the computation for the decays to gravitinos, via the capabilities introduced above, \cpp{neutralino\_}\py{i}\cpp{\_decay\_rates\_gravitino} and \cpp{chargino\_}\py{j}\cpp{\_decay\_rates\_gravitino}.

\subsection{Additions to \colliderbit}

\renewcommand\metavar\metavarf

\renewcommand\metavar\metavars
\newcommand\descwidtht{4.6cm}

\begin{table*}[tbp]
 \centering
 \scriptsize
 \begin{tabular}{l|p{\descwidtht}|l|l}
  \toprule
  \textbf{Capability}
   & \multirow{2}{*}{\parbox{\descwidtht}{\textbf{Function} (\textbf{return type}): \\  \textbf{brief description}}}
   & \textbf{Dependencies (\textbf{type})}
   & \textbf{Options} (\textbf{type})	\\ & & & \\
  \hline
  \cpp{L3_Gravitino_LLike}
   & \multirow{2}{*}{\parbox{\descwidtht}{\cpp{L3_Gravitino_LLike} (\cpp{double}): \\
    Computes the log-likelihood for an L3 search for multi-photons and MET.}}
   & \cpp{MSSM_spectrum} &\\
   & & \cpp{LEP207_xsec_chi00_11} &  \\
   & & \cpp{decay_rates} & \\
  \hline
  \cpp{HardScatteringEvent}
   & \multirow{2}{*}{\parbox{\descwidtht}{\cpp{generateEventPythia} \\
   (\cpp{Pythia8::Event}): \\
    Generates a \cpp{Pythia8::Event} using \pythiaeight.}}
   & \cpp{HardScatteringSim} & \cpp{drop_HepMC2_file}(\cpp{bool})\\
   & & \cpp{EventWeighterFunction} & \cpp{drop_HepMC3_file}(\cpp{bool}) \\
   & & & \\
   & & & \\
  \hline
  \cpp{HardScatteringEvent}
   & \multirow{2}{*}{\parbox{\descwidtht}{\cpp{generateEventPythia_HEPUtils}\\
    (\cpp{HEPUtils::Event})\\
    Generates a \cpp{Pythia8::Event} using \pythiaeight and converts it to \cpp{HEPUtils::Event}.}}
   & \cpp{HardScatteringSim} & \cpp{jet_pt_min} (\cpp{double})\\
   & & \cpp{HardScatteringEvent} & \\
   & & (\cpp{Pythia8::Event}) & \ \\
   & & \cpp{EventWeighterFunction} &  \\
   & & & \\
  \hline
  \cpp{HardScatteringEvent}
   & \multirow{2}{*}{\parbox{\descwidtht}{\cpp{generateEventPythia_HepMC}\\
    (\cpp{HepMC3::GenEvent})\\
    Generates a \cpp{Pythia8::Event} using \pythiaeight and converts it to \cpp{HEPUtils::Event}.}}
   & \cpp{HardScatteringSim} & \\
   & & \cpp{HardScatteringEvent} & \\
   & & (\cpp{Pythia8::Event}) & \ \\
   & & & \\
   & & & \\
  \hline
  \cpp{Rivet_measurements}
   & \multirow{2}{*}{\parbox{\descwidtht}{\cpp{Rivet_measurements} \\
   (\cpp{shared_ptr<ostringstream>}): \\
    Runs a \cpp{HardScatteringEvent} \\
    through \rivet. Outputs to \yoda \\
    \cpp{ostringstream} on \cpp{BASE_FINALIZE}.}}
   & \cpp{HardScatteringEvent} & \cpp{drop_YODA_file} (\cpp{bool})\\
   & & & \cpp{drop_used_analyses} (\cpp{bool}) \\
   & & & For each collider: \\
   & & & \hspace{2pt} \cpp{analyses} (\cpp{vector<string>})\\
   & & & \hspace{2pt} \cpp{exclude_analyses} (\cpp{vector<string>})\\
  \hline
  \cpp{LHC_measurements}
   & \multirow{2}{*}{\parbox{\descwidtht}{\cpp{Contur_LHC_measurements_from_} \\
    \phantom{x}\cpp{stream} (\cpp{class Contur_output}): \\
    Runs \contur on \yoda\xspace\cpp{ostringstream}.}}
   & \cpp{Rivet_measurements} & \cpp{contur_options} (\cpp{vector<string>})\\
   & & & Each option is equivalent to running\\
   & & & \contur with the \cpp{--option} flag.\\
   & & & \\
  \hline
  \cpp{LHC_measurements}
   & \multirow{2}{*}{\parbox{\descwidtht}{\cpp{Contur_LHC_measurements_from_} \\
    \phantom{x}\cpp{file} (\cpp{class Contur_output}):\\
    Runs \contur on \yoda file. }}
   &  & \cpp{contur_options} (\cpp{vector<string>})\\
   & & & Each option is equivalent to running\\
   & & & \contur with the \cpp{--option} flag.\\
  \hline
  \cpp{LHC_measurements_}
   & \multirow{2}{*}{\parbox{\descwidtht}{\cpp{Contur_LHC_measurements_} \\
    \phantom{x} \cpp{LogLike} (\cpp{double}):\\
    Extracts the log-likelihood from a \\
    \cpp{Contur_output} object. }}
   & \cpp{LHC_measurements} & \\
   \ \cpp{LogLike} & & & \\
   & & & \\
   & & & \\
  \hline 
  \cpp{LHC_measurements_}        
   & \multirow{2}{*}{\parbox{\descwidtht}{\cpp{Contur_LHC_measurements_} \\
    \phantom{x}\cpp{LogLike_perPool} \\
    (\cpp{map_str_double}): \\
    Extracts the log-likelihood contribution for each \contur analysis pool from a \cpp{Contur_output} object }}
   & \cpp{LHC_measurements} & \\
   \ \cpp{LogLike_perPool} & & & \\
   & & & \\
   & & & \\
   & & & \\
   & & & \\
  \hline 
  \cpp{LHC_measurements_}
   & \multirow{2}{*}{\parbox{\descwidtht}{\cpp{Contur_LHC_measurements_} \\ 
    \phantom{x}\cpp{histotags_perPool} (\cpp{map_str_str}): \\
    For debugging with the \cpp{cout} printer only. Extracts the tag of the dominant histogram(s) in each pool from a \cpp{Contur_output} object. }}
   & \cpp{LHC_measurements} & \\
   \ \cpp{histotags_perPool} & & & \\
   & & & \\
   & & & \\
   & & & \\
   & & & \\
   & & & \\
  \hline 
 \end{tabular}
 \caption{New \colliderbit capabilities for the \colliderbit added in this study, including a new LEP search, restructuring of the event generation capabilities and new capabilities for LHC measurements, using \rivet and \contur as backends. Note that the dependency type is only provided when there is ambiguity and that every \contur related function has a partner function with the prefix \cpp{Multi_}, which allows running different sets of \contur options in the same run.}
 \label{tab:colliderbit}
\end{table*}

\renewcommand\metavar\metavarf
\colliderbit~\cite{ColliderBit} is the \gambit module that handles anything related to collider physics,\footnote{With the exception of flavour, which is handled by \flavbit~\cite{FlavBit}.} such as LHC searches, LEP limits and Higgs measurements. It is naturally the most relevant module used in a collider-focused study such as this one. We extended \colliderbit significantly for this study, implementing new LHC analyses (Appendix~\ref{app:LHC_searches}), adding a new LEP limit, and introducing the ability to compute the likelihood for BSM models to agree with LHC SM-like measurements, using the new interfaces to \rivet and \contur.

For the most part in this study we reuse the pre-existing LEP upper limits on electroweakino production, which we re-interpret for the gravitino model. However, we added a new module function to compute the limit from an L3 multi-photon and missing energy search, which is exclusive of models such as this one. For this purpose we added a new module function, \cpp{L3_Gravitino_LLike}, providing a capability of the same name, to compute the 95\% CL upper limit on the cross section. The new capabilitiy and module function can be found in  Table~\ref{tab:colliderbit}.

The machinery for computing the likelihood for LHC measurements follows the structure of the existing \colliderbit, where a Monte Carlo event generator (e.g.\ \pythiaeight) is used to simulate hard scattering events at the LHC, each of those events is passed through a native detector simulation, then a collection of analyses, and a likelihood is computed from the resulting yields. In previous incarnations of \colliderbit, the events generated by \pythiaeight were immediately converted into a \cpp{HEPUtils::Event} type, which is needed for detector simulation and analysis of the events. However, to compute the predicted yields of a given simulated event for LHC SM measurements in \rivet, the event must be provided as a HepMC event (\cpp{HepMC3::GenEvent} type). This is done by splitting the existing module function \cpp{generateEventPythia}, which provides the capability \cpp{HardScatteringEvent}, into three parts, one that returns a \cpp{Pythia8::Event}, carrying the original name, one that converts the event to a \cpp{HEPUtils::Event}, called \cpp{generateEventPythia_HEPUtils}, and one that converts the event into a \cpp{HepMC3::GenEvent}, called \cpp{generateEventPythia_HepMC}. With this, each generated event can be converted to whichever format it is needed in for native \colliderbit LHC search analyses, or for SM measurements using \rivet. This new structure of module functions can be seen in Table~\ref{tab:colliderbit}.

Lastly, some of the most relevant additions to \colliderbit are module functions to compute the likelihood for LHC SM measurements, using \rivet and \contur. The new module function \cpp{Rivet_measurements} provides an eponymous capability and analyses a given HepMC event using \rivet's measurements. Its output, a \yoda\footnote{\href{https://yoda.hepforge.org/}{https://yoda.hepforge.org/}} analysis object written as a stream, is then used by the module function \cpp{Contur_LHC_measurements_from_stream}, which provides the capability \cpp{LHC_measurements} and that runs the \yoda analysis object through \contur to compute the likelihood. An additional module function, \cpp{Contur_LHC_measurements_from_file} provides the same capability and performs the same computation but starting from a \yoda analysis object read from a file. Both of these functions return a \cpp{Contur_object} structure which contains the output from \contur, including the total likelihood value as well as the likelihood value per pool. These respective values are extracted by the module functions \cpp{Contur_LHC_measurements_LogLike} and \cpp{Contur_LHC_measurements_LogLike_perPool}, which provide the capabilities \cpp{LHC_measurements_LogLike} and \cpp{LHC_measurements_LogLike_perPool}, respectively. For debugging purposes only, there exists an additional module function \cpp{Contur_LHC_measurements_histotags_perPool}, providing the capability \cpp{LHC_measurements_histotags_perPool} that extract the tag of the dominant histogram(s) in each pool. Descriptions of these capabilities and module functions can be found in Table \ref{tab:colliderbit}. Finally, each of the module functions that use \contur has a partner function that allows running with multiple sets of \contur options simultaneously, and these are called \cpp{Multi_Contur_LHC_measurements_from_stream}, \cpp{Multi_Contur_LHC_measurements_LogLike_perPool},  \cpp{Multi_Contur_LHC_measurements_histotags_perPool}, \cpp{Multi_Contur_LHC_measurements_LogLike_single} and \cpp{Multi_Contur_LHC_measurements_LogLike_all}. The \cpp{single} variation of the \cpp{LogLike} function is required to provide a single one of the multiple likelihoods to \colliderbit for combination with LHC search likelihoods, whereas the \cpp{all} version -- which comes under its own dedicated \cpp{LHC_measurements_LogLike_Multi} capability -- can be used to collect and save all the likelihoods for examination after the run.

\subsection{New backend interfaces: \rivet and \contur}

One the major breakthroughs of this study is the simultaneous combination of LHC searches and measurements in the analysis. This is possible due to the newly developed interfaces to \rivet 3.1.5~\cite{Buckley:2010ar}, which contains the extensive library of measurements, and \contur 2.1.1~\cite{Butterworth:2016sqg}, which takes care of the rigorous statistical combination of the results from individual measurements.

\rivet is written in \Cpp, so its interface to \gambit is auto-generated using the package \BOSS~\cite{gambit}, to allow dynamic loading of \rivet classes. Although some modifications of \BOSS were needed to interface \rivet, these will not be documented here, but in a future publication. This interface provides \gambit with access to the classes inside of \rivet and some of its global functions. In particular, the only class of interest for the interface is the \cpp{AnalysisHandler} class, which handles the analysis of the passed HepMC events. The usage of this class in \colliderbit was described above. Note that at the time of writing, \rivet 3.1.5 and, in particular, the class \cpp{AnalysisHandler}, is not threadsafe. This class must therefore be used within a \cpppragma{critical} section. In addition to the aforementioned class, \gambit also uses the global function \cpp{addAnalysisLibPath}. This function is just called once at the beginning of the scan by the backend initialisation function to inform \rivet of the location of the analysis library.

The interface to the \texttt{Python} package \contur opens the possibility to compute the likelihood for LHC measurements from a \yoda analysis object generated by \rivet. Two backend convenience functions were implemented for this purpose: \cpp{Contur_LogLike_from_file}, which reads the analysis object from a \yoda file, and \cpp{Contur_LogLike_from_stream}, which reads it from a standard \cpp{stringstream}. Both of these functions provide the same capability \cpp{Contur_Measurements}, but with different signatures, as the former takes just a string with the \yoda file path, and the latter a shared pointer to the stream. In addition, both functions take as an argument a list of options for \contur (see \cite{Butterworth:2016sqg} for a list of useful options). The value returned by both functions is an object of the class \cpp{Contur_output}, which is a simple class designed to manage the \texttt{Python} dictionary produced by \contur. Lastly, the backend convenience function \cpp{Contur_get_analyses_from_beam}, which provides the capability \cpp{Contur_GetAnalyses}, is used to inform \rivet of the analyses known to \contur.

\subsection{Adaptations to \rivet and \contur}

Besides the implementation of the interfaces to \rivet and \contur on the side of \gambit, minor modifications of each of these packages were necessary to adapt them to the \gambit workflow.

Typically, \contur is run on \yoda files that have been generated in a separate \rivet run. However, in a high-performance computing environment, the cost of writing and reading from a \yoda file at each parameter point would be prohibitive. Therefore, \rivet and \contur were adapted so that the \yoda file could be passed between them in memory via \cpp{stringstream}.

In \rivet, this means the  \cpp{AnalysisHandler} class received a new overload to the \cpp{write} method to output to \cpp{stringstream}. This is also available in the Python interface, and became available as of \rivet 3.1.4.

The changes to \contur were more significant. The \cpp{main} run function now takes a dictionary of arguments, and if the \cpp{"YODASTREAM"} term contains a Python \cpp{StringIO} --- which can be converted to and from a \texttt{C++} \cpp{stringstream} --- then \contur will run on that stream. When run like this, the \contur main will return a dictionary containing various statistics. Which outputs appear in the dictionary is controlled by the \cpp{"YODASTREAM_API_OUTPUT_OPTIONS"} argument, and the options here are summarised by Table~\ref{tab:contur_stat_types}. These features first appeared in \contur 2.1.1.

\renewcommand\metavar\metavars
\newcommand\descwidth{6cm}

\begin{table*}[tbp]
 \centering
 \scriptsize
 \begin{tabular}{l|p{\descwidth}}
  \toprule
  \textbf{Keyword} & {\textbf{Description} } \\
  \hline
  \cpp{"LLR"} & The log-likelihood ratio for the current point.  Note that the \contur convention differs from that in equation~\ref{eq:contur_lnlike}  by a factor of $-1/2$.\\
  \hline
  \cpp{"CLs"} & The CLs exclusion (in \%) for the current point. \\
  \hline
  \cpp{"Pool_LLR"} & The LLR from each individual \contur pool (as a dictionary). \\
  \hline
  \cpp{"Pool_CLs"} & The CLs exclusion from each individual \contur pool (as a dictionary). \\
  \hline
  \cpp{"Pool_tags"} & The dominant statistical contribution to each pool (i.e.\ a specific histogram bin, or specific set of correlated histogram bins or histograms), returned as a \cpp{str, str}  dictionary.  \\
  \hline 
 \end{tabular}
 \caption{Statistics that can be returned by \contur when running on a Python \cpp{StringIO}. \label{tab:contur_stat_types}}
\end{table*}

\end{appendices}


\bibliography{refs_1, refs_2, refs_contur_analyses}
\end{document}